\documentclass[a4paper,11pt]{article}
\pdfoutput=1
\usepackage[utf8]{inputenc}

\usepackage[no-natbib-sort]{jheppub}
\usepackage{enumitem}

\usepackage{hyperref}
\usepackage{fancybox}
\usepackage{float}
\usepackage{amsfonts}
\usepackage{amsmath}
\usepackage{amsbsy}
\usepackage{graphicx}
\usepackage{amssymb}
\usepackage{amsthm}
\usepackage{bm}
\usepackage{comment}
\usepackage{epsfig}
\usepackage{latexsym}
\usepackage{pdflscape}
\usepackage{cancel}
\usepackage{color}
\usepackage{cleveref}
\usepackage{bm,nicefrac}

\makeatletter
\def\@fpheader{\relax}
\makeatother

\DeclareSymbolFont{AMSa}{U}{msa}{m}{n}
\DeclareSymbolFont{AMSb}{U}{msb}{m}{n}
\DeclareMathSymbol{\fieldR}{\mathalpha}{AMSb}{"52}

\newcommand{\beq}{\begin{eqnarray}}
\newcommand{\eeq}{\end{eqnarray}}
\newcommand{\bea}{\begin{eqnarray}}
\newcommand{\eea}{\end{eqnarray}}
\newcommand{\be}{\begin{equation}}
\newcommand{\ee}{\end{equation}}
\newcommand{\bq}{\begin{equation}}
\newcommand{\eq}{\end{equation}}

\newcommand{\eqn}[1] {(\ref{#1})}

\def\phit{\tilde \phi}
\def\calE{{\mathcal{E}}}
\def\calP{{\mathcal{P}}}
\def\calPeq{{\mathcal{P}_{\rm eq}}}
\def\calPH{{\mathcal{P}}}
\def\calPinf{{\mathcal{P}_{\infty}}}

\def\calPhyd{{\mathcal{P}^{{\rm hydro}}}}

\def\Ainf{{A_\infty}}
\def\Sinf{{S_\infty}}

\def\O{{\mathcal{O}}}
\def\LIR{L_{\rm IR}}
\def\mIR{{m^2_{\rm IR}}}
\def\phiM{\phi_M}

\def\dd{{\rm d}}

\def\6{\partial}

\newcommand{\im}{\mathrm{Im}}

\def\6{\partial}

\usepackage{color}
    \definecolor{darkgreen}{rgb}{0,0.5,0}
    \definecolor{darkblue}{rgb}{0,0,0.6}
    \definecolor{purple}{rgb}{0.4,.2,0.7}

\newcommand{\sect}[1]{Sec.~\ref{#1}}

\title{Strong-coupling dynamics and entanglement in de Sitter space}

\author[a]{Jorge Casalderrey-Solana,}
\author[b]{Christian Ecker,}
\author[a,c]{David Mateos,}
\author[d]{and Wilke van der Schee}

\affiliation[a]{Departament de F\'\i sica Qu\`antica i Astrof\'\i sica and Institut de Ci\`encies del Cosmos (ICC), \\ Universitat de Barcelona, Mart\'\i\  i Franqu\`es 1, ES-08028, Barcelona, Spain}
\affiliation[b]{Institut f\"ur Theoretische Physik, Goethe Universit\"at, Max-von-Laue-Str.~1, 60438 Frankfurt am Main, Germany}
\affiliation[c]{Instituci\'o Catalana de Recerca i Estudis Avan\c cats (ICREA), Passeig Llu\'\i s Companys 23, \\ES-08010, Barcelona, Spain.}
\affiliation[d]{Theoretical Physics Department, CERN, CH-1211 Gen\`eve 23, Switzerland}

\emailAdd{jorge.casalderrey@ub.edu}
\emailAdd{dmateos@fqa.ub.edu}
\emailAdd{ecker@itp.uni-frankfurt.de}
\emailAdd{wilke.s@cern.ch}

\preprint{ICCUB-20-026, CERN-TH-2020-195}
\begin{abstract}

\end{abstract}

\abstract{
We use holography to study the dynamics of a strongly-coupled gauge theory in four-dimensional de Sitter space with Hubble rate $H$.
The gauge theory is non-conformal with a characteristic mass scale $M$.
We solve Einstein's equations numerically and determine the time evolution of homogeneous gauge theory states.
If their initial energy density is high compared with $H^4$ then the early-time evolution is well described by viscous hydrodynamics with a non-zero bulk viscosity.
At late times the dynamics is always far from equilibrium.
The asymptotic late-time state preserves the full de Sitter symmetry group and its dual geometry is a domain-wall in AdS$_5$.
The approach to this state is characterised by an emergent relation of the form $\mathcal{P}=w\,\mathcal{E}$ that is different from the equilibrium equation of state in flat space.
The constant $w$ does not depend on the initial conditions but only on $H/M$ and is negative if the ratio $H/M$ is close to unity.
The event and the apparent horizons of the late-time solution  do not coincide with one another, reflecting its non-equilibrium nature.
In between them lies an ``entanglement horizon'' that cannot be penetrated by extremal surfaces anchored at the boundary, which we use to compute the entanglement entropy of boundary regions.
If the entangling region equals the observable universe then the extremal surface coincides with a bulk cosmological horizon that just touches the event horizon, while for larger regions the extremal surface probes behind the event horizon.
}

\begin{document} 
\maketitle
\flushbottom

\section{Introduction}
Understanding the dynamics of non-Abelian gauge theories beyond the weak-coupling limit is an important challenge.
In this regime a quasi-particle description is likely not applicable and one must resort to a different intuition in order to understand the physics, especially out of equilibrium.
Holography provides a powerful framework with which a variety of theories can be analysed from first principles in this regime.
In this context the quasi-particle intuition is replaced by intuition based on higher-dimensional gravity, black hole horizons, etc.
In the case in which the gauge theory dynamics takes place in flat space, holography has provided valuable qualitative insights into the properties of Quantum Chromodynamics (QCD), especially into the far-from-equilibrium dynamics of its deconfined phase (see e.g.~\cite{CasalderreySolana:2011us} and references therein).
These properties are explored experimentally via the small drops of deconfined QCD matter that are created in Heavy Ion Collisions (HIC) (for a recent review, see \cite{Busza:2018rrf}).
Since these little fireballs are violently produced at an initial temperature just a few times the QCD deconfinement temperature, the physics immediately after the collision is non-weakly coupled and far from equilibrium.
One of the insights provided by holography is that the system becomes well described by hydrodynamics at a time at which the viscous corrections are still very large \cite{Heller:2011ju,Chesler:2010bi,Casalderrey-Solana:2013aba}.
During its subsequent evolution the fireball expands, cools down and eventually hadronises.
In QCD in equilibrium, this transition is realised as a smooth crossover \cite{Aoki:2006we}. 

Extending the analysis to the dynamics of gauge theories in curved spacetime is interesting from several viewpoints.
At a theoretical level, the spacetime curvature may lead to new effects and hence richer dynamics.
Phenomenologically, one motivation comes from Cosmology, where the dynamics of the gauge theory is coupled to an expanding spacetime.
This situation was certainly realised about one microsecond after the Big Bang, when the decreasing temperature of the Universe crossed the QCD critical temperature and quarks and gluons became bound into hadrons.
Another interesting scenario comes from the possibility that the physics beyond the Standard Model might be completed at some high-energy scale by a Grand Unified non-Abelian gauge Theory (GUT).
In this scenario there may be implications for the early Universe, and out-of-equilibrium effects may arise if, for example, the GUT theory undergoes a phase transition (see for example   \cite{Schwaller:2015tja,Caprini:2015zlo}).
Finally, it has recently been suggested that dark matter may be strongly self-interacting (recent reviews include \cite{Kribs:2016cew,Tulin:2017ara}), in which case a complete understanding of the dark sector would require going beyond perturbative methods.  

This paper is an exploratory investigation aimed at understanding the dynamics of strongly coupled matter in a cosmological context via holography.
We emphasize that, unlike in e.g.~\cite{Strominger:2001pn,McFadden:2009fg}, our goal is not to provide a dual holographic description of the cosmological gravitational field but only of the strongly coupled matter that lives in this background, as in e.g.~\cite{Koyama:2001rf, Marolf:2010tg, Ghoroku:2012vi,Buchel:2016cbj,Buchel:2017lhu,Buchel:2017pto,Buchel:2019qcq,Buchel:2019pjb}.
We will therefore assume that the four-dimensional gravitational field is prescribed a priori and use five-dimensional gravity to describe only the dynamics of the four-dimensional gauge theory.
We also stress that, at this very early stage, we are mainly motivated by theoretical curiosity, with the phenomenological motivation being mostly inspirational.
For this reason we will not be guided by an attempt to describe a realistic scenario but rather make a number of simplifying assumptions. 

The first one is that we will ignore the backreaction of the matter on the expanding metric.
The second one is that we will consider the simplest possible expanding geometry, namely de Sitter (dS) space.
The third simplification is that we will restrict our attention to spatially homogeneous states.
And the fourth one concerns the gauge theory that we will study. This will be defined by the condition that it be a four-dimensional, non-conformal theory with the simplest possible gravity dual.
The non-conformal nature of the theory is absolutely crucial in order to uncover the physics that we are interested in.
The reason is that de Sitter space is conformal to Minkowski space. Roughly speaking this means that, up to the effect of the conformal anomaly, the physics of a conformal theory in dS is the same as in flat space \cite{Apostolopoulos_2009}.  

Despite these simplifications, we will still be able to capture several novel effects.
These include the fact that at late times the apparent and the event horizons do not coincide, reflecting the non-equilibrium nature of the state, or the existence of an entanglement horizon in between them that cannot be penetrated by extremal surface anchored at the boundary.

The rest of the paper is structured as follows.
In section \ref{sec:HM} we introduce the holographic model.
In section \ref{sec:flat} we review the thermodynamic and transport properties of the model on flat space and discuss how we fix ambiguities due to anomalies in the hydrodynamic approximation.
In section \ref{sec:DdSs} we introduce the numerical algorithm we use to solve the dual gravity problem, explain how we construct initial states and  analyse our results for the time evolution of the model on de Sitter space, the properties of late time states and how they are approached.
In section \ref{sec:entropy0} we discuss entanglement and horizon entropies.
We end with a summary and discussion in section \ref{sec:discussion}.

\section{Holographic model}
\label{sec:HM}
We follow a bottom-up approach and use five-dimensional Einstein-dilaton gravity with non-trivial potential to model the dynamics of a strongly coupled field theory with broken conformal symmetry in four dimensions.
We will consider the same holographic model as in \cite{Attems:2016ugt}.
That reference explored the thermodynamics and the transport properties of the dual gauge theory in flat space.
We will review these properties in \sect{sec:flat}. 
In the current section we will focus on the extension that is needed on the holographic side in order to describe the dynamics in de Sitter space.

The action of the holographic model is given by
\begin{equation}\label{action}
S=\frac{2}{8\pi G}\int_{\mathcal{M}} d^5x\sqrt{-g}\left(\frac{1}{4}R[g]-\frac{1}{2}(\partial\phi)^2-V(\phi)\right)+\frac{1}{8\pi G}\int_{\partial\mathcal{M}}d^4x\sqrt{-\gamma}K+S_{ct}\,.
\end{equation}
Here $G$ is the five-dimensional Newton's constant, $R[g]$ is the Ricci scalar associated to the five-dimensional bulk metric $g_{\mu\nu}$ on $\mathcal{M}$, $\gamma_{ij}$ is the metric induced  on a four-dimensional slice near the boundary  $\partial \mathcal{M}$, and 
\be
K=\gamma^{ij}K_{ij}=\gamma^{ij}\nabla_i n_j
\ee
is the trace of the extrinsic curvature $K_{ij}$ associated to this slice.
The second term on the right-hand side of \eqn{action} is the familiar Gibbons--Hawking term.
The third term in \eqn{action} will be described below. The equations of motion take the form 
\begin{subequations}\label{EOM}
\begin{align}
R_{\mu\nu}-\frac{1}{2}R g_{\mu\nu}&=2\partial_\mu\phi\partial_\nu\phi-2 V(\phi) g_{\mu\nu}-(\partial\phi)^2 g_{\mu\nu}\label{EOM3}\,,\\
\nabla^2\phi&= \frac{\partial V}{\partial\phi}\label{EOM1}\,.
\end{align}
\end{subequations}

The potential $V(\phi)$ encodes the properties of the dual gauge theory.
We wish to choose the simplest possible potential with the following two properties: (i) it describes a \emph{non}-conformal theory, and (ii) the vacuum of the theory in flat space is described by a completely regular solution on the gravity side.
Following Ref.~\cite{Attems:2016ugt} we therefore choose the potential 
\be
\label{eq:pot}
L^2 V(\phi)=-3 -\frac{3}{2} \phi^2 - \frac{1}{3} \phi^4 + \left( \frac{1}{3 \phi_M^2} +  \frac{1}{2 \phi_M^4}\right) \phi^6-\frac{1}{12 \phi_M^4} \phi^8 \,,
\ee
which can be derived from the superpotential 
\be
\label{eq:defWs}
L\, W\left(\phi \right)=-\frac{3}{2} - \frac{\phi^2}{2} + \frac{\phi^4}{4 \phi^2_M} 
\ee
via the relation
\be
\label{eq:defW}
V(\phi)= -\frac{4}{3} W\left(\phi \right)^2 + \frac{1}{2} W'\left(\phi\right)^2 \,. 
\ee 
$L$ is a length scale.
The dimensionless constant $\phiM$ is a free parameter that controls the degree of non-conformality of the model, for example the maximum value of the bulk viscosity.
For concreteness, in this paper we will choose
\be
\phiM=2 \,.
\ee
Both $V(\phi)$ and $W(\phi)$ have a maximum at $\phi=0$ and a minimum at $\phi=\phiM$.
Each of these extrema yields an AdS solution of the equations of motion with constant $\phi$ and radius $L^2=-3/V(\phi)$.
In the gauge theory each of these solutions is dual to a fixed point of the Renormalisation Group (RG) with a number of degrees of freedom $N^2$ proportional to $L^3/G$.
In top-down models this relation is known precisely.
For example, in the case in which the gauge theory is $\mathcal{N}=4$ SYM with $N$ colours we would have 
\be
\label{defN}
\frac{L^3}{8\pi G}= \frac{N^2}{4\pi^2} \,.
\ee
In our bottom-up model we will take this as a definition of the number of degrees of freedom in the gauge theory, $N$, at each fixed point.

The potential \eqn{eq:pot} leads to three important properties of the model:
First, the resulting geometry is asymptotically AdS$_5$ in the UV with radius $L$, since $V(0)=-3/L^2$.
Second, the second derivative of the potential at $\phi=0$ implies that, in this asymptotic region, the scalar field has mass $m^2=-3/ L^2$.
Following the standard quantisation analysis this means that, in the UV, this field is dual to an operator in the gauge theory, $\hat{\O}$, with scaling dimension $\Delta_\textrm{UV}=3$.
The value of the source $M$ of this operator introduces a scale responsible for the breaking of conformal invariance. 
Third, the solution near $\phi=\phi_M$ is again AdS$_5$ with a different radius
\be
\label{eq:LIR}
L_{\rm IR}= \sqrt{- \frac{3}{V\left(\phi_M\right)}} = \frac{1}{1+ \frac{1}{6} \phi_M^2} L \, .
\ee
In this region the effective mass of the scalar field differs from its UV value and it is given by 
\be
\label{eq:MIR}
\mIR= \frac{12}{ L ^2} \left(1+\frac{1}{9} \phi_M^2 \right)= \frac{12}{L^2_{IR}} \frac{\left(1+\frac{1}{9} \phi_M^2 \right)}{\left(1+\frac{1}{6} \phi_M^2 \right)^2}\,. 
\ee
As a consequence, the operator $\hat{\O}$ at the IR fixed point has dimension 
\be
\label{IRdim}
\Delta_\textrm{IR}=2 + 2\sqrt{ 1+ \frac{\mIR \LIR^2}{4}}=
6\, \left( 1+\frac{\phiM^2}{9}\right) \left(1+\frac{\phiM^2}{6} \right)^{-1}\,.
\ee
As we will review in \sect{sec:flat}, when the gauge theory is placed in flat space there exists an RG flow between the UV and the IR fixed points.
The crossover takes place at the scale $M$ and the geometry dual to the entire flow is completely regular.
In most expressions below we will fix the radius of the UV AdS solution to unity, i.e., we will set $L=1$.

In order to understand the UV properties of the theory, such as anomalies, UV divergences, etc. we will solve the Einstein's equations near the boundary in a power expansion in the so-called Fefferman--Graham (FG) coordinate $\rho$.
This has dimensions of (length)$^2$ and in terms of it the near-boundary metric takes the form
\begin{equation}\label{metricFG}
ds^2= \frac{\dd\rho^2}{4\rho^2}+\gamma_{ij}(\rho,x)\dd x^i \dd x^j \,. 
\end{equation}
The boundary is located at $\rho=0$ and is parametrised by the coordinates $x^i$ with \mbox{$i=0, \ldots, 3$}.
Near the boundary the metric and the scalar field take the form
\begin{subequations}
\label{seriesVac}
\begin{align}
\gamma_{ij}(\rho,x)&=\frac{1}{\rho} \Bigg\{
g_{(0)ij}(x) + \rho\, g_{(2)ij}(x)+\rho^2 \Big[ g_{(4)ij}(x) + h_{(4)ij}(x) \log \rho \Big] 
+O(\rho^3) \Bigg\} \,,\\[2mm]
\phi(\rho,x)&=\rho^{1/2}\Bigg\{ \phi_{(0)}(x) + \rho \, \Big[ \phi_{(2)}(x) +  \psi_{(2)}(x)  \log \rho \Big] +O(\rho^2) \Bigg\}\,.
\end{align}
\end{subequations}
As we will see below, the logarithmic terms are related to the presence of anomalies.
The first term $g_{(0)ij}(x)$ is the boundary metric.
We will be interested in the case in which this is a maximally symmetric, four-dimensional spacetime with constant positive curvature $R=12H^2$, namely a dS$_4$ metric with Hubble rate $H$:
\be
\label{ds}
ds_{\mathrm{b}}^2 =g_{(0)ij}\dd x^i\dd x^j= -\dd t^2 + e^{2Ht} \dd\vec{x}^2 \,.
\ee
Similarly, we will assume that the first term in the expansion of the scalar field is a constant that defines the characteristic mass scale in the gauge theory:
\be
\phi_{(0)} = M \,.
\ee

The first term of the action \eqn{action} suffers from large-volume divergences, as can be verified by substituting the expansions \eqref{seriesVac} into the action.
These divergences can be regularised and renormalised by a procedure called holographic renormalisation (see e.g.~\cite{deHaro:2000vlm,Bianchi:2001de,Bianchi:2001kw}), which makes the action finite and the variational principle well-defined. 
This procedure is implemented  by including in \eqn{action} the counterterm action 
\begin{equation}\label{eq:Sct}
 S_{\mathrm{ct}}= \frac{1}{8\pi G} \int \dd^4x\sqrt{-\gamma}\Bigg[ \left(
-\frac{1}{8}R[\gamma]-\frac{3}{2}-\frac{1}{2}\phi^2 \right) 
+ \frac{1}{2} \left( \log \rho\right)  \mathcal{A}
+  \left( \alpha \mathcal{A} + \beta\phi^4 \right)\Bigg] \,,
\end{equation}
defined on a timelike, constant-$\rho$ hypersurface near the boundary.
The induced metric on this hypersurface is denoted $\gamma_{ij}$ and $R[\gamma]$ is the associated Ricci scalar.
The second term of \eqn{action} is also understood to be evaluated on this slice at $\rho$, the first term  of \eqn{action} is understood to be evaluated by integrating down to this slice, and the limit $\rho \to 0$ is understood to be taken at the end of the calculation. 

In \eqref{eq:Sct},  $\mathcal{A}(\gamma_{ij}, \phi)$ is the so-called conformal anomaly, which in our case is given by 
\be
\mathcal{A}=\mathcal{A}_g + \mathcal{A}_\phi 
\ee
where
\be
\label{eq:Ag}
\mathcal{A}_g = \frac{1}{16}(R^{ij}R_{ij}-\frac{1}{3}R^2)
\ee
is the holographic gravitational conformal anomaly and 
\be
\label{eq:Aphi}
\mathcal{A}_{\phi}=-\frac{\phi^2}{12}R 
\ee
is the conformal anomaly due to matter.
In these equations all the terms are functionals of the metric $\gamma_{ij}$ and of the scalar field $\phi$ induced on the $\rho$-hypersurface.
However, making use of the expansions \eqref{seriesVac} we see that the product with the determinant of the induced metric yields a finite contribution in the limit in which the cut-off is removed, since  
\be
\lim_{\rho\to 0} \, \sqrt{-\gamma}\,  
\mathcal{A} \left( \gamma_{ij}, \phi \right) = 
\lim_{\rho\to 0} \,\left[  \frac{1}{\rho^4} \,  
\sqrt{-g_{(0)}} \right] \Big[ \rho^4  \mathcal{A} \left( g_{(0)ij}, \phi_{(0)} \right) \Big] = 
\sqrt{-g_{(0)}} \, \mathcal{A} \left( g_{(0)ij}, \phi_{(0)}\right) \,.
\ee 
For this reason we will often think of the anomaly as evaluated on the boundary values of the fields, in which case  \eqref{eq:Ag} and \eqref{eq:Aphi} reduce to 
\begin{eqnarray}
\label{often}
\mathcal{A}_g  &=& -\frac{3}{4} H^4 \,,\\[1mm]
\mathcal{A}_{\phi} &=& -M^2 H^2 \,.
\end{eqnarray}
The fact that $\sqrt{-\gamma}  \mathcal{A}$ yields a finite result has two consequences.
First, it means that the logarithmic term in \eqn{eq:Sct} cancels a purely logarithmic divergence from the bulk action.
The requirement that this cancellation takes place fixes uniquely the form of the anomaly, including the values of all the numerical coefficients in \eqref{eq:Ag} and \eqref{eq:Aphi}.
The presence of this logarithmic term on the gravity side breaks diffeomorphism invariance and is dual to the presence of the conformal anomaly in the dual gauge theory. 

The second consequence is that the anomaly itself (without the $\log$) can be added to the counterterm action with an arbitrary coefficient, which we named $\alpha$ in \eqn{eq:Sct}.
It is important to note that not just the anomaly but any local, finite term that is invariant under the symmetries of the theory can be added to the counterterm action with an arbitrary coefficient. 
These terms can be constructed out of non-negative powers of the scalar field and of curvature invariants of the induced metric $\gamma_{ij}$ in such a way that their overall mass dimension is four.
The $\beta \phi^4$ term is an example of such a term. 
Other possible terms include combinations of $R_{ij}R^{ij}$,  $R^2$ and $\phi^2$ that are linearly independent of $\mathcal{A}$.
Therefore we could replace the last term of \eqn{eq:Sct} by 
\be
\label{all}
\alpha \mathcal{A} + \beta \phi^4 + \delta_1 \left( R_{ij}R^{ij} +R^2\right) + \delta_2\,  \phi^2 R + \cdots \,.
\ee
The freedom to add these terms with arbitrary coefficients $\delta_i$ is part of the general freedom in the choice of renormalisation scheme. 
The coefficient $\alpha$ plays a special role since it can be shifted by a scale transformation, which is implemented via the following rescaling of the  coordinates 
\begin{equation}
\label{rescresc}
x_i= \lambda x_i' \,, \qquad \rho=\lambda^2 \rho' \,,
\end{equation}
where $\lambda$ is a positive real number. 
It is easy to see that the effect of this transformation is to shift the counterterm action by a term of the form $(\log \lambda) \mathcal{A}$, which in turn can be absorbed through the redefinition $\alpha \to \alpha + \log \lambda$. 
The freedom to rescale $\rho$, or equivalently to shift $\alpha$, is the freedom to choose a renormalisation scale. 
We thus see that the freedom to choose a renormalisation scheme includes, but is larger than, the freedom to choose a  renormalisation scale.
This statement is well known on the gauge theory side.
In order to renormalise the theory it is not enough to choose a renormalisation scale since finite parts must also be fixed.
For example, the difference between the MS and $\overline{\mbox{MS}}$ schemes is precisely the choice of the finite parts.
Below we will discuss the effect of the anomaly on the gauge theory observables of interest to us, namely the expectation values of the stress tensor and of the scalar operator. 

The value $\beta=1/4\phi_M^2$ is special because in this case the $\beta \phi^4$ term combines with the second and the third summands in the first term of \eqn{eq:Sct} to give precisely the superpotential \eqn{eq:defWs}.
This means that, if the theory \eqn{action} is the bosonic truncation of a supersymmetric theory with superpotential $W$, then in flat space this choice of $\beta$ leads to a supersymmetric renormalisation scheme. 
Motivated by this discussion, in this paper we will set to zero all the coefficients in \eqref{all} but $\alpha$ and $\beta$. 
This implies no loss of generality since physically meaningful quantities are scheme-independent.
 
Substitution of the expansions \eqn{seriesVac} in the equations of motion \eqn{EOM} determines several coefficients \cite{Bianchi:2001kw}.
The Klein--Gordon equation for the scalar field fixes the logarithmic coefficient $\psi_{(2)}$ in terms of $g_{(0)ij}$ and $\phi_{(0)}$ as
\begin{equation}
\label{psi2}
\psi_{(2)}=\frac{1}{24} \phi_{(0)} R = \frac{1}{2} M H^2\,. 
\end{equation}
Unless otherwise indicated, in this and in subsequent equations it is understood that the curvature tensors are those associated to the boundary metric $g_{(0)ij}$.
At leading order Einstein's equations determine 
\begin{equation}
g_{(2)ij}=-\frac{1}{2}\left( R_{ij}-\frac{1}{6} R \, g_{(0)ij} \right)
-\frac{\phi_{(0)}^2}{3} g_{(0)ij}= 
- \left( \frac{1}{2}H^2 + \frac{1}{3}M^2 \right)  g_{(0)ij}\,.\\[2mm]
\end{equation}
The logarithmic part at subleading order fixes 
\begin{equation}
h_{(4)ij}=h_{(4)ij}^{\mbox{\tiny{grav}}}-\frac{1}{12}R_{ij}\phi_{(0)}^2\,,
\end{equation}
where 
\begin{align}
h_{(4)ij}^{\mbox{\tiny{grav}}}=&\frac{1}{8}R_{ikjl}R^{kl}-\frac{1}{48}\nabla_i\nabla_j R+\frac{1}{16}\nabla^2R_{ij}-\frac{1}{24}RR_{ij}
\nonumber\\[2mm]
&+\left(\frac{1}{96}R^2-\frac{1}{96}\nabla^2R-\frac{1}{32}R_{kl}R^{kl}\right)g_{(0)ij}
\end{align}
is the purely gravitational part.
For conformally flat metrics, such as \eqn{ds}, this part vanishes, hence
\be
\label{h4}
h_{(4)ij}= -\frac{1}{12}R_{ij}\phi_{(0)}^2  = -\frac{1}{4} H^2 M^2  g_{(0)ij}\,.
\ee
The subleading non-logarithmic part of  Einstein's equations fixes the trace of $g_{(4)ij}$:
\begin{subequations}
\begin{align}
\mathrm{Tr} g_{(4)ij}&=-2\phi_{(0)}{\phi}_{(2)} +\frac{5}{72}R\phi_{(0)}^2 +\frac{1}{16}(R_{ij}R^{ij}-\frac{2}{9}R^2)+\frac{2}{9}\phi_{(0)}^4\\[2mm]
&= -2 M {\phi}_{(2)}  + \frac{5}{6} M^2 H^2 + \frac{1}{4} H^4 + \frac{2}{9} M^4 
\,,
\end{align}
\end{subequations}
as well as its covariant divergence
\begin{subequations}
\begin{align}
\nabla^j g_{(4)ij}\,=\,\,&\nabla^j\Bigg\{-\frac{1}{8}\Big[ 
\mathrm{Tr} \left( g_{(2)}^{\,\,\,\,\,2}\right) - \left(\mathrm{Tr} g_{(2)} \right)^2 \Big]g_{(0)ij}+\frac{1}{2}\left( g_{(2)}^{\,\,\,\,\,2}\right)_{ij} -\frac{1}{4}g_{(2)ij}\, \mathrm{Tr}\,g_{(2)}
\nonumber\\
&\qquad\, -\frac{3}{2}h_{(4)ij} -g_{(0)ij}\phi_{(0)}\Big( \phi_{(2)}+\psi_{(2)} \Big)\Bigg\} \\[2mm]
\,=\,\,& \nabla^j \Bigg\{ \left( \frac{1}{4} H^4 + \frac{5}{24} H^2M^2 + \frac{1}{9} M^4 
- M {\phi}_{(2)} \right) g_{(0)ij} \Bigg\}
\,.
\end{align}
\end{subequations}

The coefficients in \eqref{seriesVac} determine the holographic stress tensor as
\begin{align}\label{EMT}
\langle \hat T_{ij} \rangle&=\lim\limits_{\rho \to 0} \frac{2\rho^{-2}}{\sqrt{-\gamma}}\frac{\delta S}{\delta \gamma^{ij}}\nonumber\\[2mm]
&=2 \left( \frac{N^2}{4\pi^2} \right) 
\Bigg\{g_{(4)ij}+\frac{1}{8}\left[\mathrm{Tr}g_{(2)}^2-(\mathrm{Tr}g_{(2)})^2\right]g_{(0)ij}-\frac{1}{2}g_{(2)}^2+\frac{1}{4}g_{(2)ij}\mathrm{Tr}g_{(2)}\nonumber\\
& \qquad \,\,\, +\phi_{(0)}\left(\phi_{(2)}-\frac{1}{2}\psi_{(2)}\right)g_{(0)ij}+\alpha\left(\mathcal{T}^{g}_{ij}+\mathcal{T}_{ij}^\phi\right)+\left(\frac{1}{18}+\beta \right)\phi_{(0)}^4g_{(0)ij}\Bigg\}\,. 
\end{align}
In this and in subsequent equations we have made use of \eqref{defN} to replace $G$ in favour of $N$, which makes the expected $N^2$-scaling of the stress tensor manifest. 
The contributions $\mathcal{T}^{g}_{ij}$ and $\mathcal{T}_{ij}^\phi$ come from the variation of $\mathcal{A}_g$ and $\mathcal{A}_\phi$ in \eqn{eq:Sct}, respectively, and are given by:
\begin{align}
\label{eq:Tg}
\frac{1}{2} \mathcal{T}^{g}_{ij}\, =\,\,&h_{(4)ij}^{\mbox{\tiny{grav}}} = 0 \,,
\\[2mm]
\mathcal{T}^{\phi}_{ij}\, =\,\,& -\frac{1}{6} \phi_{(0)}^2 \left( R_{ij} - \frac{1}{2} R g_{(0)ij} \right)
= \frac{1}{2} M^2 H^2 g_{(0)ij}
\label{eq:Tphi}\,.
\end{align}
As we mentioned above, the first equation follows from the conformal flatness of the dS metric \eqn{ds}.
The expectation value of the scalar operator in the field theory is given by
\begin{equation}\label{VEV}
\langle\hat{\O}\rangle=\lim\limits_{\rho \to 0} \frac{\rho^{-\Delta_{\mathrm{UV}}/2}}{\sqrt{-\gamma}}\frac{\delta S}{\delta \phi}=
2 \left( \frac{N^2}{4\pi^2} \right)
\left\{ -2\phi_{(2)}+(1-4\alpha)\psi_{(2)}-4\beta\phi_{(0)}^3\right\}\,.
\end{equation}

In the presence of external sources the holographic stress tensor satisfies anomaly-corrected Ward identities.
These can be obtained from the variation of the renormalised on-shell action
\begin{equation}
\delta S[\delta g_{(0)},\delta \phi_{(0)}]=\int \dd^4x \sqrt{g_{(0)}}\left( \frac{1}{2}\langle \hat{T}_{ij}\rangle \delta g_{(0)}^{ij}+\langle \hat{\O}\rangle \delta \phi_{(0)}\right)\,.
\end{equation}
Invariance of the action under diffeomorphisms
\begin{equation}
\delta g_{(0)}^{ij}=-(\nabla^i\xi^j+\nabla^j\xi^i)\,,\quad \delta\phi_{(0)}=\xi^i\nabla_i \phi_{(0)}\,,
\end{equation}
leads to the diffeomorphism Ward identity 
\begin{equation}
\nabla^i\langle \hat T_{ij}\rangle=-\langle\hat{\O}\rangle\nabla_j \phi_{(0)}\,.
\end{equation}
Weyl transformations
\begin{equation}
\delta g_{(0)}^{ij}=-2\sigma g^{ij}\,,\quad \delta\phi_{(0)}=-(d-\Delta_{\mathrm{UV}})\sigma\phi_{(0)}\,,
\end{equation}
give the anomaly-corrected conformal Ward identity
\begin{equation}\label{eq:Ward}
\langle \hat T^i_i\rangle=-(d-\Delta_{\mathrm{UV}}) \phi_{(0)}\langle\hat{\O}\rangle+
\left( \frac{N^2}{4\pi^2} \right)\left(\mathcal{A}_{g}+\mathcal{A}_{\phi}\right)\,,
\end{equation}
where $d$ is the spacetime dimension of the boundary theory and $\mathcal{A}_{g}$ and $\mathcal{A}_{\phi}$ are given in \eqref{eq:Ag} and \eqref{eq:Aphi}, respectively.
In our case the Ward identities reduce to
\begin{equation}
\label{wardward}
\nabla^i\langle \hat T_{ij}\rangle= 0 \,,\qquad 
\langle \hat T^i_i\rangle=-M \langle\hat{\O} \rangle 
-\left( \frac{N^2}{4\pi^2} \right) \left(\frac{3}{4} H^4 +  M^2H^2 \right) \,.
\end{equation}

We are now ready to discuss the effect of the anomaly on physical observables such as the stress tensor and the scalar operator.
To see this consider again the rescaling \eqref{rescresc}.
In the gauge theory this is equivalent to rescaling  $H$ and $M$ as
\begin{equation}\label{eq:rescale}
H' = \lambda H \,, \qquad M' = \lambda M \,.
\end{equation}
Following \cite{Bianchi:2001de}, we note that the rescaling above leaves the FG form of the metric \eqref{metricFG} invariant and transforms all the expansion coefficients homogeneously,
\begin{equation}
g_{(0)ij}'= g_{(0)ij} \,, \qquad g_{(2)ij}'= \lambda^2 g_{(2)ij} \,, \qquad 
h_{(4)ij}'= \lambda^4 h_{(4)ij} \,,
\end{equation}
except for $g_{(4)ij}$, which acquires an inhomogeneous piece due to the logarithmic term in \eqref{seriesVac}:
\begin{equation}
g_{(4)ij}' = \lambda^4 g_{(4)ij} + 2\lambda^4 \log \lambda \, h_{(4)ij}
\,.
\end{equation}
Similarly, the coefficients $\phi_{(0)}$ and $\psi_{(2)}$ in the expansion of the scalar field transform homogeneously, whereas $\phi_{(2)}$ acquires an inhomogeneous piece: 
\begin{equation}
\phi_{(2)}' = \lambda^3 \phi_{(2)} + 2\lambda^3 \log \lambda \, \psi_{(2)} \,.
\end{equation}
It follows that the stress tensor and the scalar expectation value transform as
\begin{eqnarray}
\langle \hat T_{ij}' \rangle &=& \lambda^4 \langle \hat T_{ij}\rangle + 
4 \left( \frac{N^2}{4\pi^2}\right)\lambda^4 \log \lambda \, h_{(4)ij} \,, \\[2mm]
\langle \hat{\O}' \rangle &=& \lambda^3 \langle \hat{\O} \rangle - 
8 \left( \frac{N^2}{4\pi^2}\right) \lambda^3 \log \lambda \, \psi_{(2)} \,,
\end{eqnarray}
namely
\begin{eqnarray}
\langle \hat T_{ij}(\lambda H, \lambda M) \rangle &=& 
\lambda^4 \langle \hat T_{ij}(H, M)\rangle - \lambda^4 \log \lambda \, 
\left( \frac{N^2}{4\pi^2}\right)\, H^2 M^2
 \, g_{(0)ij} \,, \\[2mm]
 \label{mu2}
\langle \hat{\O} (\lambda H, \lambda M) \rangle &=& \lambda^3 \langle \hat{\O} (H, M)\rangle 
- \lambda^3 \log \lambda \, \left( \frac{N^2}{4\pi^2}\right) \, 4 M H^2  \,,
\end{eqnarray}
where we have made use of \eqref{psi2} and \eqref{h4}. 
This immediately implies that these expectation values must take the form
\begin{eqnarray}
\label{stress}
\langle \hat T_{ij}( H,  M) \rangle &=& 
H^4 \, t_{ij} \left( \frac{H}{M} \right) - \log \left( \frac{H}{\mu} \right) \, 
\left( \frac{N^2}{4\pi^2}\right) \, H^2 M^2
 \, g_{(0)ij} \,, \\[2mm]
 \label{scalar}
\langle \hat{\O} ( H, M) \rangle &=& H^3 \, f \left( \frac{H}{M} \right) 
- \log \left( \frac{H}{\mu} \right)\, 
\left( \frac{N^2}{4\pi^2}\right) \, 4 M H^2 \,,
\end{eqnarray}
where $\mu$ is some arbitrary reference scale, a remnant of the renormalisation process much like the renormalisation scale in QFT.
The first and second terms on the right-hand sides transform homogeneously and inhomogeneously under the rescaling \eqref{eq:rescale}, respectively.
Needless to say, one could rewrite the first terms in a variety of forms, for example as $M^4 t_{ij}(H/M)$ for the stress tensor, etc.
Also, one could replace $\log (H/\mu)$ by $\log(H/M) + \log(M/\mu)$, thus redefining
\begin{equation}
t_{ij} \to t_{ij} + \log \left( \frac{H}{M} \right) \, \left( \frac{N^2}{4\pi^2}\right)\, 
\frac{M^2}{H^2} \, g_{(0)ij}\,.
\end{equation}
Note also that there is no loss of generality in assuming that the scale $\mu$ is the same in both equations, since the difference can again be absorbed in a redefinition of the homogeneous terms. 

The key conclusion is that, because of the anomaly, expectation values in the field theory do not only depend on the ratio $H/M$, but on the two independent dimensionless ratios that can be built from $M, H$ and $\mu$.
Put differently, in order to specify the theory it is not enough to specify the ratio between $H$ and $M$, but instead both scales must be specified independently with respect to some arbitrary reference scale $\mu$.
The freedom in the choice of this scale is part of a bigger freedom in the choice of renormalisation scheme, as we discussed around \eqref{all}.
Throughout this paper we will measure all dimensionful quantities in units of $M$ and, when necessary, we will fix the renormalisation scheme by specifying $\alpha$ and $\beta$. 

In the flat-space limit, namely if $H=0$, both the anomaly \eqref{often} and its contributions \eqref{eq:Tg} and \eqref{eq:Tphi} to the stress tensor vanish identically.
This means that in this case both the stress tensor and the scalar operator transform covariantly under scale transformations.
In other words, the non-homogeneous terms in the equations above vanish.
Moreover, the only non-zero finite term among all the possible ones in \eqref{all} is the $\beta \phi^4$ term.
This produces a contribution to the stress tensor \eqref{EMT} that shifts its value by a term proportional to the boundary metric $g_{(0)ij}=\eta_{ij}$, namely it shifts the energy density and the pressure by opposite amounts.
Therefore the choice of scheme in the flat-space case reduces entirely to fixing the energy or the pressure of some reference state, for example that of the vacuum.
We will come back to this point in the next section.  

In this work we will only consider states that are homogeneous and isotropic, for which the associated energy momentum tensor takes the  diagonal form
\begin{equation}\label{eq:EMTdiag}
\langle \hat T^i_j \rangle =\mathrm{diag}\left\{-\epsilon(t),p(t),p(t),p(t)\right\}\,.
\end{equation}
When plotting numerical results we will often use ``reduced'' quantities such as reduced energy density, reduced pressure and reduced expectation value of the scalar operator defined as 
\begin{equation}
\mathcal{E}(t)\equiv  \frac{2\pi^2}{N^2}\, \epsilon(t) \,, \qquad 
\mathcal{P}(t)\equiv  \frac{2\pi^2}{N^2}\, p(t) \,, \qquad
\O(t)\equiv \frac{2\pi^2}{N^2}\,\langle\hat{\mathcal{O}}(t)\rangle \,.
\end{equation}
In terms of these variables, the trace Ward identity  \eqref{wardward} takes the form
\begin{equation}\label{eq:WardO}
\mathcal{E}(t)-3\mathcal{P}(t)=M \O(t) + \frac{1}{2} \left( \frac{3}{4}H^2 + M^2 H^2 \right) \,.
\end{equation}

\section{Dynamics in flat space}
\label{sec:flat}
\subsection{Thermodynamics and transport}
\label{subsec:Thermo}
In this section we review the most salient thermodynamic and transport properties of the holographic model on flat space, studied in detail in \cite{Attems:2016ugt}.
This is useful because later we will use thermal equilibrium states on flat space to initialise the time evolution of non-equilibrium states on dS$_4$ and compare their evolution to viscous hydrodynamics with transport coefficients presented in this section.

The gauge/gravity correspondence maps thermodynamic equilibrium states on the gauge theory side to equilibrium black  brane geometries on the gravity side.
In our case these are homogeneous and isotropic solutions of the equations of motion \eqref{EOM} with a regular horizon and asymptotically AdS boundary conditions for the metric and appropriate asymptotic scaling for the scalar field.
A convenient gauge to construct these solutions is one where the holographic coordinate is identified with the scalar field\footnote{The function $H(\phi)$ appearing in this section should not be confused with the Hubble rate $H$ appearing throughout the whole paper.}
\be\label{eq:thermal_line_element}
ds^2= e^{2 A(\phi)}\left(-H(\phi) \dd\tau^2 + \dd\vec{x}^2 \right) -2 e^{A(\phi)+B(\phi)} \, \dd\tau \dd\phi \,.
\ee
In this gauge the boundary is located at $\phi=0$ and the value of the scalar field at the horizon $\phi_{\mathrm{h}}$ is determined by the condition $H(\phi_{\mathrm{h}})=0$. 
After introducing a master field 
\be
G(\phi)= \frac{\dd}{\dd\phi} A(\phi) \,,
\ee
the equations of motion \eqref{EOM} can be rewritten in terms of a single master equation\footnote{As in \cite{Attems:2016ugt}, we normalise the scalar field differently than in \cite{Gubser:2008ny}, which is the reason why some of the coefficients in \eqref{eq:master} differ from those in the corresponding master equation in \cite{Gubser:2008ny}.} \cite{Gubser:2008ny,Attems:2016ugt}
\be
\label{eq:master}
\frac{G'(\phi )}{G(\phi )+\frac{4 V(\phi )}{3 V'(\phi )}}
=
\frac{\dd}{\dd\phi} \log \left(
\frac{1}{3 G(\phi )}-2 G(\phi )+\frac{G'(\phi )}{2 G(\phi )}-\frac{G'(\phi )}{2 \left(G(\phi )+\frac{4 V(\phi )}{3 V'(\phi )}\right)} 
\right) \,.
\ee
Close to the horizon a solution to the master equation can be expressed as a power series:
\be
\label{eq:masterNH}
G(\phi)=
   -\frac{4
   V(\phi_{\mathrm{h}})}{3 V'(\phi_{\mathrm{h}})} + 
 \frac{2}{3} ( \phi - \phi_{\mathrm{h}})
   \left(\frac{V(\phi_{\mathrm{h}}) V''(\phi_{\mathrm{h}})}{V'(\phi_{\mathrm{h}})^2}-1\right) + O\left(\left(\phi-\phi_{\mathrm{h}}\right)^2\right) \,.
\ee
Close to the boundary, $\phi\to 0$, the master field can be expanded as
\begin{equation}
G(\phi)=-\frac{1}{\phi}+\cdots \,.
\end{equation}

\begin{figure}[t]
\center
 \includegraphics[width=0.33\linewidth]{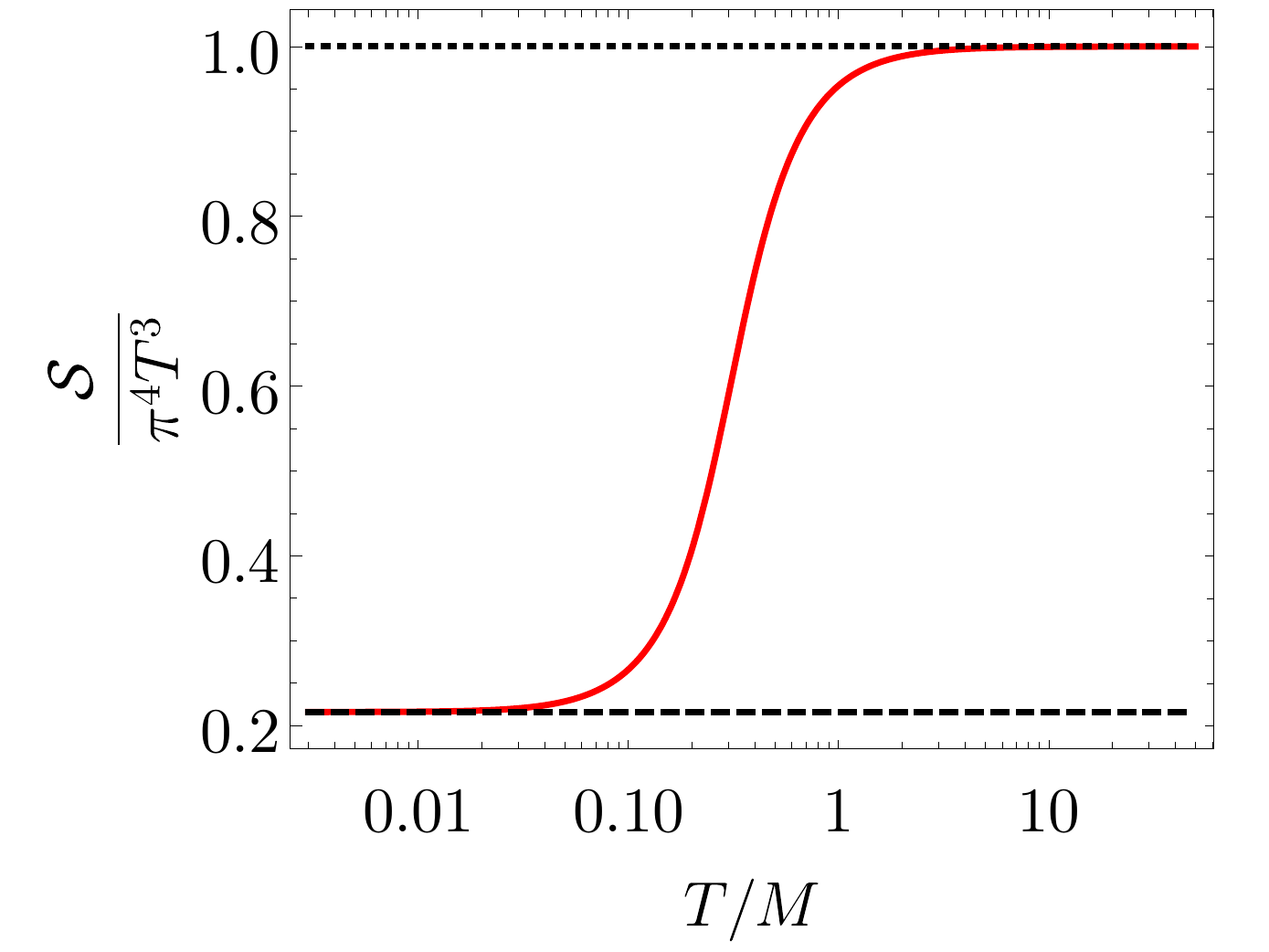}\includegraphics[width=0.33\linewidth]{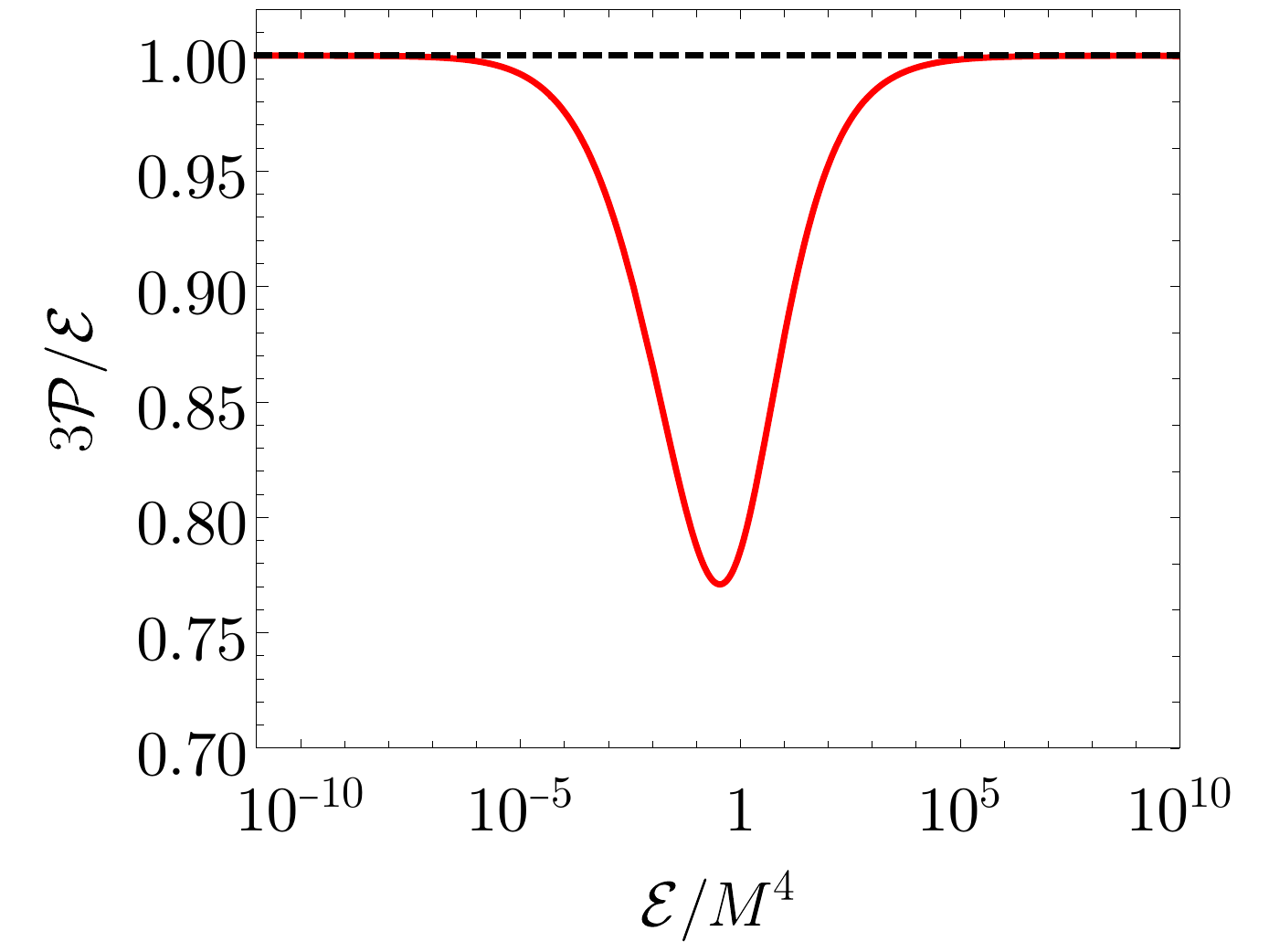}\includegraphics[width=0.33\linewidth]{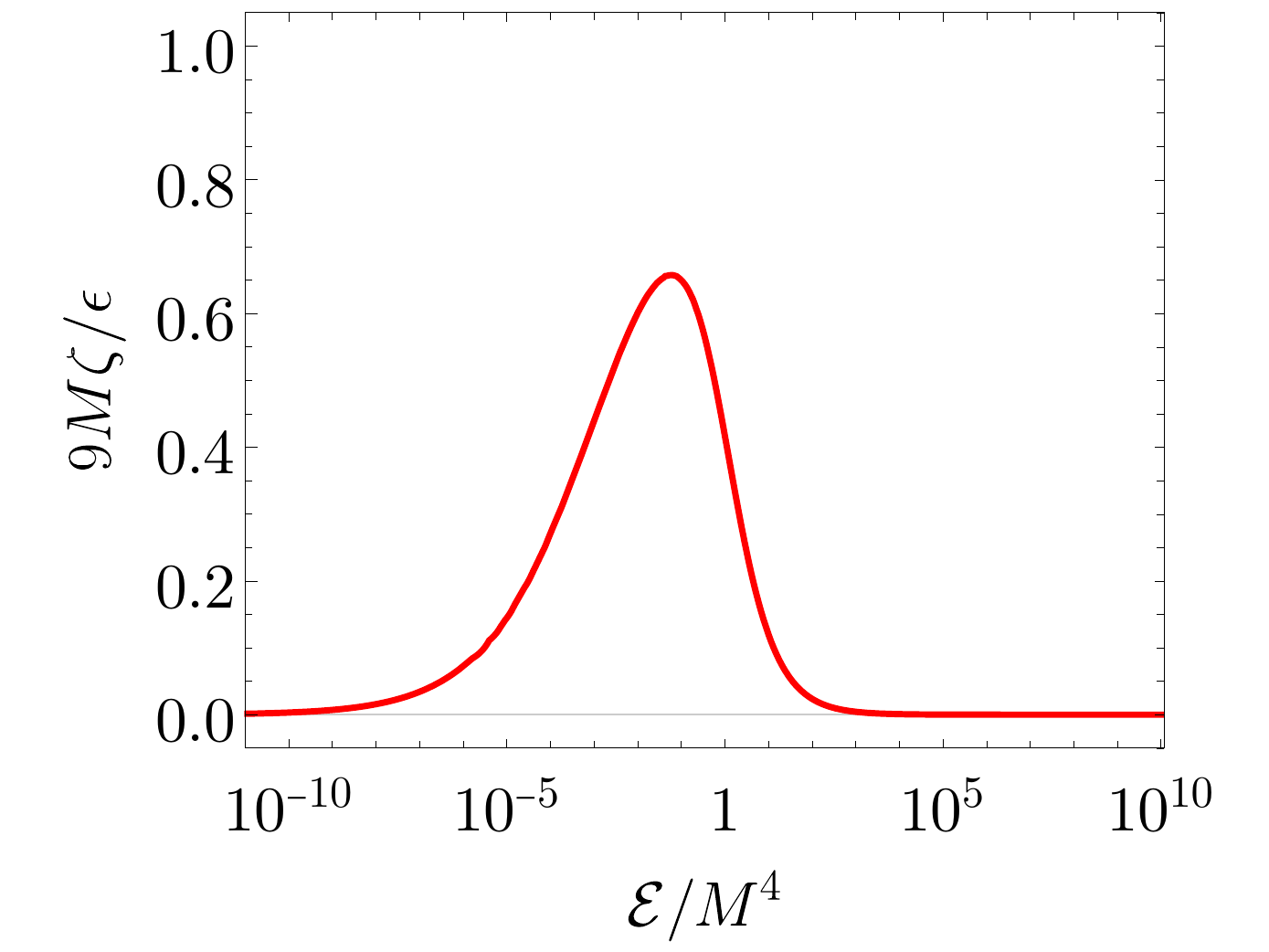}
\caption{(Left) Entropy density as function of temperature for $\phiM=2$.
The black dashed and dotted lines indicate respectively the IR- and the UV-limits \eqref{limits}. 
(Center) Ratio of reduced pressure to energy density as a function of the reduced energy density.
(Right) Ratio of bulk viscosity to energy density as a function of the reduced energy density.
}
\label{fig:eos}
\end{figure}

In practice we obtain a one--parameter family of solutions for $G(\phi)$, parametrised by the value of $\phi_{\mathrm{h}}$, by numerically integrating \eqref{eq:master} from a value of $\phi$ close to the horizon to a value close to the boundary using boundary conditions for $G$ and $G'$ constructed from \eqref{eq:masterNH}. 
The metric functions in \eqref{eq:thermal_line_element} can then be obtained through the relations
\beq
\label{eq:A}
A(\phi)&=&-\log \left(\frac{\phi}{M}\right) + \int_0^\phi \dd\phit \left( G(\phit) + \frac{1}{\phit }\right) \, ,
\\[2mm]
\label{eq:B}
B(\phi)&=& \log \left( \left| G(\phi) \right|\right) + \int_0^\phi \dd\phit \frac{2}{3 G(\phit)} \, ,\\[2mm]
\label{eq:h}
H(\phi)&=&-\frac{e^{2 B(\phi )} \left(4V(\phi )+3 G(\phi )V'(\phi )\right)}{3 G'(\phi )} \,.
\eeq
The temperature and the entropy density of field theory states dual to these numerically constructed geometries can be expressed in terms of \eqref{eq:A} and \eqref{eq:B} evaluated at the horizon  (see \cite{Gubser:2008ny,Attems:2016ugt} for details):
\be
\label{eq:TandsG}
T= \frac{A(\phi_{\mathrm{h}})-B(\phi_{\mathrm{h}})}{4 \pi}\,,   \quad \quad  
s= 2 \pi \left( \frac{N^2}{4\pi^2}\right)  e^{3 A(\phi_{\mathrm{h}})} \,.
\ee
In Fig.~\ref{fig:eos} (left) we show the reduced entropy density $\mathcal{S}\equiv 2\pi^2 s/N^2$ divided by $T^3$ as a function of $T/M$ for $\phiM=2$.
The dotted and the dashed black lines indicate, respectively, the infinite- and the zero-temperature limits (recall that we are setting $L=1$)
\be
\label{limits}
\lim\limits_{T \to \infty}\, \frac{\mathcal{S}}{\pi^4T^3}=1 \,, \qquad
\lim\limits_{T \to 0}\, \frac{\mathcal{S}}{\pi^4T^3}=L_{\mathrm{IR}}^3 = \frac{27}{125}\,, 
\ee
with $L_{\mathrm{IR}}$ given by \eqref{eq:LIR}. 
As explored in detail in \cite{Attems:2016ugt}, for real values of $\phiM$ the model has a smooth crossover between the IR and UV fixed point.

The energy density and the pressure can be extracted from the stress tensor discussed in \sect{sec:HM}.
In thermal equilibrium they can be equivalently obtained from the knowledge of the entropy density as a function of the temperature  via the thermodynamic relations 
\begin{subequations}
\label{eq:Pthermo}
\begin{align}
p_{\mathrm{eq}}&\equiv p_0+\int_0^T \dd T' \, s(T') \,,\label{eq:p}\\
h &\equiv \epsilon+p_{\mathrm{eq}}= T s\, ,\label{eq:enthalpy}
\end{align}
\end{subequations}
where $p_0$ is the pressure of the vacuum state in the $T\rightarrow 0$ limit and $h$ is the enthalpy density.
The value of $p_0$ is not fixed by thermodynamic considerations or by the equations of motion.
In contrast, $T$ and $s$ are uniquely defined via \eqref{eq:TandsG}, and hence so is the enthalpy density. 
It follows that the energy density is also defined up to an arbitrary constant equal to $-p_0$.  
As we saw towards the end of \sect{sec:HM}, the choice of renormalisation scheme in flat space boils down precisely to the choice of this constant.
Thus one way to fix the scheme in flat space and uniquely determine the stress tensor is to impose that the vacuum energy density and pressure vanish.
As discussed in \sect{sec:HM}, this corresponds to the choice  $\beta=1/4\phiM^2$ and this is what we implicitly assume in the rest of this section.

In Fig.~\ref{fig:eos} (center) we show the ratio of reduced pressure and energy density.
The equilibrium values for $p$ and $\epsilon$ constitute the equation of state (EoS) $p_{\mathrm{eq}}(\epsilon)$.
We use the deviation of the ratio $w\equiv p/\epsilon=\calP/\calE$ from 1/3 as a measure of the amount of conformal symmetry breaking.
Similarly, we could measure it via the deviation of the speed of sound squared $c_{\mathrm{s}}^2\equiv \dd p_{\mathrm{eq}}/\dd \epsilon$ from its conformal value $c_\mathrm{s,CFT}^2\equiv 1/(d-1)=1/3$.
One advantage of $w$ over $c_s^2$ is that the former can be computed without taking derivatives, and hence it is defined instantaneously, which will be useful when we study the model in de Sitter space.
In flat space, $w$ asymptotes to the conformal value $w=1/3$ in the low- and high-energy density regimes.
In between, at energy densities comparable to the scale of the theory, $\calE\approx M^4$, $w$ deviates significantly from its conformal value.
 
At leading order in the hydrodynamic expansion (see below) the transport properties in flat space are determined by two coefficients: the shear viscosity and the bulk viscosity.
The ratio of shear viscosity over entropy density, $\eta/s=1/4\pi$, is universal in all holographic theories with an Einstein gravity dual \cite{Kovtun:2004de}.
This means that knowledge of the entropy density \eqref{eq:Pthermo} is sufficient to determine the shear viscosity in our case.
The bulk viscosity can be obtained from the logarithmic derivative of the entropy density with respect to the value of the scalar field at the horizon \cite{Eling:2011ms}:
\be
\frac{\zeta}{s}= \frac{1}{\pi} \left(\frac{\dd \log s}{\dd \phi_{\mathrm{h}}}\right)^{-2} \,.
\ee
In Fig.~(\ref{fig:eos}) (right) we plot the ratio of bulk viscosity and energy density as a function of the reduced energy density. 
As we will see in Sec.~\ref{sec:dS4Hydro}, the specific combination $9M\zeta/\epsilon$ measures the viscous contribution to $w$ in dS$_4$.
The bulk viscosity vanishes at small and large energy densities where the model is conformal.
In between, at energy densities comparable to the scale of the theory, $\calE\approx M^4$, $\zeta$ is non-zero.

\subsection{Hydrodynamics}
\label{sec:hydro}
The thermodynamic analysis of the previous section only applies to static equilibrium states.
It serves as a starting point for describing the long wavelength dynamics of the system in a hydrodynamic approximation.
The modern interpretation of hydrodynamics as an effective theory in terms of a gradient expansion has been reviewed many times (see e.g.~\cite{romatschke_romatschke_2019} and references therein).
The purpose of this section is to review some of the basic definitions in order to fix notation and to show how scheme dependence manifests itself in the hydrodynamic expansion.

In the absence of other conserved charges, long-wavelength excitations in the gauge theory are solely controlled by the dynamics of the energy-momentum tensor.
In this limit the stress tensor can be approximated in terms of a derivative expansion
\begin{eqnarray}\label{eq:hydro}
\langle \hat{T}^{ij}\rangle & = & \epsilon\,u^i u^j+p_{\mathrm{eq}}(\epsilon)\Delta^{ij}-\eta(\epsilon)\,\sigma^{ij}-\zeta(\epsilon)\Delta^{ij}\overline{\nabla}^{k}u_{k}+O(\overline{\nabla}^{2})\,,\label{eq:hydro-constituive}\\[2mm]
\sigma^{ij} & = & \Delta^{ik}\Delta^{jl}(\overline{\nabla}_{k}u_{l}+\overline{\nabla}_{l}u_{k})-\frac{2}{3}\Delta^{ij}\overline{\nabla}_k u^k\,,\label{eq:sigma}
\end{eqnarray}
where $u^{i}$ is the fluid velocity and $\overline{\nabla}^{i}\equiv \Delta^{ij}\nabla_{j}$, with $\Delta^{ij} = g^{ij}_{(0)}+u^{i}u^{j}$, is the projection of the covariant derivative to the spatial components in the local rest frame of the fluid.
The EoS $p_{\mathrm{eq}}(\epsilon)$ and the transport coefficients $\eta(\epsilon)$ and $\zeta(\epsilon)$ are functions of the energy density that depend on the microscopic details of the theory. 
The hydrodynamic approximation involves a choice of hydrodynamic variables and specifying these variables is called a choice of frame.
We choose the Landau frame, in which the velocity $u^i$ and energy density $\epsilon$ are defined as the timelike eigenvector and the eigenvalue, respectively, of the stress tensor, namely 
$\langle \hat{T}^{ij}\rangle u_j=-\epsilon u^i$.

The leading term in the expansion \eqref{eq:hydro} is called the ideal hydrodynamic part.
It describes the flow of energy and momentum in terms of a locally equilibrated ensemble locally boosted to non-vanishing velocity $u^i$.
Higher-order terms are expressed in terms of gradients of energy density and fluid velocity.
In the following we will neglect terms of $O(\overline{\nabla}^{2})$ and only consider the leading ideal and sub-leading viscous part of \eqref{eq:hydro}.
We will write this term as 
\be
\label{eq:Thydroform}
\langle \hat{T}^{ij}\rangle = \Big( \epsilon + p_{\mathrm{eq}}(\epsilon)  \Big) u^i u^j + p_{\mathrm{eq}}(\epsilon) \, g_{(0)}^{ij} + \Pi^{ij} \,,
\ee
where the viscous tensor $\Pi^{ij}$ in general contains contributions due to bulk and shear viscosity.\footnote{The appearance of shear and bulk viscosity as leading contributions in $\Pi^{ij}$ is specific to our choice of Landau frame. In other frames these contributions are different \cite{Kovtun:2019hdm}.
}
In this work we will only consider homogeneous and isotropic flows without shear stresses, in which case the viscous tensor simplifies to 
\be
\label{pibulkdef}
\Pi^{ij} = - \left( g_{(0)}^{ij} + u^{i} u^{j}\right) \, \zeta(\epsilon) \overline{\nabla}_k u^k  \,. 
\ee

The possible scheme dependence of the microscopic energy-momentum tensor also manifests itself in the hydrodynamic approximation.
As discussed in Sec.~\ref{subsec:Thermo}, $p_{\mathrm{eq}}$ given in \eqn{eq:Pthermo} is not uniquely defined but contains an arbitrary contribution $p_0$ identified as the vacuum pressure.
As explained above, in flat space the entire freedom in the choice of renormalisation scheme reduces to the choice of this constant.
Thus one way to proceed is to make an explicit choice and perform all calculations in that scheme. 
Alternatively, we may work with manifestly scheme-independent quantities as follows. 
We first define the excess pressure and the excess energy density over the vacuum as
\be
\Delta \epsilon\equiv \epsilon+p_0 \,,\qquad 
\Delta p_{\mathrm{eq}}=p_{\mathrm{eq}}-p_0 \,.
\ee
These are scheme-independent, and we may then view the EoS as a relation of the form $\Delta p_{\mathrm{eq}}=\Delta p_{\mathrm{eq}}(\Delta \epsilon)$.
Next we rewrite the second term in \eqn{eq:Thydroform} as
\be
p_{\mathrm{eq}} g_{(0)}^{ij} =\Delta p_{\mathrm{eq}} g_{(0)}^{ij} + \langle \hat{T}^{ij}_{\mathrm{vac}}\rangle \,,  \quad \langle \hat{T}^{ij}_{\mathrm{vac}}\rangle\equiv p_0 g_{(0)}^{ij} \,. 
\ee
This separates \eqref{eq:Thydroform} into a scheme-dependent vacuum contribution $\langle \hat{T}^{i j}_{\mathrm{vac}}\rangle$ and a scheme-independent contribution 
\be
\langle\Delta \hat{T}^{ij}\rangle\equiv \langle \hat{T}^{ij}\rangle - \langle \hat{T}^{ij}_{\mathrm{vac}}\rangle \,.
\ee
Note that the fact that $\langle \hat{T}^{ij}_{\mathrm{vac}}\rangle$ is proportional to the background metric $ g_{(0)}^{ij}$ is consistent with the expectation that the vacuum must respect the symmetries of this background. 
Since $\langle\Delta \hat{T}^{ij}\rangle$ and $\Delta \epsilon$ are scheme-independent, we can now define a scheme-independent  velocity field defined through the relation 
\be
\langle\Delta \hat{T}^{ij}\rangle u_j=-\Delta \epsilon \, u^i \,. 
\ee
The scheme-independent part of \eqref{eq:Thydroform} is then given by
\be\label{eq:hydroDe}
\langle \Delta \hat{T}^{ij}\rangle = \Big( \Delta\epsilon + \Delta p_{\mathrm{eq}}(\Delta\epsilon)  \Big)u^i u^j + \Delta p_{\mathrm{eq}}(\Delta\epsilon)  g_{(0)}^{ij} - 
\left( g_{(0)}^{ij} + u^{i} u^{j}\right) \, \zeta(\Delta\epsilon) \overline{\nabla}_k u^k  \,.
\ee

\section{Dynamics in de Sitter space}
\label{sec:DdSs}
\subsection{The dual gravity problem}
We now turn to the main subject of this work: the far-from-equilibrium dynamics of our strongly coupled non-conformal gauge theory on a time-dependent background geometry.
For this we numerically solve the fully non-linear equations of motion \eqref{EOM} of the dual gravity theory, following the method reviewed in e.g.~\cite{Chesler:2013lia,vanderSchee:2014qwa,Ecker:2018jgh}, and extract the time evolution of the expectation values of the energy momentum tensor and the scalar operator from the solution near the boundary.

We are ultimately interested in the evolution of field theory observables on dS$_4$. 
However, it is useful to set up the problem with a slightly more general boundary metric of Friedmann--Lemaître--Robertson--Walker type
\begin{equation}
\label{eq:sFRW}
ds_{\mathrm{b}}^2=g_{(0)ij}\dd x^i \dd x^j=- \dd t^2 + S_0(t)^2 \dd\vec{x}^2\,.
\end{equation}
For $S_0(t)=e^{H t}$ the boundary metric \eqref{eq:sFRW} becomes the dS$_4$ \eqref{ds} with curvature scalar $R=12H^2$.

We use generalised Eddington--Finkelstein (EF) coordinates to parametrise the bulk geometry and the scalar field
\begin{equation}
\label{metricEF}
ds^2=-A(r,t)\dd t^2+2\dd r\dd t+S(r,t)^2\dd\vec{x}^2\,,\quad \phi=\phi(r,t)\,, 
\end{equation}
where the asymptotic boundary is located at $r= \infty$.
The line element \eqref{metricEF} has a residual gauge freedom in the radial coordinate
\begin{subequations}
\begin{align}
r&\rightarrow \bar{r} \equiv r+\xi(t)\label{gauger} \,,\\[1mm]
A(r,t)&\rightarrow\bar{A}(\bar{r},t)\equiv A(\bar{r}-\xi(t),t)+2\partial_t \xi(t)\,,\\[1mm]
S(r,t)&\rightarrow\bar{S}(\bar{r},t)\equiv S(\bar{r}-\xi(t),t)\,,
\end{align}
\end{subequations}
which we exploit in our numerical scheme to fix the coordinate value $r_{\mathrm{AH}}$ at the apparent horizon, defined by the condition $\dot{S}(r_{\mathrm{AH}},t)=0$, to a constant.

Using \eqref{metricEF} the equations of motion \eqref{EOM} result in the following set of equations
\begin{subequations}\label{CharEOM}
\begin{align}
 S''&=-\frac{2}{3} S \left(\phi '\right)^2 \,,\label{eq:Spp}\\
 \dot{S}'&=-\frac{2 \dot{S} S'}{S}-\frac{2 S V}{3}\,, \label{eq:Sd}\\
 \dot{\phi }'&=\frac{V'}{2}-\frac{3\dot{S} \phi '}{2S}-\frac{3S'\dot{\phi }}{2S}\,, \label{eq:phid} \\
 A''&=\frac{12 \dot{S} S'}{S^2}+\frac{4 V}{3}-4 \dot{\phi } \phi'\,,\label{eq:App} \\
 \ddot{S}&=\frac{\dot{S} A'}{2}-\frac{2 S \dot{\phi }^2}{3} \label{eq:Sdd}\,,
\end{align}
\end{subequations}
where a prime denotes a radial derivative, $f'\equiv \partial_r f$, and an overdot is short-hand for the modified derivative $\dot{f}\equiv \partial_t f +\frac{1}{2}A \partial_r f $.

Imposing the  metric \eqref{eq:sFRW}  and the asymptotic behaviour of the scalar field 
\be
\lim\limits_{r\to \infty}\phi(r,t)=\frac{M}{r^{d-\Delta_\mathrm{UV}}}=\frac{M}{r} 
\ee
as boundary conditions, solutions of \eqref{CharEOM} can be expressed near the boundary as generalised power series:
\begin{subequations}\label{BoundarySeries}
\begin{align}
 A(r,t)=& \,\, r^2+2r\xi(t)+\xi(t)^2-2\xi'(t)+\frac{S_0'(t)^2-2 S_0(t)S_0''(t)}{S_0(t)^2}-\frac{2 M^2}{3} +\frac{a_{(4)}(t)}{r^2}\nonumber\\
       &+\frac{2 M^2 S_0''(t)}{3 S_0(t)}\frac{\log(r)}{r^2} +O(r^{-3}) \,, \\
 S(r,t)= &\,\, S_0(t)r+S_0'(t)+\xi(t) S_0(t)-\frac{M^2 S_0(t)}{3r}+\frac{M^2(3\xi(t)S_0(t)-S_0'(t))}{9r^2}\nonumber\\
       &+\frac{M(4 M^3 S_0(t)^2-72\bar{\phi}_{(2)}(t)S_0(t)^2+48 M\xi(t) S_0(t) S_0'(t)+9M(S_0'(t)^2+S_0(t)S_0''(t)))}{216S_0(t)r^3}  \nonumber\\
       &+\frac{M^2(S_0'(t)^2+S_0(t)S_0''(t))}{6S_0(t)}\frac{\log(r)}{r^3}+O(r^{-4}) \,, \\
\phi(r,t)=& \,\, \frac{M}{r}-\frac{M\xi(t)}{r^2}+\frac{\bar{\phi}_{(2)}(t)}{r^3}-\frac{M \left(S_0'(t)^2+S_0(t) S_0''(t)\right)}{2S_0(t)^2}\frac{\log(r)}{r^3}+O(r^{-4})\,.
\end{align}
\end{subequations} 
Note that the fall-off coefficient $\bar{\phi}_{(2)}$ in EF coordinates is generically different from the one in FG coordinates ${\phi}_{(2)}$. 
The coefficients $a_{(4)}(t)$ and $\bar{\phi}_{(2)}(t)$ in these series cannot be determined from the near-boundary analysis but need to be extracted from the full bulk solution.
The equations of motion impose the following relation on these coefficients
\begin{eqnarray}
\label{Econs}
a_{(4)}'(t)+  4 a_{(4)}(t)\frac{ S_0'(t)}{S_0(t)} &=&-\frac{4}{3}M\left( 2 \bar{\phi}_{(2)}(t)  \frac{S_0'(t)}{ S_0(t)} +  \bar{\phi}_{(2)}'(t)\right) +\frac{16 M^4 S_0'(t)}{27 S_0(t)}  \\
 & &+\frac{2}{3} M^2\left(\frac{S_0'(t) S_0''(t)}{S_0(t){}^2}+\frac{4 S_0{}^{(3)}(t)}{3 S_0(t)}+\frac{4\xi(t)^2S_0'(t)}{S_0(t)}+4\xi(t)\xi'(t) \right)\,. \nonumber
\end{eqnarray}
This relation follows from the momentum constraint \eqref{eq:Sdd} and implies covariant conservation of the holographic stress tensor in the boundary theory, namely the first Ward identity in \eqref{wardward}.

The EF coordinate system \eqref{metricEF} is useful to  obtain time-dependent solutions of the equations of motion numerically.
However, our expressions for the expectation values of the stress tensor \eqref{EMT} and the scalar operator \eqref{VEV} assume the FG coordinate system \eqref{metricFG}.
Although we could  recompute the corresponding expressions in EF gauge, it will prove more convenient to relate the EF and the FG coefficients.
For this purpose we need to find the asymptotic coordinate transformation between the EF and the FG coordinate systems.
We first write a series ansatz for the EF coordinates $r_{EF}$ and $t_{EF}$ in powers of the radial FG coordinate $r_{FG}$, related to $\rho$ in \eqref{metricFG} through $\rho=1/r_{FG}^2$:
\begin{subequations}\label{eq:EFFGansatz}
\begin{align}
 r_{EF}(r_{FG},t_{FG})&=\sum_{n=1}^\infty \Big[ 
 r_n(t_{FG}) +\rho_n(t_{FG})\log(r_{FG}) + \cdots \Big](r_{FG})^n \,,\\
 t_{EF}(r_{FG},t_{FG})&=t_{FG}+\sum_{n=1}^\infty \Big[
 t_n(t_{FG}) + \tau_n(t_{FG})\log(r_{FG})+\cdots \Big] (r_{FG})^{-n} \,,
\end{align}
\end{subequations}
where dots stand for therms with higher powers of $\log(r_{FG})$. 
All these logarithmic terms appear because we are working with a curved boundary metric.
The metric transforms as follows
\begin{equation}\label{CoordTrafo}
g^{FG}_{\mu\nu}=\frac{\partial x_{EF}^\alpha}{\partial x_{FG}^\mu} \frac{\partial x_{EF}^\beta}{\partial x_{FG}^\nu}g_{\alpha\beta}^{EF}\,,
\end{equation}
where $x_{EF}^\mu=(r_{EF},t_{EF})$ and $x_{FG}^\mu=(r_{FG},t_{FG})$.
Substituting  the general expressions for metrics in EF and FG coordinates
\begin{equation}
g^{EF}_{\mu\nu}=\begin{pmatrix} 0 & 1 \\ 1 & g^{EF}_{1,1} \end{pmatrix}\,, \qquad g^{FG}_{\mu\nu}=\begin{pmatrix} r_{FG}^2 & 0 \\ 0 & g^{FG}_{1,1} \end{pmatrix}
\end{equation}
into the transformation law \eqref{CoordTrafo} leads to a set of two equations 
\begin{subequations}
\begin{align}
 0&=\frac{\partial{t_{EF}}}{\partial{r_{FG}}} \frac{\partial{t_{EF}}}{\partial{t_{FG}}}g^{EF}_{1,1}+\frac{\partial{r_{EF}}}{\partial{r_{FG}}}\frac{\partial{t_{EF}}}{\partial{t_{FG}}}+\frac{\partial{t_{EF}}}{\partial{r_{FG}}}\frac{\partial{r_{EF}}}{\partial{t_{FG}}}\,,\\
 0&=r_{FG}^2-2\frac{\partial{r_{EF}}}{\partial{r_{FG}}}\frac{\partial{t_{EF}}}{\partial{r_{FG}}}-\left(\frac{\partial{t_{EF}}}{\partial{r_{FG}}}\right)^2 g^{EF}_{1,1}\,,
\end{align}
\end{subequations}
which can be solved order by order in $r_{FG}$ and $r_{FG}\log(r_{FG})$ for the expansion coefficients in \eqref{eq:EFFGansatz}.
For the leading behaviour of the metric in FG coordinates we find the following expressions
\begin{subequations}\label{MetricFG}
\begin{align}
 g^{FG}_{1,1}&=-r^2+\frac{\bar{\phi}_{(0)}^2}{3}-\frac{S_0'(t)^2-2S_0(t) S_0''(t)}{2S_0(t)^2}-\frac{\bar{\phi}_{(0)}^2 S_0''(t)}{2S_0(t)}\frac{\log(r)}{r^2}-\left(\frac{\bar{\phi}_{(0)}^4}{36}+\frac{3 a_{(4)}(t)}{4}\right)\frac{1}{r^2}\nonumber\\
  & +\frac{2\bar{\phi}_{(0)}^2S_0(t)^2\big(2S_0'(t)^2-3S_0(t)S_0''(t)\big)-3(S_0'(t)^2-2S_0(t)S_0''(t))^2)}{48S_0(t)^4}\frac{1}{r^2}+O(r^{-4})\,,\\
 g^{FG}_{2,2}&=r^2S_0(t)^2-\frac{1}{3}\bar{\phi}_{(0)}^2S_0(t)^2-\frac{1}{2}S_0'(t)^2+\frac{\bar{\phi}_{(0)}^2}{6}\left(2S_0'(t)^2+S_0(t)S_0''(t)\right)\frac{\log(r)}{r^2}\nonumber\\
  &+S_0(t)^2\left(\frac{19}{108}\bar{\phi}_{(0)}^4-\frac{1}{4}a_{(4)}(t)-\frac{2}{3}\bar{\phi}_{(0)}\bar{\phi}_{(2)}(t)\right)\frac{1}{r^2}\nonumber\\
  &+\left(\frac{27S_0'(t)^4+186\bar{\phi}_{(0)}^2S_0(t)^3S_0''(t)+288\bar{\phi}_{(0)}^2S_0(t)^4\xi(t)^2}{432S_0(t)^2}\right)\frac{1}{r^2}+O(r^{-4})\,,
\end{align}
\end{subequations}
where we have set $r\equiv r_{FG}$ and $t \equiv t_{FG}$ to shorten notation.
After replacing the radial coordinate by $\rho\equiv 1/r^2$ we obtain explicit expressions for the non-vanishing metric components in FG coordinates (\ref{metricFG}) in terms of the coefficients in EF gauge
\begin{equation}
g_{(0)tt}=-1,\quad g_{(0)xx}=S_{0}(t)^2\,,
\end{equation}
\begin{equation}
g_{(2)tt}=\frac{\bar{\phi}_{(0)}^2}{3}-\frac{S_0'(t){}^2-2 S_0(t) S_0''(t)}{2 S_0(t){}^2},\quad g_{(2)xx}=-\frac{2\bar{\phi}_{(0)}^2 S_{0}(t)^2+3S_{0}'(t)^2}{6}\,,
\end{equation}
\begin{subequations}
\begin{align}
g_{(4)tt}&=-\frac{3a_{(4)}}{4}-\frac{\bar{\phi}_{(0)}^4}{36}+\frac{\bar{\phi}_{(0)}^2 \left(2 S_0'(t){}^2-3 S_0(t) S_0''(t)\right)}{24 S_0(t){}^2}-\frac{\left(S_0'(t){}^2-2 S_0(t) S_0''(t)\right){}^2}{16 S_0(t){}^4}\,,\\
g_{(4)xx}&=-\frac{ a_{(4)}(t)S_0(v){}^2}{4} -\frac{2\bar{\phi}_{(0)} \bar{\phi}_{(2)}(t) S_0(t){}^2}{3}+\frac{2}{3}\bar{\phi}_{(0)}^2\xi(v)^2S_0(v)^2\nonumber\\
         &+\frac{31\phi_{(0)}^2 S_0(t) S_0''(t)}{72} +\frac{19\bar{\phi}_{(0)}^4 S_0(v){}^2}{108} +\frac{S_0'(v){}^4}{16 S_0(v){}^2}\,,
\end{align}
\end{subequations}
\begin{equation}
h_{(4)tt}=\frac{\bar{\phi}_{(0)}^2S_{0}''(t)}{4S_{0}(t)},\quad h_{(4)xx}=-\frac{\bar{\phi}_{(0)}^2 \left(2 S_0'(v){}^2+S_0(t) S_0''(t)\right)}{12} \,.
\end{equation}
The corresponding near-boundary expansion of the scalar field reads
\begin{subequations}\label{BoundarySeries}
\begin{align}
\phi(r,t)&=\bar{\phi}_{(0)}\rho^{1/2}+\left(\bar{\phi}_{(2)}(t) -\frac{\bar{\phi}_{(0)}^3}{6}+ \frac{\bar{\phi}_{(0)} \left(S_0'(t){}^2-2 S_0(t) S_0''(t)\right)}{4 S_0(t){}^2}-\bar{\phi}_{(0)}\xi(t)^2\right)\rho^{3/2}\nonumber\\
&+\frac{\bar{\phi}_{(0)} \left(S_0'(t)^2+S_0(t) S_0''(t)\right)}{4S_0(t)^2}\rho^{3/2}\log(\rho)+O(\rho^{5/2})\,.
\end{align}
\end{subequations}
From the above expression one can now read off the relations between the expansion coefficients in FG and EF coordinates
\begin{eqnarray}
\phi_{(0)} &=& \bar{\phi}_{(0)}=M \,,\\[2mm]
\psi_{(2)}(t) &=& -\frac{1}{2}\bar{\psi}_{(2)}(t)\,,\\[1mm]
\phi_{(2)}(t) &=& \bar{\phi}_{(2)}(t)-\frac{\bar{\phi}_{(0)}^3}{6}+\frac{\bar{\phi}_{(0)} \left(S_0'(t){}^2-2 S_0(t) S_0''(t)-4\xi(t)^2S_0(t)^2\right)}{4 S_0(t){}^2}\,.
\end{eqnarray}
Using these relations we can express the non-vanishing components of the holographic stress tensor in terms of the coefficients in the near boundary expansion in EF gauge:
\begin{subequations}\label{eq:EP}
\begin{align}
\mathcal{E}(t)=&\,\,-\frac{3 a_{(4)}(t)}{4}-M \bar{\phi}_{(2)}(t)+\frac{3 S_0'(t)^4}{16 S_0(t)^4} +M^2 \left(\xi(t)^2+\frac{S_0'(t)^2}{8 S_0(t)^2}+\frac{2 S_0''(t)}{3S_0(t)}\right) \nonumber\\
&-M^2 \alpha\frac{S_0'(t)^2}{2 S_0(t)^2}- M^4\left(\beta-\frac{7}{36}\right)\,,\\[2mm]
\mathcal{P}(t)=&\,\,-\frac{a_{(4)}(t)}{4} +\frac{1}{3} M\bar{\phi}_{(2)}(t) +\frac{S_0'(t)^2\left(S_0'(t)^2-4 S_0(t)S_0''(t)\right)}{16 S_0(t)^4}\nonumber\\
 &-\frac{M^2}{3}\left(\xi(t)^2+\frac{S_0'(t)^2}{8 S_0(t)^2}+\frac{13 S_0''(t)}{12 S_0(t)}\right)+M^2\alpha\left(\frac{S_0'(t)^2}{6 S_0(t)^2}+\frac{S_0''(t)}{3S_0(t)}\right)+M^4\left(\beta-\frac{5}{108}\right)\,.
\end{align}
\end{subequations}
Similarly, the expectation value of the operator $\hat{\O}$ is  given by
\begin{align}\label{eq:O}
\O(t)=&\,\, -2 \bar{\phi}_{(2)}(t)+M\left(2\xi (t)^2-\frac{S_0'(t)^2}{4 S_0(t)^2}+\frac{5S_0''(t)}{4 S_0(t)}\right) \nonumber \\[2mm]
&-M\alpha\left(\frac{S_0'(t)^2}{S_0(t)^2}+\frac{S_0''(t)}{S_0(t)}\right)-M^3\left(4\beta-\frac{1}{3}\right)\,.
\end{align}
As mentioned above, the coefficients $\alpha$ and $\beta$ in \eqref{eq:EP} and \eqref{eq:O} encode the scheme-dependence of $\mathcal{E}, \mathcal{P}$ and $\mathcal{O}$.

\subsection{Numerical procedure}
Our main interest is to compute the time evolution of $\mathcal{P}$, $\mathcal{E}$ and $\O$ in a dynamical background geometry.
For this purpose we have to solve the set of equations \eqref{CharEOM} with consistent initial and boundary conditions for the metric and the scalar field.
The preparation of initial states and our specific choice of time-dependent boundary conditions for the metric will be discussed in the next section.
Here we concentrate on the evolution algorithm assuming that the initial data and the boundary metric are known.

The set of equations \eqref{CharEOM} has a nested structure that allows us to treat them on every slice of constant EF time as ordinary differential equations in the radial coordinate for the functions $S$, $\dot S$, $\phi$, $\dot{\phi}$ and $A$.
Knowing these functions on a single time slice is sufficient to evolve them to the next slice.
In practice we do not directly solve \eqref{CharEOM} but rather a set of equations for a new set of functions $\{\tilde{S},\tilde{\dot S},\tilde{\phi},\tilde{\dot{\phi}},\tilde{A}\}$.
These are obtained from the original variables by subtracting all the divergent terms, as well as some finite ones, that are explicitly known from the near-boundary analysis \eqref{BoundarySeries}.
The new functions are therefore manifestly finite and take the form
\begin{subequations}\label{eq:finiteVarianles}
\begin{align}
 A(r,t) = &\,\, r^2+2r\xi(t)+\xi(t)^2-2\xi'(t)+\frac{S_0'(t)^2-2 S_0(t)S_0''(t)}{S_0(t)^2}
 -\frac{2 M^2}{3} \nonumber\\
       &+\log(r)\left(\frac{A_{\log,1}(t)}{r^2}+\frac{A_{\log,2}(t)}{r^3}\right)+\frac{\tilde{A}(r,t)}{r^2}\,,\\
 S(r,t)=&\,\,S_0(t)r+S_0'(t)+\xi(t) S_0(t)-\frac{M^2 S_0(t)}{3r}+\frac{M^2(3\xi(t)S_0(t)-S_0'(t))}{9r^2}\nonumber\\
       &+\log(r)\left(\frac{S_{\log,1}(t)}{r^3}+\frac{S_{\log,2}(t)}{r^4}+\frac{S_{\log,3}(t)}{r^5}\right)+\frac{\log(r)^2}{r^5}S_{\log,4}(t)+\frac{\tilde{S}(r,t)}{r^3} \,, \\
\phi(r,t)=&\,\,\frac{M}{r}-\frac{M\xi(t)}{r^2}+\log(r)\left(\frac{\phi_{\log,1}(t)}{r^3}+\frac{\phi_{\log,2}(t)}{r^4}\right)+\frac{\tilde{\phi}(r,t)}{r^3}\,,\\
\dot{S}(r,t)=&\,\,\frac{1}{2} r^2 S_0(t)+r\left(S_0(t) \xi(t)+S_0'(t)\right)+\frac{3\left(S_0(t)\xi(t)+S_0'(t)\right)^2-M^2 S_0(t)^2}{6 S_0(t)}-\frac{2 M^2 S_0'(t)}{9 r}\nonumber\\
       &+\log(r)\left(\frac{\dot{S}_{\log,1}(t)}{r^2}+\frac{\dot{S}_{\log,2}(t)}{r^3}\right)+\frac{\tilde{\dot{S}}(r,t)}{r^2}\,,\\
\dot{\phi}(r,t)=&\,\,-\frac{M}{2}+\log(r)\left(\frac{\dot{\phi}_{\log,1}(t)}{r^2}+\frac{\dot{\phi}_{\log,2}(t)}{r^3}\right)+\frac{\tilde{\dot{\phi}}(r,t)}{r^2}\,.
\end{align}
\end{subequations} 
The functions $A_{\log,1}(t)$, $A_{\log,2}(t)$, etc.~are explicitly known from the near-boundary analysis but are too long to be displayed here.
Furthermore, we switch to the inverse radial coordinate $z\equiv 1/r$ where the boundary is located a $z=0$. 

We now implement the following procedure to solve the initial value problem:
\begin{enumerate}
\item For a given radial profile of the scalar field $\tilde{\phi}(z,t_0)$ at some initial time $t_0$ we first solve the second-order Hamiltonian constraint equation \eqref{eq:Spp} for $\tilde{S}(z,t_0)$.
In principle this differential equation requires boundary conditions, but in our (subtracted) formulation this equation (and the following except the one for $\tilde{\dot{S}}$) has regular singular points, and demanding regularity of the solution fixes the boundary condition.
The simplest way to achieve this is to use spectral methods (see below), which are by construction regular.
\item Next we use $\tilde{\phi}(z,t_0)$ and $\tilde{S}(z,t_0)$ in \eqref{eq:Sd} and solve for $\tilde{\dot{S}}(z,t_0)$. 
The boundary condition for this function reads
\begin{eqnarray} 
\tilde{\dot{S}}(z=0,t_0)& =&\frac{1}{36} S_0(t_0) \Big[18 M \left(\bar{\phi }_2(t_0)-M \xi(t_0)^2\right)+18 a_4(t_0)-5 M^4\Big]\nonumber\\[1mm]
&&+\frac{1}{144} M^2 \Big[32 \xi(t_0) S_0'(t_0)-61 S_0''(t_0)\Big]+\frac{3 M^2 S_0'(t_0)^2}{16 S_0(t_0)}\,.
\end{eqnarray}
Evaluating this requires knowledge of $a_4(t_0)$ and $\bar{\phi }_2(t_0)$, whereby $a_4(t_0)$ is required as a separate initial condition and $\bar{\phi }_2(t_0)$ needs to be extracted from the initial data for the scalar field.
\item Once $\tilde{\phi}(z,t_0)$, $\tilde{S}(z,t_0)$ and $\tilde{\dot{S}}(z,t_0)$ are known we  solve \eqref{eq:phid} for $\tilde{\dot{\phi}}(z,t_0)$.
\item Once $\tilde{\phi}(z,t_0)$, $\tilde{S}(z,t_0)$, $\tilde{\dot{S}}(z,t_0)$ and $\tilde{\dot{\phi}}(z,t_0)$ are known we can solve \eqref{eq:App} for $\tilde{A}(z,t_0)$. 
\item Once $\tilde{A}(z,t_0)$ and $\tilde{\dot{\phi}}(z,t_0)$ are known, we use the definition of the dot-derivative 
\be
\tilde{\dot{\phi}}(z,t_0)=\partial_t\tilde{\phi}(z,t_0)+\frac{1}{2}z^2\tilde{A}(z,t_0)\partial_z\tilde{\phi}(z,t_0)
\ee
to solve for time derivative of the initial data, $\partial_t\tilde{\phi}(z,t_0)$.
This equation depends on the gauge choice $\xi'(t_0)$, which we fix by demanding that the apparent horizon stays at a constant value of the radial coordinate.
This is done by solving $\partial_t \dot{S}(z_{\rm AH}, t) = 0$ for $\xi'(t_0)$.
\item Finally, we  compute $a_4'(t_0)$ from \eqref{Econs}, whereby we obtain $\phi_2'(t_0)$ from the near-boundary expansion of $\partial_t\tilde{\phi}(z,t_0)$.
We subsequently evolve the initial data to the next time slice $t_1=t_0+\Delta t$ for example via
\be
a_4(t_1)=a_4(t_0)+a_4'(t_0)\Delta t \,,\qquad  \tilde{\phi}(z,t_1)=\tilde{\phi}(z,t_0)+\partial_t\tilde{\phi}(z,t_0)\Delta t \,,
\ee
and the start over at the first entry in this list.
\end{enumerate}

The only element missing in the above discussion is the initial value of the gauge function $\xi(t)$. 
We use this freedom to fix the initial apparent horizon (AH) at $z_{\rm AH}=1/2$. 
This is done by computing $\dot{S}$ for the gauge $\xi_0(t_0)=0$ first, and then numerically solving the apparent horizon equation $\dot{S}(z_{\rm hor},\,t_0) = 0$ (see also \eqref{eq:ah}).
We then update 
\be
\xi_1(t_0) = r_{\rm AH} - r_{\rm hor} = \xi_0(t_0) + 2 - 1/z_{\rm hor} \,, 
\ee
where $z_{\rm hor} = 1/r_{\rm hor}$ is the current location of the apparent horizon. 
To increase precision this procedure is repeated a few times, each time using an updated $\xi(t_0)$.

We solve the resulting equations numerically with a Chebyshev pseudo-spectral method (see e.g.~\cite{boyd01}) using typically $N=60$ grid points in the radial direction.
For the stepping between time slices (sixth entry in the list above) we use a fourth-order Adams--Bashforth method \cite{press2007numerical} with a time step $\Delta t=1/(10 N^2)$.
After each time step we evaluate the momentum constraint \eqref{eq:Sdd} to monitor the accuracy of the numerical evolution.

\subsection{Initial states and time evolution}
\label{sec:InialEvolution}
We are interested in studying the time evolution of gauge theory states in dS$_4$.
There are of course many ways to construct initial states for this evolution. 
The strategy that we will follow is to start with thermal equilibrium states in flat space. 
We will then smoothly increase the value of $H$ at the boundary from $H=0$ to its desired value in each case.
This leads to a transient period of time in which $\dd H /\dd t \neq 0$ and the boundary geometry interpolates between flat space and dS$_4$.
After this time $H$ becomes constant and we are in the desired situation of studying the dynamics of an excited gauge theory state in de Sitter space. 
A natural question that arises is whether the resulting state at late times is sensitive to the specific way in which we prepare the initial state.
One of our main results is that the answer to this question is negative. In this sense the way in which we initialise the evolution implies no loss of generality. 

The thermal equilibrium states in flat space could be constructed in FG coordinates with the procedure presented in Sec.~\ref{subsec:Thermo} and then numerically transformed to the EF gauge \eqref{metricEF}, which is better suited for time evolution. 
However, in practice it is simpler to construct the solutions directly in EF coordinates by relaxation. 
To do this, we start with some initial guess for $a_4(t_0)$ and for the scalar field profile $\tilde\phi(z,t_0)$.
This is in general an excited state, which we then evolve with flat boundary conditions for the metric ($S_0=1$) using the algorithm outlined in the previous section.
After a few units of $tM$ the state relaxes to thermal equilibrium. 
\begin{figure}[t]
\center
 \includegraphics[width=0.33\linewidth]{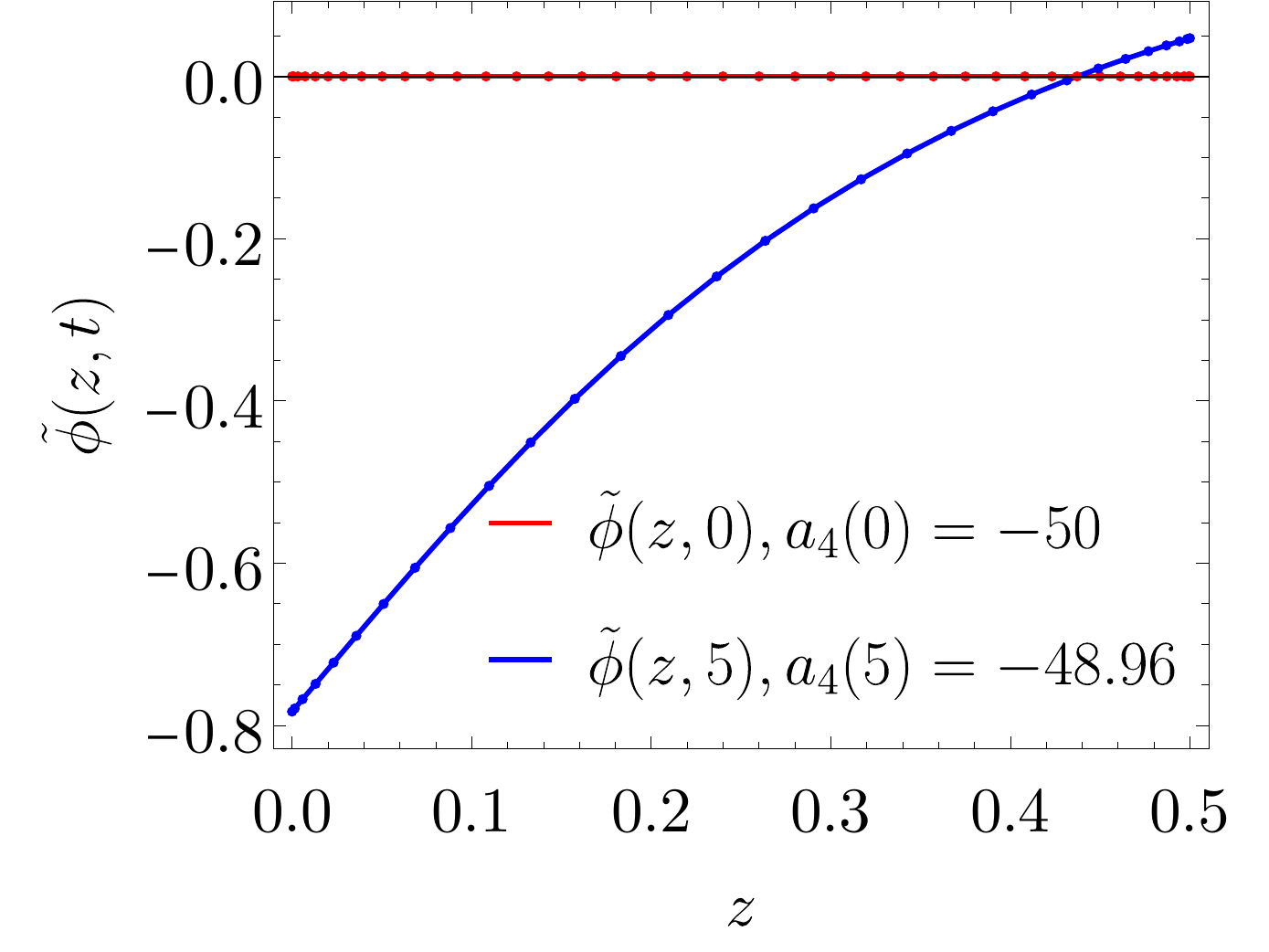}\includegraphics[width=0.33\linewidth]{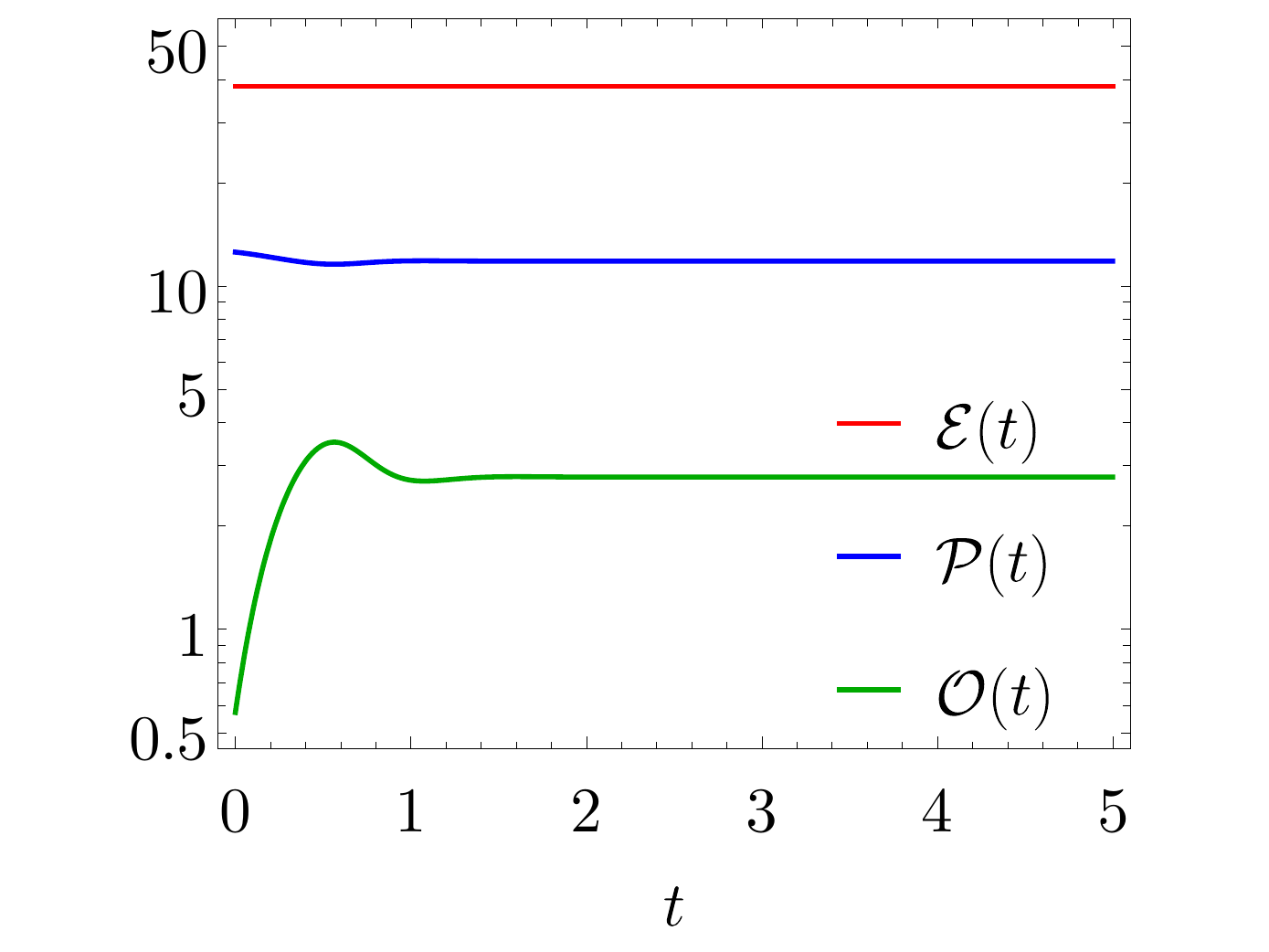}\includegraphics[width=0.33\linewidth]{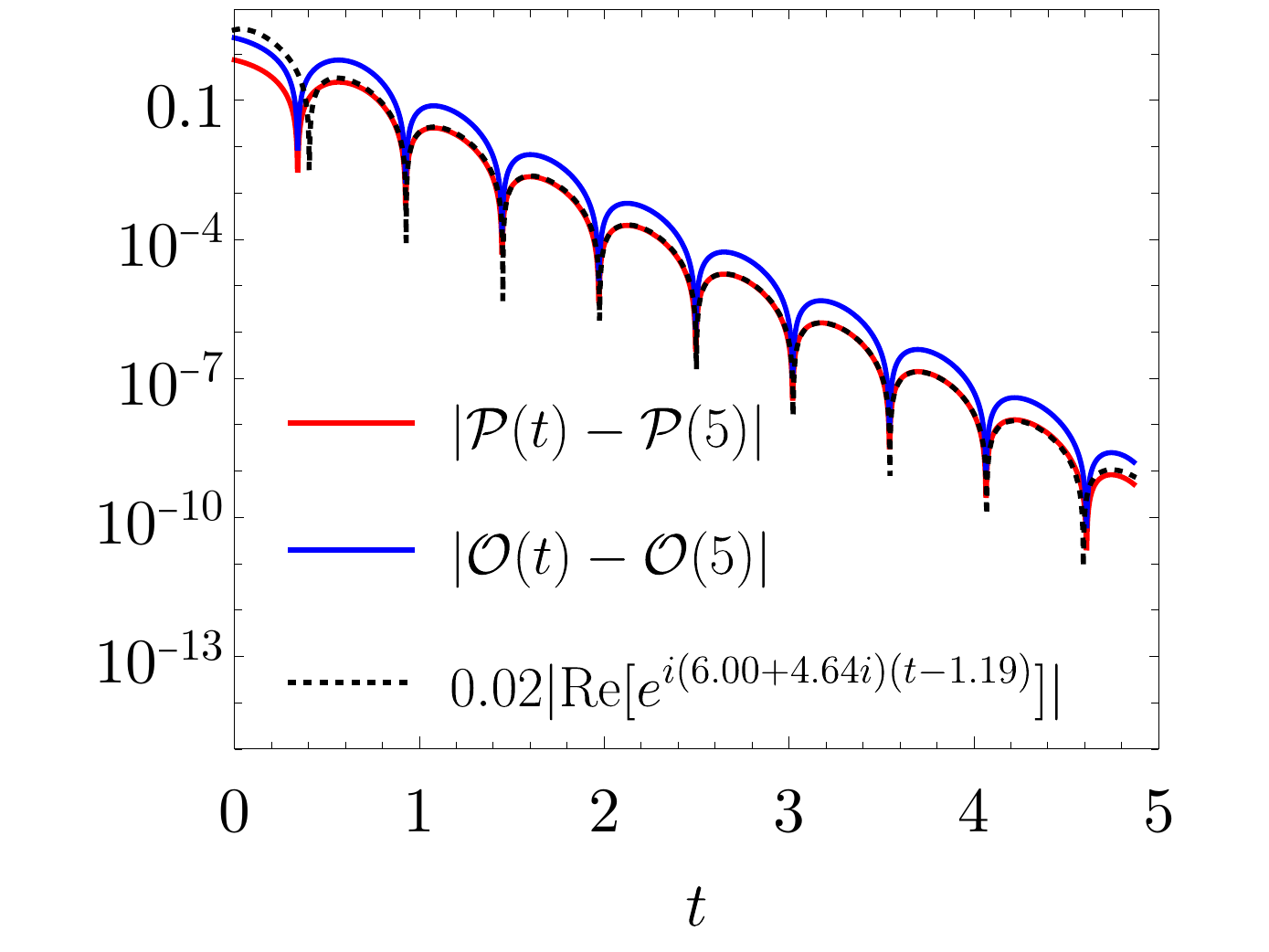}
\caption{Construction of initial states in flat space.
All dimensionful quantities are given in units of $M$ and we choose the renormalisation scheme $\beta=1/16$. 
(Left) Initial guess (red) and resulting thermalised equilibrium configuration (blue) for the scalar field profile.
(Center) Evolution of $\calE(t)$, $\calP(t)$ and $\O(t)$.
(Right) Ring-down of $|\O(t)-\O(5)|$ (blue) and $|\calP(t)-\calP(5)|$ (red) together with an exponential fit for the lowest quasinomal mode $\omega_1\approx 6.00+4.64i$ (doted black). 
}
\label{fig:initialState}
\end{figure}
An example of this procedure is shown in Fig.~\ref{fig:initialState}, where all dimensionful quantities are given in units of $M$ and we have chosen $\beta=1/16$. 
In the left plot we show the initial guess $\tilde\phi(z,t_0)=0$ (red) together with the thermalised equilibrium result $\tilde\phi(z,t=5)$ (blue).
Dots indicate the non-equidistant distribution of points on the Chebyshev grid in the radial direction.
For the subsequent evolution with time-dependent boundary metric at least 60 grid points together with 70-digit-accurate arithmetic is used.
In the middle plot we show how $\calE$, $\calP$ and $\O$ evolve towards their equilibrium values.
Energy conservation in flat space and homogeneity of the state imply that the energy density $\mathcal{E}$ remains constant during the evolution.
The evolutions of $\calP$ and $\O$ are not independent but are constrained by the Ward identity \eqref{eq:WardO}.
Close to the equilibrium state both oscillate according to the quasinormal ring-down shown in the plot on the right.
Our numeric evolution allows us to extract an estimate for this lowest quasinormal mode at zero momentum given by $\omega_1\approx 6.00 +4.64i$.

\begin{figure}[t]
\center
 \includegraphics[height=0.31\linewidth]{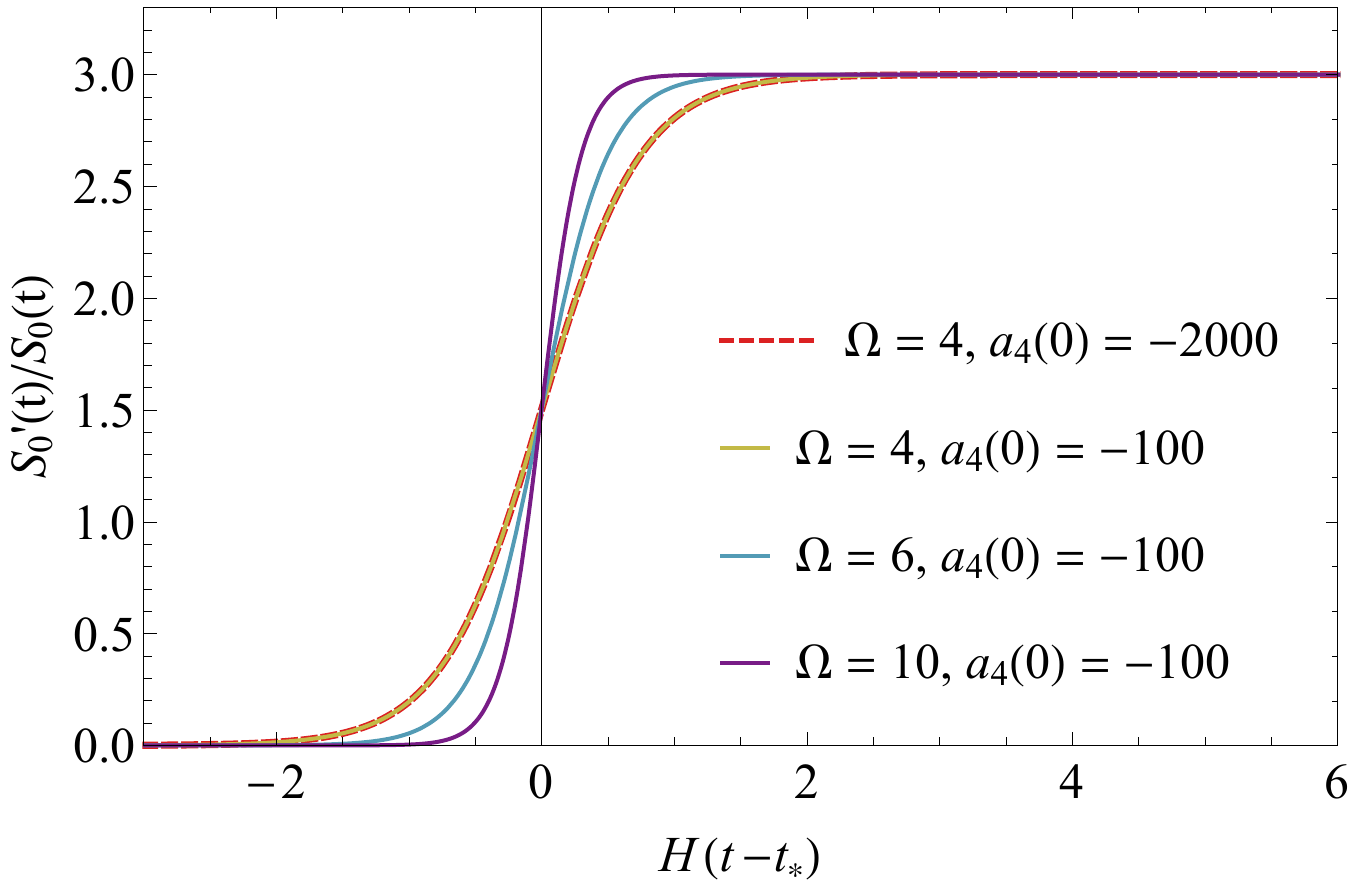}\quad
 \includegraphics[height=0.31\linewidth]{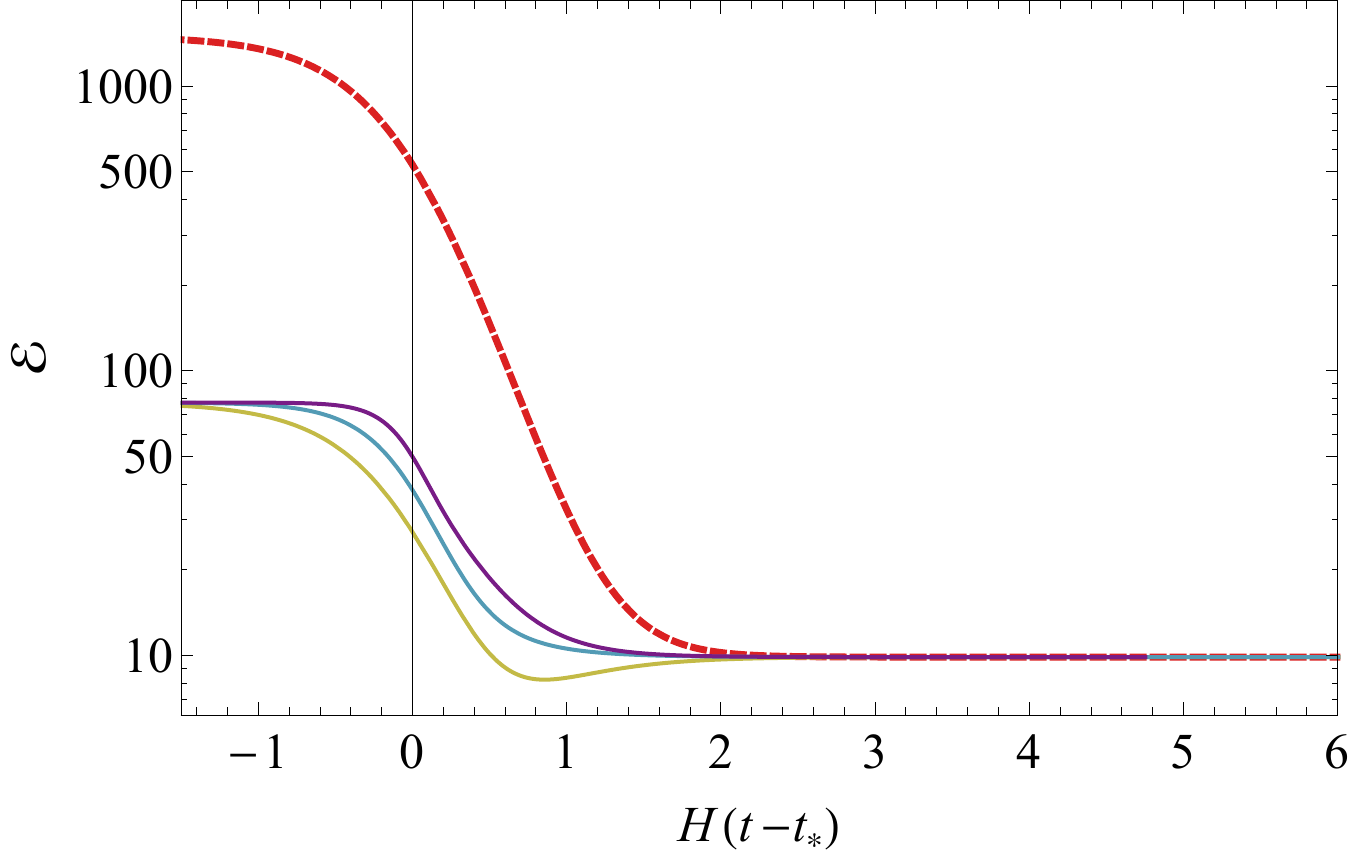}
\caption{(Left) Different protocols for $S_0'(t)/S_0(t)$, all using our maximal value of $H=3$. (Right) All protocols lead to the same late-time value of $\mathcal{E}$.
}
\label{Fig:protocol}
\end{figure}

We now turn on the Hubble rate at the boundary so that the boundary metric changes smoothly from Minkowski to dS$_4$.
We implement this by imposing the following relation on the function $S_0(t)$ in the boundary metric \eqref{eq:sFRW}
\be\label{eq:protocol1}
\frac{S_0'(t)}{S_0(t)}= H \, \frac{\Big[ 1+\tanh \Big(\Omega\,(t-t_\ast) \Big)\Big] }{2}.
\ee
This relation mimics a time-dependent Hubble rate that changes from zero at $t\ll t_\ast$ to $H$ at $t\gg t_\ast$ in a time of order $1/\Omega$.
We will refer to this transient period as ``the quench''.   
The corresponding form of  $S_0(t)$ follows from integrating \eqref{eq:protocol1} subject to the initial condition $S_0(t=0)=1$
\begin{equation}\label{eq:S0}
S_0(t)=e^{\frac{H t}{2}}\left[\frac{\cosh \Big( \Omega(t-t_\ast) \Big)}{\cosh(\Omega t_\ast)}\right]^{\frac{H}{2\Omega}}\,.
\end{equation}
In Fig.~\ref{Fig:protocol} we illustrate this procedure for several choices of parameters. 
We will refer to each of these choices as  a ``protocol''.
We choose a theory with $H/M=3$, we measure all dimensionful quantities in the figure in units of $M$ and we fix the renormalisation scheme by choosing $\alpha=0, \beta=1/16$. 
On the left plot we show the ratio $S_0'(t)/S_0(t)$ and on the right the evolution of the energy density. 
We vary both the ``quench parameter'' $\Omega$, which controls the length of the transient period, and the parameter $a_4$, which controls the initial energy density. 
We find that the state at $t\gg t_\ast$ does not depend on the values of these parameters.  
In particular, we see in Fig.~\ref{Fig:protocol} (right) that the energy density at late times always approaches the same value.
This convergence to the same late-time state is remarkable in view of the vastly different values of the initial energy density, and it means that the late-time state only depends on $H$ and $M$.
We will explore this dependence in our subsequent simulations.
Given the independence of the initial conditions, in most simulations we will  use the same values $\Omega = 4$, $t_\ast = 1$ and $a_4(t=0) = -100$. 
The only exception will be when $H \geq 2$, in which case we will use $\Omega = 8$ in order to have a quench time $\Omega^{-1}$ that is sufficiently shorter than the expansion rate $H^{-1}$. 

\begin{figure}[t]
\center
 \includegraphics[width=0.48\linewidth]{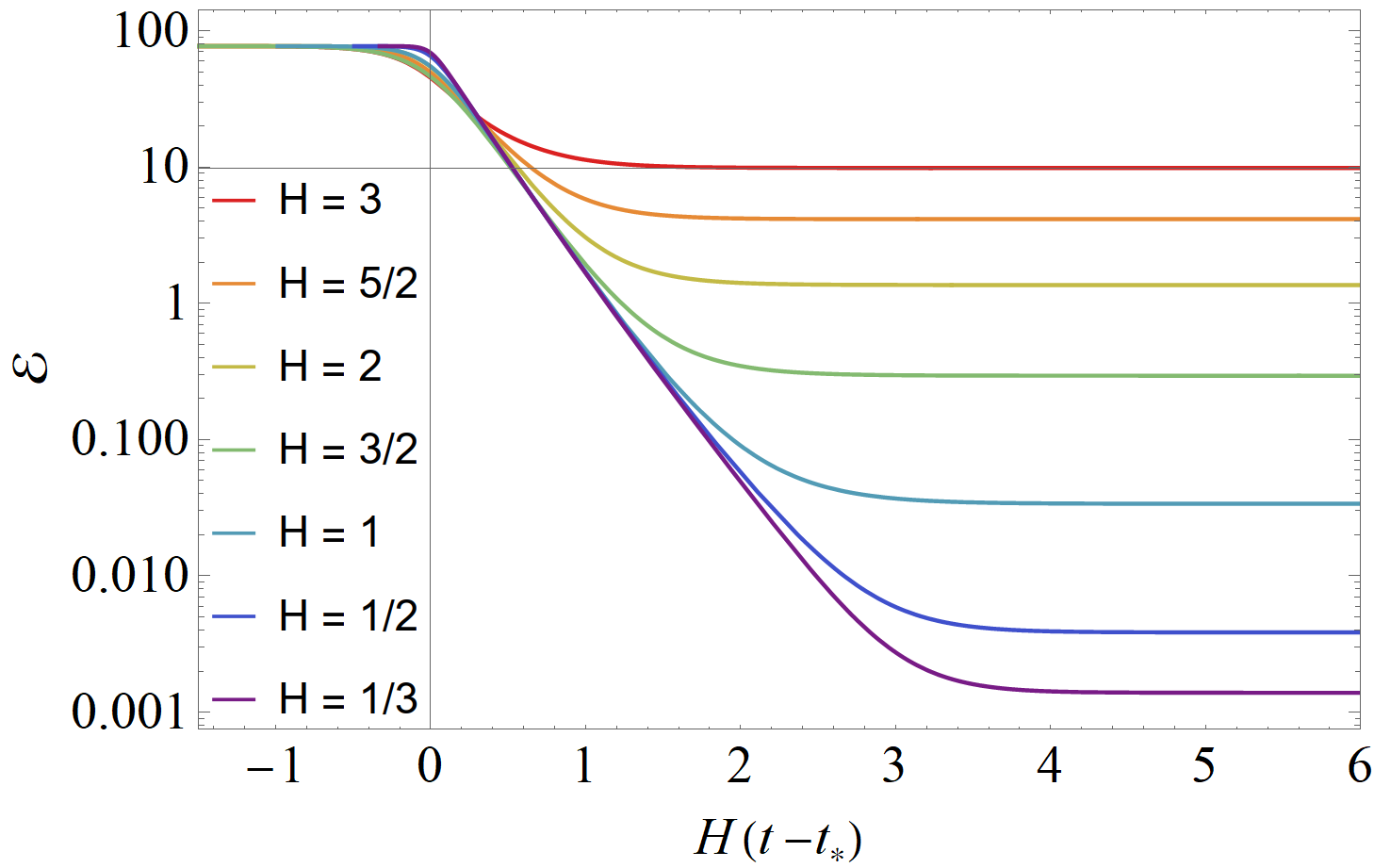}
 \includegraphics[width=0.48\linewidth]{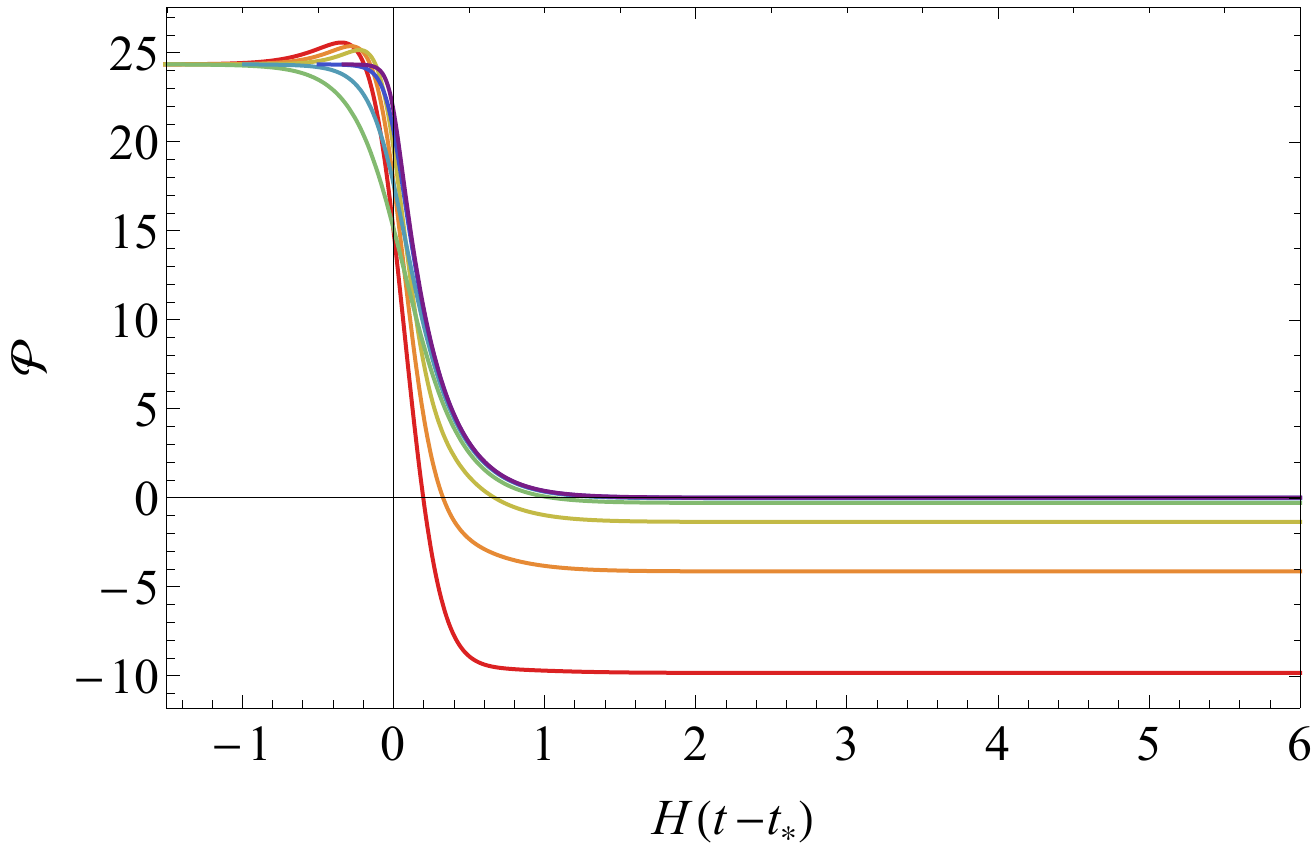}
 \includegraphics[width=0.48\linewidth]{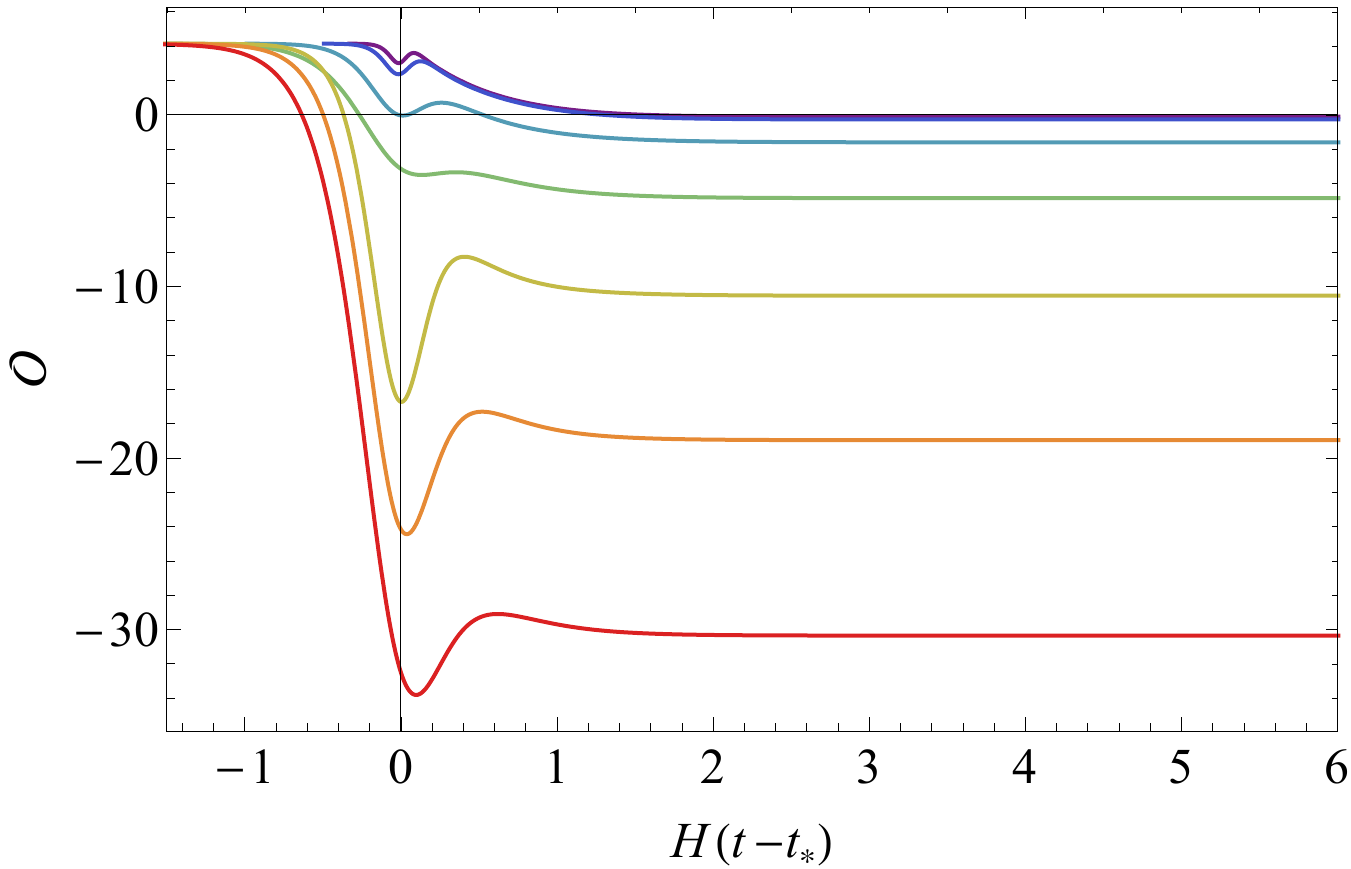}
 \includegraphics[width=0.48\linewidth]{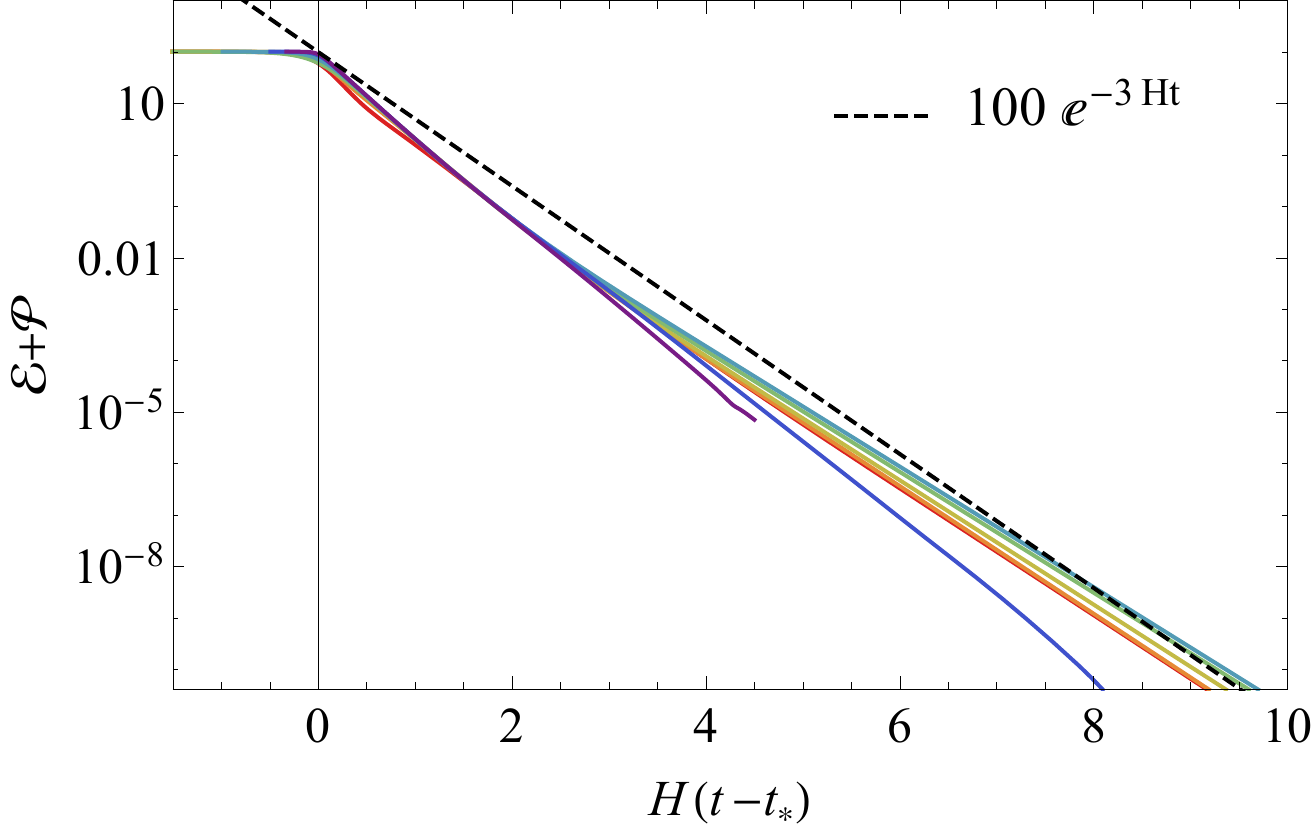}
\caption{(Top) Energy density (left) and pressure (right) as functions of time for different values of $H$.
(Bottom) Expectation value of the scalar  operator (left) and the combination $\mathcal{E}+\mathcal{P}$ as functions of time for different values of $H$. 
The black dashed line is the expansion factor $e^{-3 H t}$ of a fluid with zero pressure. 
All dimensionful quantities are measured in units of $M$ and we set $\alpha=0, \beta=1/16$.
}
\label{Fig:numevol}
\end{figure}

After a time $t-t_\ast \gg \Omega^{-1}$ the boundary geometry settles down to dS$_4$ and we can analyse the evolution of field theory states on this expanding background.
In Fig.~\ref{Fig:numevol} we show the evolution of $\mathcal{E}$, $\mathcal{P}$, $\mathcal{O}$ and $\mathcal{E}+\mathcal{P}$ (which in equilibrium would be enthalpy) for a number of different values for the  expansion rate $H$.
Shortly after the quench, the energy density and the pressure start to decrease rapidly due to the expansion of space.
At some later time the initial energy of the plasma is almost entirely depleted.
From this point onward the energy and the pressure are  dominated by their (scheme-dependent, see below) vacuum contributions equivalent to a pure cosmological constant, meaning that at late-times $\mathcal{E}=-\mathcal{P}$.
The evolution of $\mathcal{O}$ is entirely fixed in terms of the energy density and the pressure by the Ward identity \eqref{eq:Ward}. 
The scheme-dependence of the late-time limits
\begin{equation}
\mathcal{E}_\infty\equiv\lim\limits_{t\to\infty}\mathcal{E}(t)\,, \qquad \mathcal{P}_\infty\equiv\lim\limits_{t\to\infty}\mathcal{P}(t)\,, \qquad \mathcal{O}_\infty\equiv\lim\limits_{t\to\infty}\mathcal{O}(t)
\end{equation}
is encoded in the dependence on $\alpha$ and $\beta$ of Eqs.~\eqref{eq:EP}.
As mentioned above, in order to fix the scheme we choose $\beta=1/4\phiM^2=1/16$ throughout the paper.
Moreover, in this section (but not elsewhere) we set $\alpha=0$.
Note that the enthalpy $\mathcal{E}+\mathcal{P}$ is scheme-independent because the dependence on $\alpha$ and $\beta$ cancels out in this combination.
In Fig.~\ref{Fig:numevol} (bottom right) we see that $\mathcal{E} + \mathcal{P}$ decays for any value of $H$, where to guide the eye we have added a black dashed line proportional to $e^{-3 H t}$ (appropriate for a pressureless fluid).

As mentioned above, we set $\alpha=0$ only in this section.
In the following we will use variables which have their late time values subtracted.
In other words, we will measure energy, pressure, etc with respect to their late-time asymptotic values.
These subtracted quantities have the advantage that they are scheme-independent because the dependence on $\alpha,\beta$ cancels out.
Conceptually, the reason for this is that the scheme dependence is the same for any state.
For any given values of $H, M$, working with subtracted variables is equivalent to choosing a scheme in which
\be
\alpha = -\frac{2}{M^2 H^2}\mathcal{E}_{\infty}|_{\alpha=0},\label{eq:alpha}
\ee
with $\mathcal{E}_{\infty}|_{\alpha=0}$ the final numerical value of $\mathcal{E}$ in Fig.~\ref{Fig:numevol}. 
Note that we treat different values of $H, M$ as corresponding to different theories for each of which a separate renormalization scheme can be chosen.
None of our later results will hence directly compare stress tensors for different values of $H$ or $M$.
The logic behind the choice \eqref{eq:alpha} is similar to the logic behind the choice $\beta=1/4\phi_M^2$. 
In the latter case we require the energy and pressure of the vacuum state to vanish. 
In the case of \eqref{eq:alpha} we require the energy and the pressure of the late-time, asymptotic state to vanish. 
In this scheme, the values of the energy density and the pressure during the evolution can be interpreted as those of excitations on top of the late-time state.
In the next subsection we will study the dynamics of these excitations.

\subsection{Hydrodynamic regime}
\label{sec:dS4Hydro}
One of the lessons of holographic studies is that hydrodynamics becomes a good description of a plasma once the quasi-normal modes of the system have decayed.
At strong coupling, the decay time is of order of the inverse temperature $1/T$ in both conformal \cite{Heller:2011ju,Chesler:2009cy,Chesler:2010bi} and non-conformal \cite{Buchel:2015saa,Attems:2016ugt} theories. 
Since the expansion rate is of order $1/H$, we expect that hydrodynamics will provide a good description provided that $T > H$. 
In this regime the expansion can then be seen as an almost-adiabatic process in which local properties of matter are close to those in equilibrium.
Under these circumstances the expectation value of the energy-momentum tensor of the system can be approximated in terms of the gradient expansion discussed in Sec.~\ref{sec:hydro}.
If $H\gg M$ or $H\ll M$ then the energy density at which the hydrodynamic description ceases to be valid is close to the UV or to the IR fixed point of the gauge theory.
In these regions the dynamics is quasi-conformal, the bulk viscosity is close to zero, and the relation between energy and pressure is essentially determined by symmetry.
It follows that the most interesting range of parameters is $H \sim M$, which in our units means $H \sim 1$. 
Therefore we will focus on the evolution of states with initial energy densities of order 1 and we will vary the value of $H$ from a few times larger to a few times smaller than 1.
As suggested by the discussion above, we will find a qualitative change in the applicability of hydrodynamics around $H = 1$. 

In order to compare the evolution of the holographic energy-momentum tensor to the hydrodynamic approximation it is convenient to work with hydrodynamic variables that are manifestly scheme-independent, namely independent of $\alpha$. 
To do this we start with the general form of the energy-momentum tensor for homogeneous and isotropic states
\be
\label{TmunuHgen}
\langle \hat{T}^{ij}\rangle=\frac{N^2}{2\pi^2}\left\{\left(\calE(t) + \calPH (t)\right) u^i u^j +  \calPH (t) g^{i j}_{(0)}\right\} \,,
\ee
where $u^i=\left(1, \, \vec{0}\right)$ is a future pointing time-like vector and $\calE(t)$ and $\calPH(t)$ are reduced energy density and pressure in the rest frame defined by $u^i$, respectively.
This general form is independent of the hydrodynamic approximation and applies in particular to states preserving the symmetries of dS$_4$, which in addition to \eqref{TmunuHgen} also satisfy the relation $\calE(t)=-\calP(t)$.
As shown in Fig.~\ref{Fig:numevol} (bottom right) and studied in more detail in Sec.~\ref{sec:latetimesol}, all our states are attracted to states preserving the symmetries of dS$_4$ such that the energy-momentum tensor of these late-time states can be written as
\be\label{eq:Tlate}
\langle \hat{T}^{ij}_\infty\rangle = \frac{N^2}{2\pi^2} \, \calPinf g^{ij}_{(0)}\,,
\ee
where $\calPinf$ is scheme-dependent because it depends on $\alpha$.
In the spirit of Sec.~\ref{sec:hydro} we build scheme-independent combinations of the energy-momentum tensor components in which all dependencies on $\alpha$ cancel out:
\be\label{eq:DeltaT}
 \langle\Delta \hat{T}^{ij}\rangle = \langle \hat{T}^{ij}\rangle - \langle \hat{T}^{ij}_\infty\rangle\,, \quad \Delta \calE= \calE+ \calPinf\,, \quad \Delta \calP= \calP- \calPinf\,.
\ee
These are the variables we use to build the hydrodynamic benchmark we compare to the evolution obtained from solving the time-dependent dual gravity problem.
Using \eqref{eq:DeltaT} in \eqref{eq:hydroDe} we obtain the hydrodynamic approximation for the pressure in the Landau frame\footnote{Both bulk viscosities in \eqref{eq:Phyd} and \eqref{eq:hydro} are manifestly scheme independent.
A subtle question regarding scheme dependence is however if these two functions are truly the same functions. 
Even though they have both been obtained by the same prescription (defining energy densities with respect to the maximally symmetric state), these backgrounds are not precisely equal.
Nevertheless, since the scheme dependence has the form $\alpha H^2$ any difference will scale as $H^2$ and can hence be attributed to higher order viscous effects that we do not take into account in this paper.
}
\be
\label{eq:Phyd}
\Delta \calPhyd (t)\equiv \frac{N^2}{2\pi^2}\, 
\Big\{\Delta p_{\mathrm{eq}}(\Delta \calE(t)) - 3H\zeta(\Delta \calE (t))\Big\}  + O(H^2)\,, 
\ee 
where we have used $\overline{\nabla}_i u^i= 3H$. 
This expression shows that $H$ controls the size of spatial gradients in the velocity field, where $H=0$ gives the leading term in the gradient expansion which only depends on the equilibrium pressure on flat space. 
In the following we study a number of specific examples to see how well this approximation with and without the viscous contribution agrees with the exact solution obtained from the full holographic simulation.

\begin{figure}[t]
\center
 \includegraphics[width=0.99\linewidth]{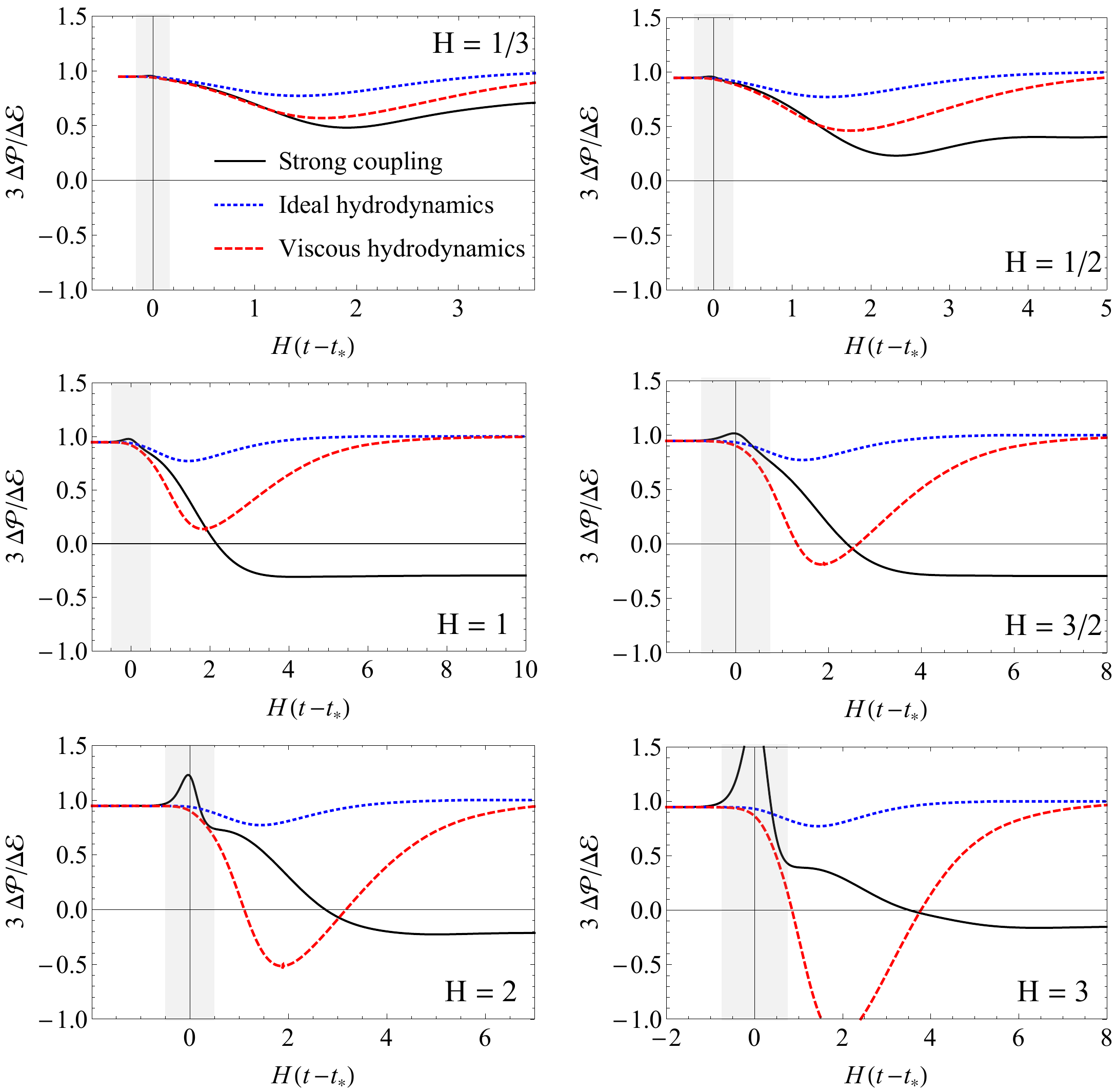}
\caption{Scheme-independent ratio of the excess pressure over the excess energy density defined in \eqref{eq:DeltaT} as a function of time for the same initial state but different values of $H$.$^{\ref{more}}$
The vertical grey bands indicate the duration of the quench. The solid black curves labelled ``strong coupling'' correspond to the exact result obtained holographically.  
Dotted blue and dashed red curves correspond to the ideal and viscous hydrodynamic approximations, respectively. 
In the plots with $H<1$ viscous hydrodynamics provides a good approximation to the exact result for some time after the quench. During part of this period the difference between ideal and viscous hydrodynamics is of order unity, meaning that gradient corrections are as large as the ideal terms.
In the plots with $H\geq 1$ viscous hydrodynamics never provides a good approximation to the exact result.  
}
\label{Fig:hydrocomp}
\end{figure}

Fig.~\ref{Fig:hydrocomp} shows the time evolution of  the excess pressure  divided by the excess energy density, which in equilibrium is determined by the equation of state, for six different values of $H$.\footnote{More precisely, what is shown is actually $3 \mathcal{P}/\mathcal{E}$ for the value of $\alpha$ specified in \eqref{eq:alpha} which, after the quench, coincides with $3 \Delta \mathcal{P}/\Delta \mathcal{E}$.\label{more}} 
Recall that all dimensionful quantites are measured in units of $M$. 
The black lines show the exact, strongly coupled evolutions obtained holographically, the red dashed lines show the viscous hydrodynamic approximation according to \eqref{eq:Phyd} and the blue dotted lines show the ideal hydrodynamic approximation according to the equation of state $\Delta \calPeq(\Delta \calE(t))$.
Note that the latter approaches unity at very early and very late times, as expected from the existence of the UV and IR fixed points in the gauge theory (see Fig.~\ref{fig:eos}), but deviates from this value in between due to the non-conformal nature of the gauge theory.
Before the quench all states are in thermal equilibrium on Minkowski and as expected all curves agree. 
Although after the quench the gradients due to the dS expansion become important, for $H < 1$ there is a period of time during which the results are well described by viscous hydrodynamics. 
For example, for $H=1/3, 1/2$ this agreement extends to times around $H(t-t_*) \sim 1.4$. 
It is remarkable that around these times the difference between ideal and viscous hydrodynamics is of order one, indicating that the first-order viscous corrections are as large as the ideal terms. 
This provides another example of hydrodynamics working with large gradients \cite{Heller:2011ju,Chesler:2010bi,Casalderrey-Solana:2013aba}, in this case in a dynamical spacetime.
For $H>1$ the evolution after the quench is never well described by hydrodynamics and we conclude that the evolution is far from equilibrium.

\begin{figure}[t]
\center
 \includegraphics[width=0.95\linewidth]{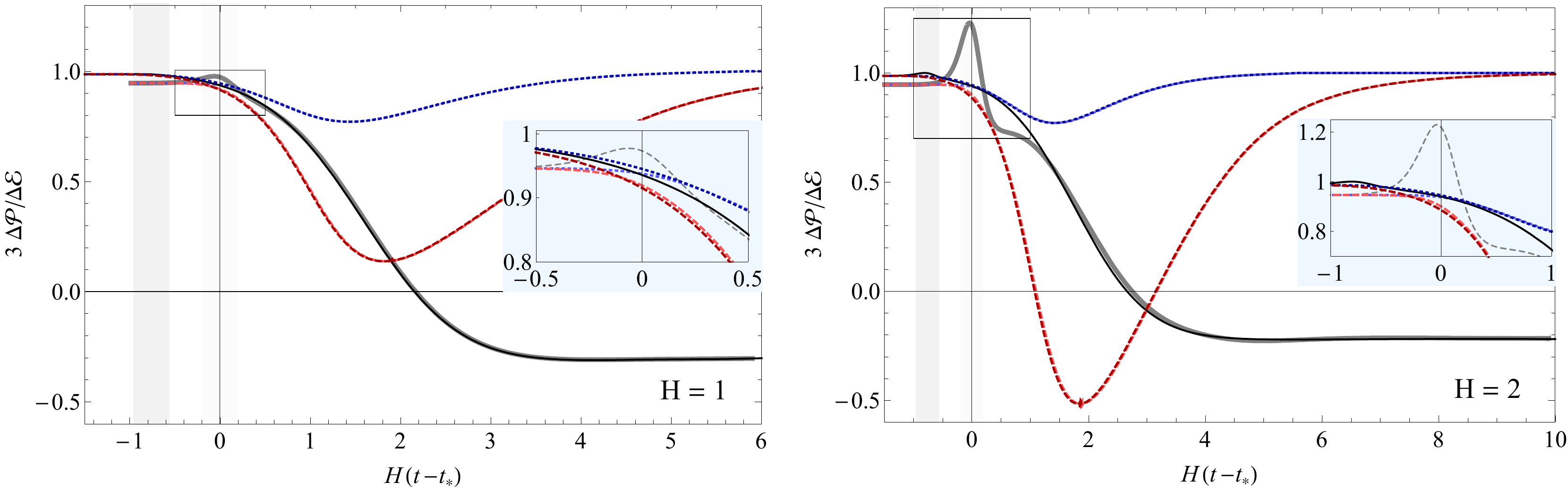}
\caption{For $H=1$ (left) and $H=2$ (right) we compare the  result of Fig.~\ref{Fig:hydrocomp} with an equivalent run that starts with an energy density 19.3 times higher.
The original run is displayed in lighter colors and gray, whereas the high-energy run is displayed in dark colors and black. 
The high-energy run is shifted by an extra time of $H \Delta t = 0.76$ for both figures, in order to guarantee that the energy evolution of both simulations agree at late times.
After this shift in time there is virtually no difference between the results, except small differences before the quenches (displayed gray and dark gray).
}
\label{Fig:hydrocomp2000}
\end{figure}

It is  interesting to ask how the results of Fig.~\ref{Fig:hydrocomp} would change when using a higher initial energy density, which for $H=1, 2$ and a 19.3 times higher initial energy density is shown in Fig.~\ref{Fig:hydrocomp2000}.
In that case the quench has a more moderate effect on the state and the energy density will take longer to cool down under the exponential expansion. 
Both these effects can at late times be incorporated by an extra shift of $H \Delta t\approx 0.76$, after which the $\Delta \mathcal{E}$ evolutions agree and hence also the viscous hydrodynamic results.
Before the quench the energy densities are different and correspondingly the equation of state, as visible in the figure.
It is also visible that the lower energy density has a larger effect from the quench (small wiggle around $t=0$), even though the effect is modest. 
This comparison clearly shows that the statement that viscous hydrodynamics provides a good approximation at early times for small expansion rates $H$ is valid independent of the initial energy density, but it has to be interpreted at a time where viscous corrections are sizeable, whereby this time can depend on the initial energy density.

Importantly, as time passes the energy density in the expanding background decreases.
At some point the energy density will become of the same scale as the gradients, and hydrodynamics need no longer apply.
Indeed, we find that at later time as $\Delta \calE \rightarrow 0$ the full evolution always deviates from the hydrodynamic evolution and can characteristically develop a negative pressure excess for $H\gtrsim 1$.
Note that since we compare the pressure excess the sign of this pressure is unrelated to the negative pressure of a cosmological constant contribution, rather it means that the final pressure in Fig.~\ref{Fig:numevol} is approached from below.
We stress that this negative pressure is a completely out-of-equilibrium effect which may not be inferred directly from the equilibrium dynamics of the plasma in flat space.
The next subsection will describe this late time evolution in more detail.

\subsection{Effective equation of state and quasinormal modes}
\label{sec:ffeq}
In the previous section we have shown that the late-time evolution of our system is not well described by hydrodynamics.
In this section we will demonstrate that it is instead possible to describe the dynamics in terms of small perturbations around the maximally symmetric late-time state.
Because the late-time state has a dual description in terms of a geometry with a horizon, the dynamics of small perturbations in the gauge theory is determined by ``quasi-normal modes'' (QNM) of the dual black brane.
We have used quotation marks in the previous sentence to emphasize that we are slightly abusing the nomenclature, because the term QNM is normally used in the context of stationary solutions, whereas the late-time, black brane solution of to us is not stationary.
Nevertheless, the fact that the time dependence of our solution is only ``along the spatial gauge theory directions'' leads to many familiar properties for excitations that are homogeneous along these directions.
The spectrum of such excitations for a different non-conformal theory on de Sitter space was computed holographically in \cite{Buchel:2017lhu} via a perturbative expansion around the conformal limit.
The low-lying QNMs were found to be purely imaginary, with the mysterious exception of the third mode, which was found to have non-vanishing real and imaginary part.
While we will not attempt a direct calculation of the QNM spectrum of our model in this paper, we will extract an estimate for the lowest QNM from the numerical solution and confirm its purely dissipative nature.

\begin{figure}[t]
\center
\includegraphics[width=0.48\linewidth]{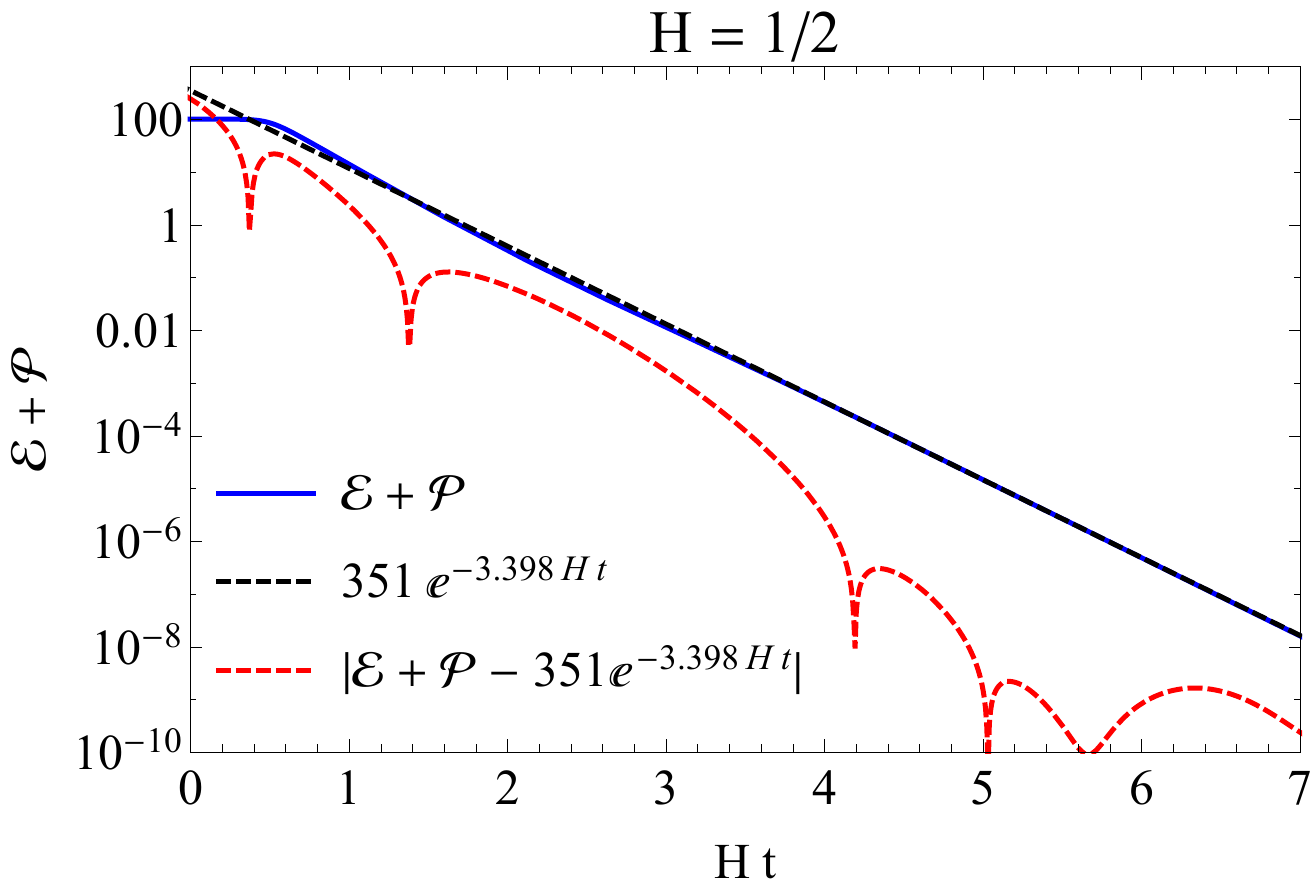}\quad
\includegraphics[width=0.48\linewidth]{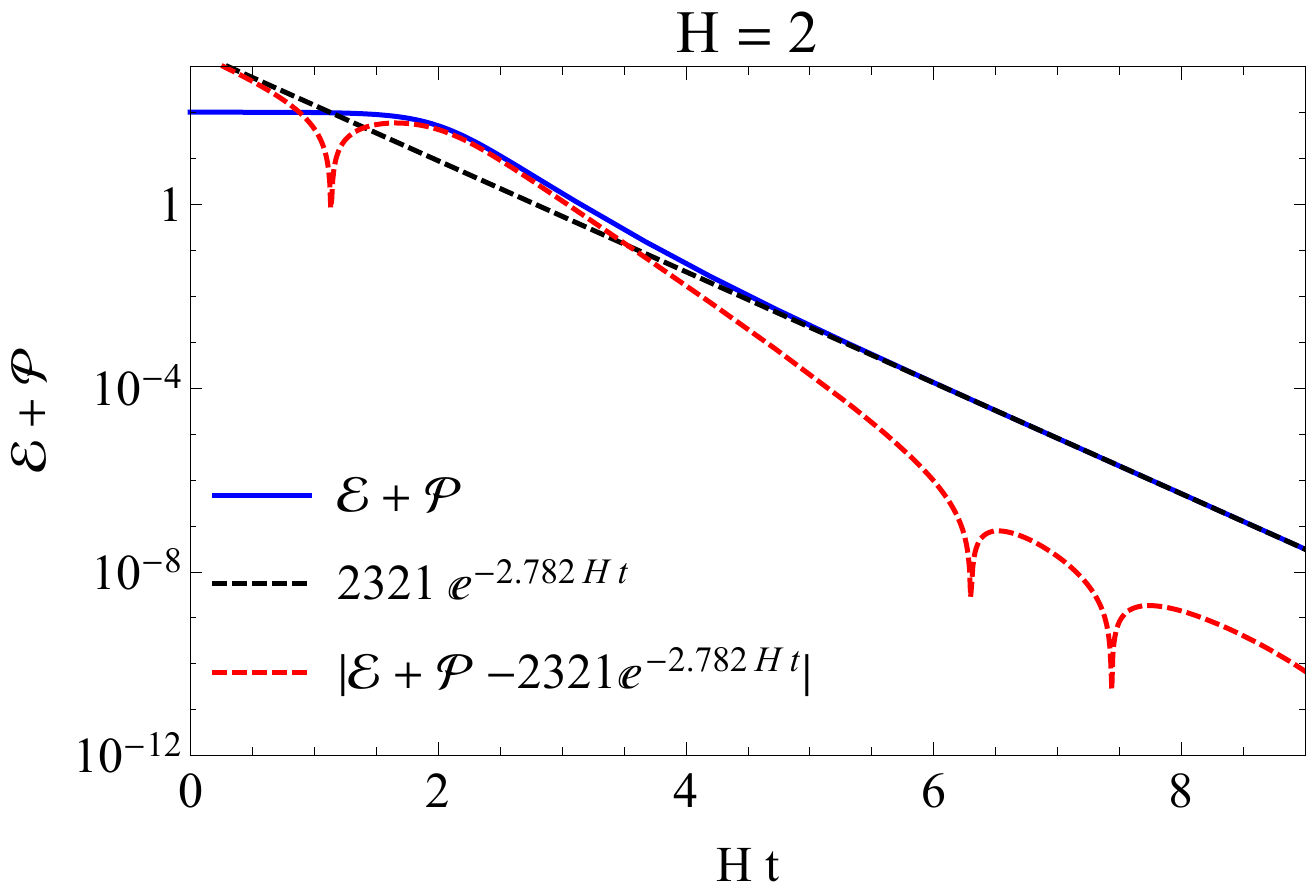}
\caption{Decay of $\calE+\calP$ (solid blue) for $H=1/2$ (left) and $H=2$ (right), together with the QNM fit (dashed black) and their absolute difference (dashed red).}
\label{Fig:qnm}
\end{figure}

Once the excess energy density as a function of time is known we can obtain the pressure from the covariant conservation of the energy-momentum tensor $\nabla^i \langle\Delta \hat{T}_{ij}\rangle=0$, which for the dS$_4$ metric \eqref{ds} evaluates to
\begin{equation}\label{eq:EMTcons}
\Delta\mathcal{P}(t)=-\Delta\mathcal{E}(t)-\frac{\Delta\mathcal{E}'(t)}{3H}\,.
\end{equation}
At late times $\Delta \calE(t)$ is well described by 
\be
\Delta \calE(t) = A\, e^{-i\omega t}\,,
\ee
where $A$ is the amplitude of the fluctuation and $\omega$ is a purely imaginary quasinormal mode with $\im \omega <0$.
By \eqref{eq:EMTcons}, this implies that $\Delta\calP(t)$ and $\Delta\calE(t)$ satisfy at sufficiently late times an EoS-like relation  of the form
\be
\Delta\calP(t)=w_{\rm eff}\Delta\calE(t) 
\ee
with 
\be
w_{\rm eff} = -1 + \frac{i\omega}{3H} = 
-1 + \frac{\left( -\im \omega\right)}{3H}\,. 
\ee
This effective EoS explains the constant late-time ratio $3\Delta\calP(t)/\Delta\calE(t)$ in Fig.~\ref{Fig:hydrocomp}.
We can extract an estimate for $w_{\rm eff}$ from our numerical results at late times. 
In Fig.~\ref{Fig:qnm} we show two examples for this.  
We find $\omega = -3.398H\,i$ $(-2.782H\,i)$ for $H = 1/2$ $(2)$ which gives $3 w_{\rm eff} = 0.398$ $(-0.218)$ in precise agreement with the final values of $3\Delta\calP(t)/\Delta\calE(t)$ shown in the middle plot in the top (bottom) row of Fig.~\ref{Fig:hydrocomp}.

Since the dS expansion dilutes the energy density of the initial state, one can think of the time evolution in dS as a dynamical implementation of the gauge theory RG group. 
It may therefore seem surprising that, in general, $w_{\rm eff}\neq 1/3$ at late times, given that our gauge theory is conformal in the infrared when formulated in flat space.
The reason is that, when the theory is placed in a dS spacetime, the Hubble rate $H$ acts as an IR cut-off \cite{Ghosh:2017big}, in a way similar to the effect of placing the theory at finite temperature. 
As a consequence, the IR behaviour of the gauge theory on dS is approximately conformal only in the limits $H\ll M$ or $H \gg 1$. 
In the first case the RG flow of the gauge theory in dS is similar to that in flat space, in the sense that it explores almost all possible energy scales and is only cut-off very close to the IR fixed point of the theory. 
This behaviour is consistent with the top row of Fig.~\ref{Fig:hydrocomp} (recall that $M=1$), where we see from the late-time behaviour that $w_{\rm eff}$ grows towards 1 as we move from the right plot ($H=1$) to the left plot ($H=1/3$). 
In the second case the IR cut-off is so close to the UV fixed point that the entire RG flow is approximately conformal.
This is supported by the perturbative analysis of \cite{Buchel:2017lhu}, which suggests that conformal symmetry is restored at late times in the limit of large expansion rates $H\gg M$.
We indeed find evidence for this behaviour in Fig.~\ref{Fig:hydrocomp}: $w_{\rm eff}$ obtains its minimal value $w_{\rm eff}^{\rm min}=-0.298$ for $H=1$ (top-right plot) and grows monotonically for $H>1$ (bottom row).
Although we were not able to obtain numerical results for $H\gg 3$ because the numerics becomes increasingly challenging, we expect $w_{\rm eff}$ to approach the value $1/3$ in the limit of large $H$.

\begin{figure}[t]
\center
 \includegraphics[height=0.31\linewidth]{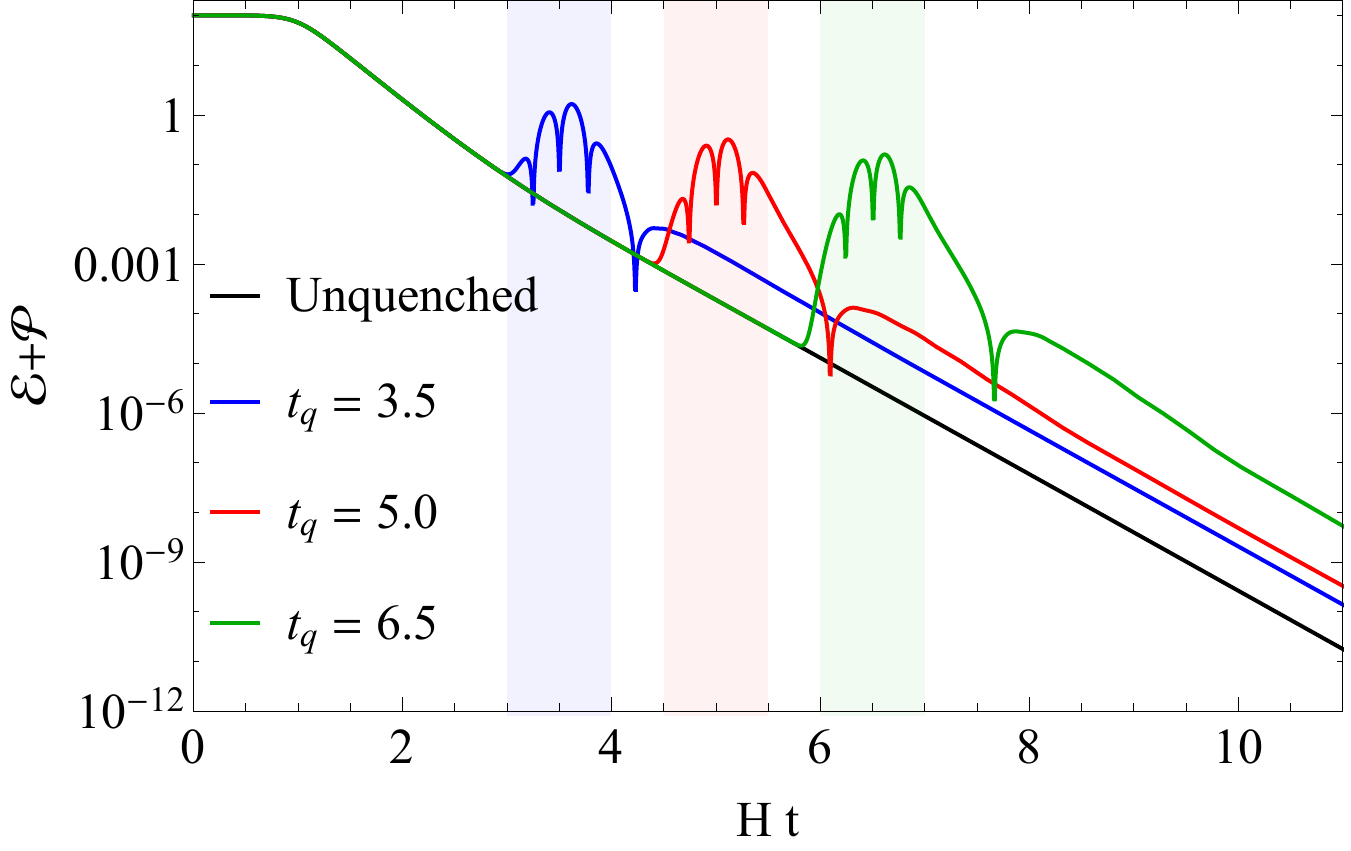}\quad
  \includegraphics[height=0.31\linewidth]{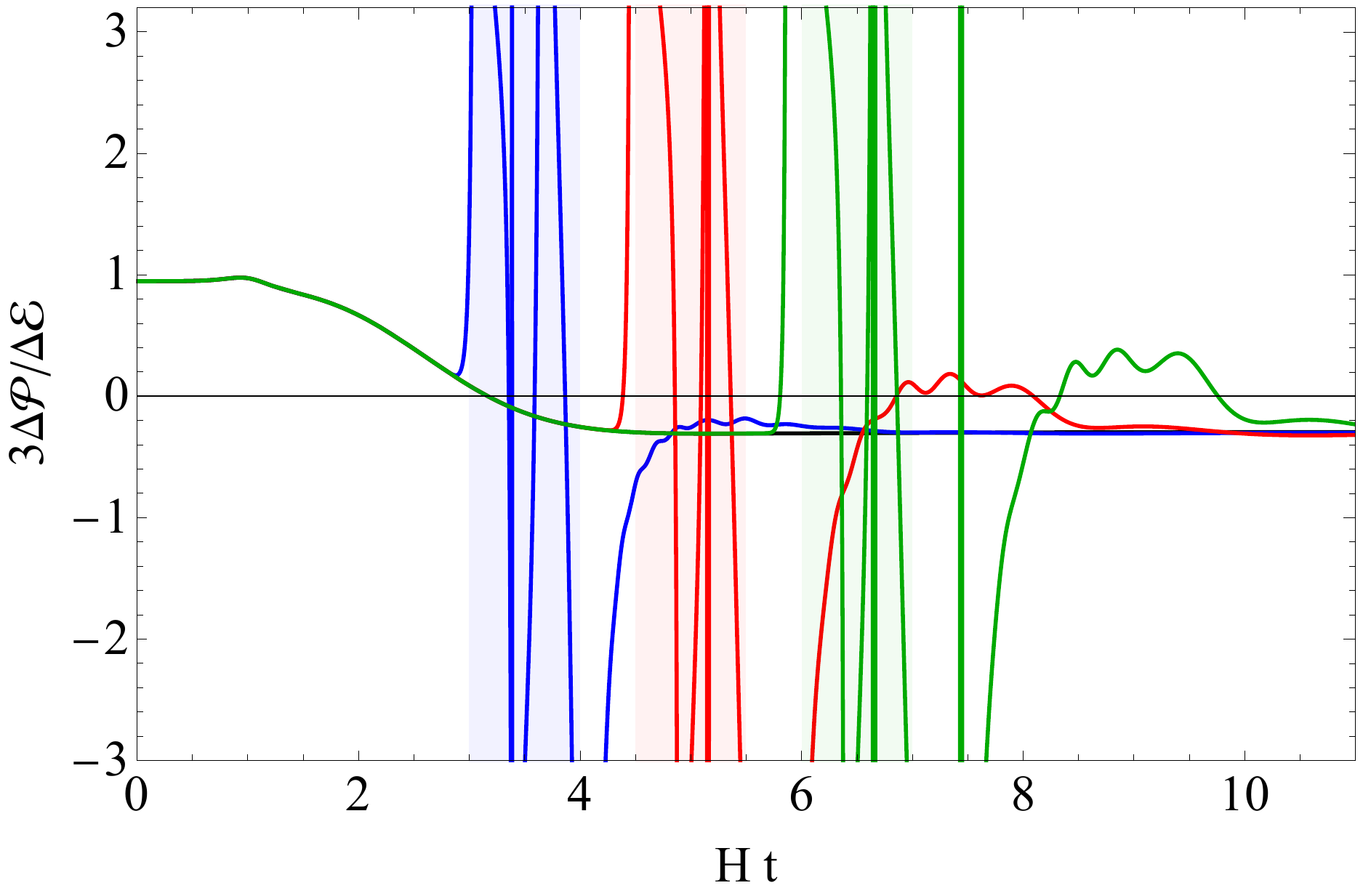}
\caption{Evolution of $\calE+\calP$ (left) and $3\Delta\calP/\calE$ (right) for $H=1$ quenched with different amplitude at different times. 
The quenches are realised by exciting the boundary metric with Gaussian perturbations whose widths are indicated by the red, green and blue bands. 
The quench performs work on the system, which leads to an increase of $\calE$ and explains the increase of $\calE+\calP$.
After the quench the ratio of pressure and energy density quickly returns to the late-time value of the unquenced evolution (the small oscillations are verified to be numerical artefacts).
}
\label{Fig:universalityquench}
\end{figure}

To strengthen our conclusion that the late-time geometry is determined by quasi-normal modes we show in Fig.~\ref{Fig:universalityquench} three evolutions for $H=1$ in which we quench the boundary metric function $S_0(t)$ by a factor $1+A e^{-32 (t - t_q)^2}$ with amplitudes $A=0.05,0.01$ and $0.005$ at quenching times $t_q=3.5,5$ and $7$, respectively.
The quenches shown perform work on the system, which leads to an increase of energy and, as a consequence, also to an increase of $\calE+\calP$.
Quenching the late-time dynamics of the system excites higher QNMs.
The curves in Fig.~\ref{Fig:universalityquench} show the decay of these modes within a time of order $1/H$, after which the system returns to a state whose late-time dynamics is entirely characterized by $w_{\rm eff}$, i.e. by the first QNM.
In the next section we discuss the maximally symmetric late-time state in more detail.

\subsection{Late-time solution}
\label{sec:latetimesol}
At late times the geometry and the scalar field approach the following form (see also \cite{Buchel:2017pto})
\begin{equation}\label{eq:ds2late}
ds_\infty^2=-A_\infty(r)\dd t^2+2 \dd r \dd t+e^{2H t}S_\infty(r)^2\dd\vec{x}^2\,, \quad \phi=\phi_\infty(r)\,, 
\end{equation}
where $A_\infty$, $S_\infty$ and $\phi_\infty$ are the time independent late-time limits of the metric functions and the scalar field defined as 
\begin{equation}
\Ainf(r)\equiv\lim\limits_{t\to \infty} A(t,r)\,,\quad \Sinf(r)\equiv\lim\limits_{t\to \infty} e^{-H t}S(t,r)\,,\quad \phi_\infty(r)\equiv\lim\limits_{t\to \infty} \phi(t,r)\,.\label{eq:latetimeEF}
\end{equation}
In the following we will drop the index ``$\infty$'' and implicitly assume this limit in all appearances of $A$, $S$ and $\phi$.
Under these conditions the equations of motion simplify to
\begin{subequations}
\begin{align}
0&=A \,\phi''+\left(A'+3 \left(\frac{A\, S'}{S}+H\right)\right)\phi'-V'\,,\\
0&=\phi '^2+\frac{3 S''}{2 S}\,,\label{eq:eomlate2}\\
0&=A''+\frac{4 \left(S' \left(A'+3 H\right)+A\, S''\right)}{S}+\frac{2 A\, S'^2}{S^2}+2 A\, \phi'^2+4 V\,,\label{eq:eomlate3}\\
0&=A^2 \left(\frac{2 \left(S\, S''+S'^2\right)}{S^2}+\frac{2 \phi '^2}{3}\right)+\frac{A\, S' \left(A'+8 H\right)}{S}+H \left(2 H-A'\right)+\frac{4 A\, V}{3}\,,\label{eq:eomlate4}\\
0&=\frac{3 \left(S' \left(A'+6 H\right)+2 A\, S''\right)}{2 S}+\frac{3 A\, S'^2}{S^2}+A \phi'^2+2 V\label{eq:eomlate5}\,.
\end{align}
\end{subequations}

The late time geometry \eqref{eq:ds2late} possesses an event horizon located at $r=r_{\mathrm{EH}}$ defined by the condition
\begin{equation}
A(r_{\mathrm{EH}})=0\,.
\end{equation}
Using this condition in \eqref{eq:eomlate4} one finds that the surface gravity at the event horizon equals the Hubble rate
\begin{equation}
\kappa_{\mathrm{EH}}=\frac{1}{2}A'(r_{\mathrm{EH}})=H\,.\label{eq:kappaeh}
\end{equation}
Because the geometry depends on time, the apparent horizon, i.e. the outermost trapped lightlike surface, does not coincide with the event horizon.
The radial position of the apparent horizon $r_{\mathrm{AH}}$ is determined by $\dot{S}|_{r_{\mathrm{AH}}}=0$, which in the coordinate system \eqref{eq:latetimeEF} gives the condition
\begin{equation}
0=S(r_{\mathrm{AH}})\frac{\dd}{\dd t}e^{H t}+\frac{1}{2} e^{H t} A(r_{\mathrm{AH}})S'(r_{\mathrm{AH}})\,.
\label{eq:ah}
\end{equation}
Using this condition in \eqref{eq:eomlate4} together with \eqref{eq:eomlate5} allows us to express surface gravity at the apparent horizon
\begin{equation}\label{eq:kappah}
\kappa_{\mathrm{AH}}=\frac{1}{2}A'(r_{\mathrm{AH}})=-H\,.
\end{equation}
We arrive at the conclusion that surface gravity at the apparent horizon equals minus surface gravity at the event horizon
\begin{equation}\label{eq:kappaboth}
\kappa_{\mathrm{AH}}=-\kappa_{\mathrm{EH}}\,.
\end{equation}

Above, we used the ansatz \eqref{eq:ds2late} and the equations of motion to derive the surface gravity of the event and of the apparent horizon.
This ansatz can be seen as educated guess motivated by our numerical results which at late times agree with \eqref{eq:ds2late} very accurately.
However, it is possible to arrive at \eqref{eq:kappaeh} without invoking the equations of motion or taking guidance from numerical analysis.
For this we use as starting point the assumption that the system evolves towards a vacuum state that by definition obeys the symmetries of the background geometry.
On de Sitter space this state is known as Bunch--Davis vacuum \cite{Bunch:1978yq}.
By the holographic duality the bulk metric dual to this vacuum state has to satisfy the isometries of dS$_4$ as well.
A domain wall parametrization in FG coordinates makes these isometries manifest\footnote{Solutions of this type have been constructed in \cite{Ghosh:2017big} for a variety of gauge theories.}
\begin{equation}\label{eq:ds2DW}
ds_\mathrm{DW}^2 = \Ainf(\rho) \left(-\dd t^2_{\mathrm{FG}} +  e^{2 H t_{\mathrm{FG}}} \dd\vec{x}^2\right) + \dd\rho^2 \,,
\end{equation}
where the time and radial coordinate are related to those in \eqref{eq:ds2late} by
\begin{equation}
\rho(r)=-\int^\infty_r \dd r'\Ainf(r')^{-1/2}\,, \quad t_\mathrm{FG}(r,t)=t-\int^\infty_r \dd r' \Ainf(r')^{-1}  \,.
\end{equation}
Demanding that \eqref{eq:ds2late} can be transformed to the manifestly dS$_4$-isometric domain wall form \eqref{eq:ds2DW} gives the relation  
\begin{equation}\label{eq:SinfofA}
\Sinf(r)= \sqrt{\Ainf(r)}\, e^{-H\int_r^\infty \dd r'\,\Ainf (r')^{-1}}\,.
\end{equation}
Beyond the event horizon ($r<r_\mathrm{EH}$) the metric function $A_\infty(r)$ is negative and naively one could expect the right-hand side of \eqref{eq:SinfofA} to acquire an imaginary part.
The exponent, however, diverges as $r$ approaches the event horizon, which leads to a term $a \log(r-r_{\rm EH})$ when assuming a regular near horizon expansion of the form $A_\infty(r)= a(r - r_{\rm EH})+{O\left((r - r_{\rm EH})^2\right)}$.
The imaginary parts exactly cancel and $\Sinf$ remains real inside the event horizon if
\be\label{eq:Ahor}
\Ainf(r)= 2 H \left( r -r_{\mathrm{EH}}\right) + O\left( \left( r -r_{\mathrm{EH}}\right)^2\right) \,.
\ee
This gives precisely the same result for $\kappa_{\mathrm{EH}}=\lim\limits_{r\to r_{\mathrm{EH}}}\frac{1}{2}A'(r)$ as \eqref{eq:kappaeh}, only by demanding dS$_4$ symmetry and regularity of the solution at the horizon.
It is possible to derive \eqref{eq:kappah} in an analogous way as well.

\begin{figure}
\center
 \includegraphics[height=0.33\linewidth]{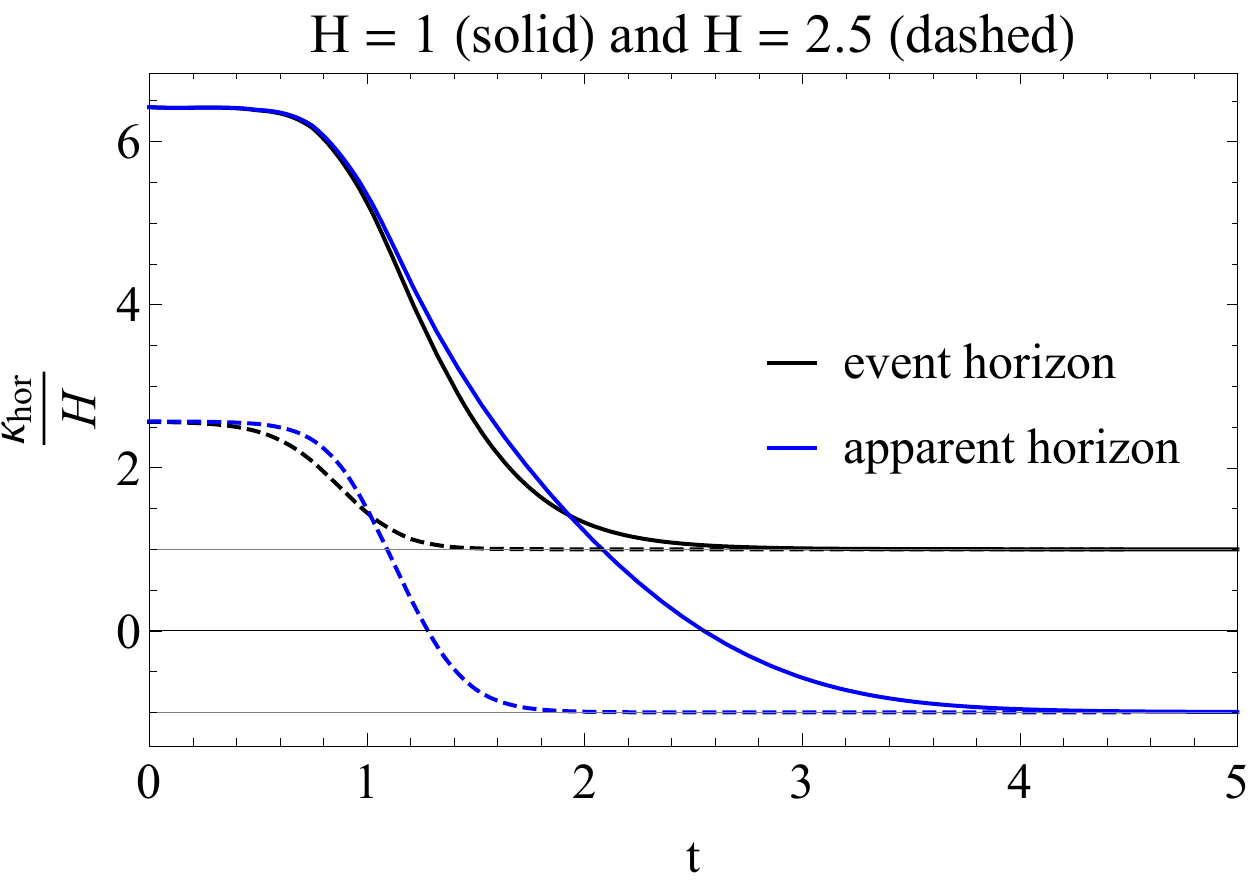}\quad
 \includegraphics[height=0.33\linewidth]{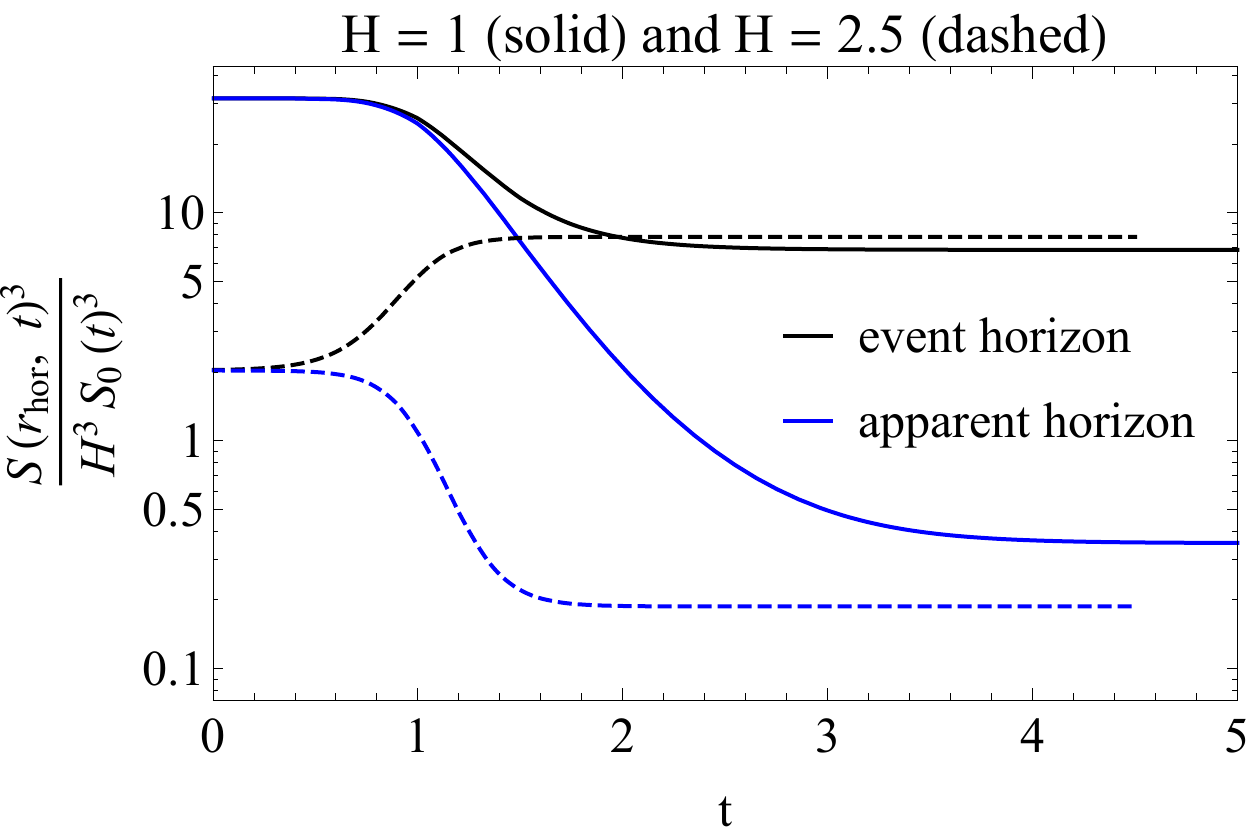}
\caption{Surface gravity $\kappa$ (left) and area density (right) of event and apparent horizons for two different values of the Hubble rate.} 
\label{Fig:Shor}
\end{figure}

In Fig.~\ref{Fig:Shor} we show surface gravity (left) and area density (right) of event and apparent horizon for $H=1$ and $H=2.5$.
Initially, when the geometry is static and dual to a thermal state on flat space, the event and apparent horizon coincide and therefore have equal surface gravity and area density that depends on the choice of the initial temperature.
On the expanding background the locations of apparent and event horizon deviate and so do the respective surface gravities and area densities.
In accordance with the analytic analysis, the numeric evolution evolves towards a solution where surface gravity of event and apparent horizon precisely satisfy  $\kappa_{\mathrm{EH}}=-\kappa_{\mathrm{AH}}=H$.
For the gravity dual of $\mathcal{N}=4$ SYM theory on de Sitter space apparent and event horizon densities can be straightforwardly computed because the bulk geometry is known explicitly \cite{Apostolopoulos:2008ru,Buchel:2016cbj}. 
This geometry has no apparent horizon but an event horizon whose area density is given by 
\be
S(r_{\rm EH},\,t)^3 = 8 H^3 S_0(t)^3 = 8 \kappa_{\mathrm{EH}}^3 S_0(t)^3\,.
\ee
For comparison, a thermal state in $\mathcal{N}=4$ SYM on flat space has $S(r_{\rm EH})^3 = \kappa_{\mathrm{EH}}^3/8$.

Indeed, we show in Fig.~\ref{Fig:Shor} (right) that as $H$ increases from $1$ to $2.5$ and thereby approaching the conformal UV fixed point ($M/H\to 0$), the apparent horizon area density ($ S(r_{\rm hor},t)^3/S_0(t)^3$) decreases and the normalised event horizon area density approaches the conformal value at late times.
The properties of the late time solution are solely determined by the value of $H$ and $M$, but the way they are approached depends on the initial conditions.
Thermal initial states with large entropy, and therefore with large initial event and apparent horizons area, lead to decreasing area densities of both, apparent and event horizons. 
An example for this is the simulation for $H=1$ shown in Fig.~\ref{Fig:Shor} (right).
However, if the initial entropy is such that the corresponding horizon area is smaller than the area density of the late-time horizon, the area density grows.
This is precisely what we find for example for the event horizon area density for $H=2.5$ (dashed black) shown in Fig.~\ref{Fig:Shor} (right).
It may seem surprising that the event horizon area density increases even though the surface gravity decreases substantially.
This, however, can qualitatively be explained by comparing the analytic area densities of $\mathcal{N}=4$ SYM in flat space to de Sitter space, where indeed at fixed surface gravity the area density is much smaller on flat space.
The area density of the apparent horizon (blue) decreases monotonically in all cases shown.

This is a clear indication that the holographic interpretation of bulk horizons areas as entropy in the dual field theory is subtle when the boundary theory is expanding.
We will elaborate on this in the discussion.
However, we emphasize that the comoving area density of the apparent horizon $S(r_{\rm hor},t )^3$ is a monotonously growing function of time in accordance with Hawinkgs area theorem \cite{hawking1972} and the more recent discussion\footnote{We are grateful to Alex Buchel for bringing this to our attention.} \cite{Buchel:2017pto,Buchel:2019pjb} in the context of a holographic model similar to the one we use here.

In the next section we analyse the entanglement properties of the de Sitter vacua constructed in this section and comment on the subtleties involved in the assignment of an effective entropy to their horizon area densities.

\section{Entanglement and horizon entropies}
\label{sec:entropy0}
The holographic duality maps the area of event horizons in time independent bulk geometries to the thermodynamic entropy of thermal states in the dual field theory.
In time-dependent geometries this mapping is obscured, firstly because the dual field theory is not in thermal equilibrium and thermodynamic concepts of entropy and temperature do not apply and secondly because the mapping of the horizon to the boundary is not necessarily unique.
To study the time evolution of non-equilibrium states in the field theory it can nevertheless be useful to define an effective temperature in terms of surface gravity of the event horizon\footnote{We will also look at the apparent horizon, even though the physical interpretation of temperatures defined in terms of apparent horizons is problematic, because it in general depends on the slicing of spacetime.}.

In time independent geometries event and apparent horizon coincide and the entropy density in the dual field theory is uniquely defined in terms of the area density of the event horizon in the bulk.
Because de Sitter space is expanding it is non-trivial to map points in the boundary to points on the horizon in the bulk whose area density changes due to the exponentially growing scale factor (see also \cite{Buchel:2019qcq} for related difficulties in interpreting the apparent horizon area).

A robust and gauge independent notion of entropy is given by entanglement entropy \cite{Holzhey:1994we} of spatial subregions $\mathcal{R}$ in the QFT defined as
\begin{equation}\label{eq:SA}
S_{\mathcal{R}}=-\mathrm{Tr}_\mathcal{R} \hat{\rho}_\mathcal{R}\log\hat{\rho}_\mathcal{R}\,,
\end{equation}
where $\hat{\rho}_\mathcal{R}=\mathrm{Tr}_{\bar{\mathcal{R}}}\hat{\rho}$ denotes the reduced density matrix obtained by performing on the full density matrix $\hat{\rho}$ a partial trace over the degrees of freedom outside $\mathcal{R}$.
For simplicity we will assume these subregions to be spatial balls at some fixed time $t=t_0$ with radius $\ell$
\begin{equation}\label{eq:region}
\mathcal{R}=\{t=t_0, 0 \leq r \leq \ell, 0 \leq \theta \leq \pi, 0 \leq \varphi < 2\pi \} \,.
\end{equation}
The coordinate $r$ in this section is the radial coordinate at the boundary and should not be confused with the holographic coordinate. 
Since the density matrix $\hat{\rho}$ can be time-dependent, i.e.~defined in terms of time-dependent states, entanglement entropy is also well defined for states that are not in thermal equilibrium.
For $H=0$ and $\ell\to \infty$ the entangling region $\mathcal{R}$ covers an entire spacelike slice of Minkowski space.
In this limit $\hat{\rho}_{\mathcal{R}}=\hat{\rho}$ and \eqref{eq:SA} equals the von Neumann entropy of the full density matrix, i.e., the thermodynamic entropy of a quantum state in thermal equilibrium.
For $H\neq 0$ the QFT inherits the causal structure of de Sitter background which has a cosmological event horizon located at $r=H^{-1}$.
Although spatial regions of size $\ell>H^{-1}$ are causally disconnected, quantum states on such regions can be entangled \cite{Maldacena:2012xp}.

Direct field theory computations of entanglement entropy are only possible in exceptional cases like for example in two-dimensional CFTs \cite{Calabrese:2004eu} and in dimensions higher than two only in non-interacting QFTs \cite{Srednicki:1993im}.
The holographic duality replaces the field theory computation of entanglement entropy by a much simpler extremisation problem for the surface area $\mathcal{A}_\mathcal{R}$ of a codimension two surface, homologous to $\mathcal{R}$, in the bulk theory \cite{Ryu:2006bv,Hubeny:2007xt}
\begin{equation}
S_\mathcal{R}=\frac{\mathcal{A}_\mathcal{R}}{4G}\,.
\end{equation}
The boundary of the relevant surface coincides with the boundary of the entangling region $\mathcal{R}$ in the field theory and extremises the area functional in the bulk theory 
\begin{equation}
\mathcal{A}_\mathcal{R}[X]=\int \dd^{3}\sigma\sqrt{\mathrm{Det}\left(\partial_a X^\mu \partial_b X^\nu g_{\mu\nu}\right)}\,.
\end{equation}
The surface embedding $X^\mu=X^\mu(\sigma^a)$ is parametrised by three intrinsic coordinates for which we choose  $\sigma^a=\{r,\theta,\varphi\}$.
The entangling regions \eqref{eq:region} do not break spherical symmetry in the boundary theory. We can then parametrise the bulk surface with
\begin{equation}
X^\mu(z)=\{z(r),t(r),r,\theta,\varphi\}\,.
\end{equation}
This choice simplifies the area functional considerably.
Integration over the angular coordinates $\theta$ and $\phi$ can be performed explicitly and the remaining expression takes the form of a geodesic action
\begin{equation}\label{eq:AreaFunctionalSphere}
\mathcal{A}_\mathcal{R}[X]=4\pi \int \dd r \sqrt{\bar{g}_{\alpha\beta}(z(r), t(r)) \frac{\dd X^\alpha}{\dd r} \frac{\dd X^\beta}{\dd r}} \quad \mathrm{s.t.} \quad X^\mu(0)=\{0,t_0,\ell,\theta,\varphi\}\,,
\end{equation}
where the metric $\bar{g}_{\alpha\beta}$ is related by a conformal factor to a three dimensional subspace ($\alpha,\beta=\{t,z,r\}$) of the bulk metric \eqref{eq:ds2late} (see also \cite{Ecker:2015kna})
\begin{equation}
\bar{ds}^2=\bar{g}_{\alpha\beta}\dd x^\alpha\dd x^\beta =\left(r\,S(z)\,a(t)\right)^4 g_{\alpha\beta}\dd x^\alpha\dd x^\beta\,.
\end{equation}
where we use $a(t)\equiv e^{Ht}$ in the following to simplify notation.
The equations of motion that follow from $\delta \mathcal{A}_{\mathcal{R}}=0$ take the form of a non-affine geodesic equation
\begin{equation}\label{eq:GeodesicsEq}
\frac{\dd^2 X^\alpha}{\dd r^2}+\Gamma^\alpha_{\beta\gamma}\frac{\dd X^\beta}{\dd r}\frac{\dd X^\gamma}{\dd r}=J\frac{\dd X^\alpha}{\dd r}\,,
\end{equation}
where $\Gamma^\alpha_{\beta\gamma}$ is the Levi-Civit\'a connection associated to $\bar{g}_{\alpha\beta}$ and is meant to be evaluated at the location of the surface $X^\alpha(r)$; the viscous friction term on the right hand side includes the Jacobian $J=\frac{\dd^2 \tau(r)}{\dd r^2}/\frac{\dd^2 \tau(r)}{\dd r^2}$ that originates from transforming from the affine parameter $\tau$ defined by $\frac{\dd X^\alpha(\tau)}{\dd\tau}\frac{\dd X^\beta(\tau)}{\dd\tau}\bar{g}_{\alpha\beta}=1$ to the non-affine parameter $r$.   

For $\phi=0$ the field theory is $\mathcal{N}=4$ SYM theory and \eqref{eq:GeodesicsEq} has a simple analytic solution $z(r) = \sqrt{\ell^2 - r^2}$ and $t(r) = -\log\left(1+z(r)\right)$ in a gauge where $A(z) = z^{-2} - 1$.
In this case the de Sitter vacuum can be mapped by a conformal transformation to the Minkowski vacuum \cite{Buchel:2017pto} and the holographic entanglement entropy is equal to the area of a hemisphere in AdS$_5$ \cite{Ryu:2006bv}.

For the non-conformal case with $\phi\neq 0$ the geodesic equations are too long to display here and no closed solutions are available.
At the turning point of the bulk surface located at $z_\ast\equiv z(r=0)$ the equations simplify by symmetry $z'(r=0)=t'(r=0)=0$ to
\begin{subequations}
\begin{align}
t''(r)&=-3 z(r)^2 a(t(r))^2 S(z(r)) S'(z(r)) \\
z''(r)&=3 z(r)^2 a(t(r)) S(z(r)) \left[z(r)^2 a(t(r)) A(z(r)) S'(z(r))-S(z(r)) a'(t(r))\right]\,.\label{eq:ddz}
\end{align}
\end{subequations}
The sign of $z''(r=0)$ determines if the bulk surface can reach the boundary ($z''(r=0)<0$) or not ($z''(r=0)>0$).
The condition $z''(r)=0$ defines the `entanglement horizon', i.e., a barrier in the bulk that extremal surfaces attached to the boundary do not cross whose location is determined by
\begin{equation}\label{eq:EEhor}
0=z^2 a(t) A(z) S'(z)-S(z) a'(t)=z^2  A(z) S'(z)-H S(z)\,.
\end{equation}
Plugging this relation into the definition of $\dot{S}$ gives
\begin{equation}\label{eq:SdotEE}
\dot{S} + \frac{1}{2} z(r)^2 A(z(r)) a(t(r)) S'(z(r)) = 0\,.
\end{equation}
Except for the second term, this relation is equal to equation \eqref{eq:ah} that determines the location of the apparent horizon.
Since $z(r)^2A(z(r)) S'(z(r))$ is positive and monotonic in $z(r)$ the entanglement horizon is located between event and apparent horizon.
For boundary geometries that are de Sitter the entanglement horizon is a Lagrangian surface of the bulk geometry, or in other words a surface with zero surface gravity (see also \cite{Engelhardt:2013tra} for similar results on extremal surface barriers).
This can be shown by combining \eqref{eq:EEhor} with the Einstein equations \eqref{eq:eomlate2} to \eqref{eq:eomlate5} which gives
\begin{equation}\label{eq:Adr}
\kappa_{\rm Ent}=-\frac{z^2}{2}A'(z)=0\,.
\end{equation}
Furthermore, when the geometry is static ($H=0$), we know from \eqref{eq:ah} that the apparent horizon ($\dot{S}=0$) and event horizon ($A=0$) coincide.
From \eqref{eq:SdotEE} we then see that this also solves the entanglement horizon equation, so that all three horizons coincide as expected \cite{Hubeny:2012ry}.

We compute the entanglement entropy numerically by shooting, i.e., by integrating \eqref{eq:GeodesicsEq} from the turning point $z_\ast$ of the geodesics to a cutoff-value for the holographic coordinate $z_{\mathrm{cut}} = \epsilon/(1-\xi \epsilon)$ close to the boundary located at $z=0$.
Note that the gauge freedom $\xi$ in $S(r) = r + \xi + \mathcal{O}(1/r)$ near the boundary leads to an additional $1/\epsilon$ dependent divergence in the entanglement entropy. 
We eliminate the $1/\epsilon$ divergence with the gauge transformation $r\rightarrow r - \xi$ that ensures $S(r) = r + \mathcal{O}(1/r)$ and in addition fixes the residual gauge freedom ($r \rightarrow r + \xi$) in the radial coordinate.
The cutoff regulated value for the entanglement entropy can then be  obtained by numerically solving the following integral
\begin{equation}
S_{\mathcal{R}}^{\mathrm{cut}}=\frac{\pi}{G}\int_{z_{\mathrm{cut}}}^{z_\ast}\dd z\left(S(z)a(t(z))r(z)\right)^2\sqrt{-A(z)t'(z)^2-\frac{2t'(z)}{z^2}+\left(S(z)a(t(z))r'(z)\right)^2} \,.
\end{equation}

\begin{figure}[t]
\center
 \includegraphics[width=0.49\linewidth]{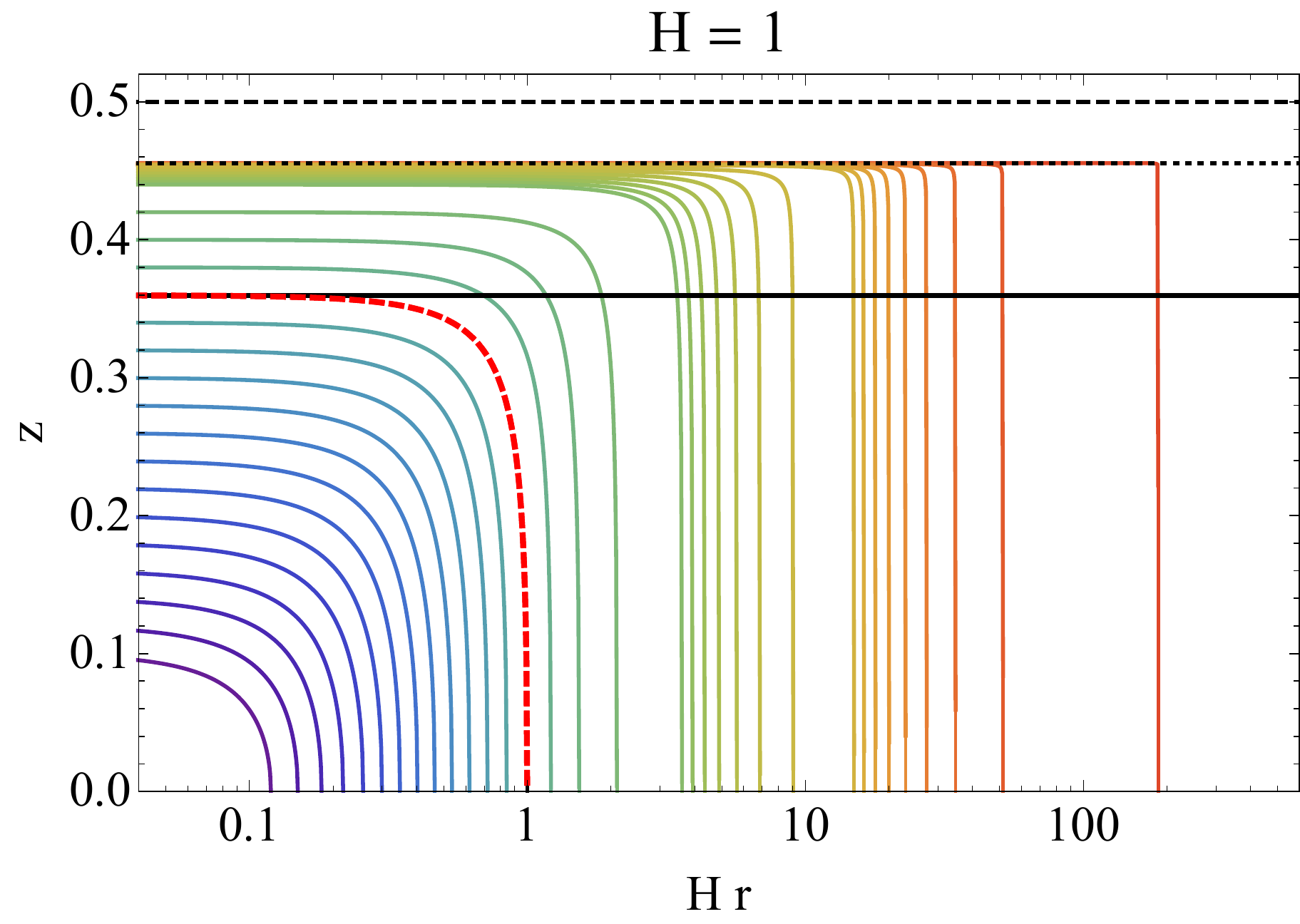}
 \includegraphics[width=0.49\linewidth]{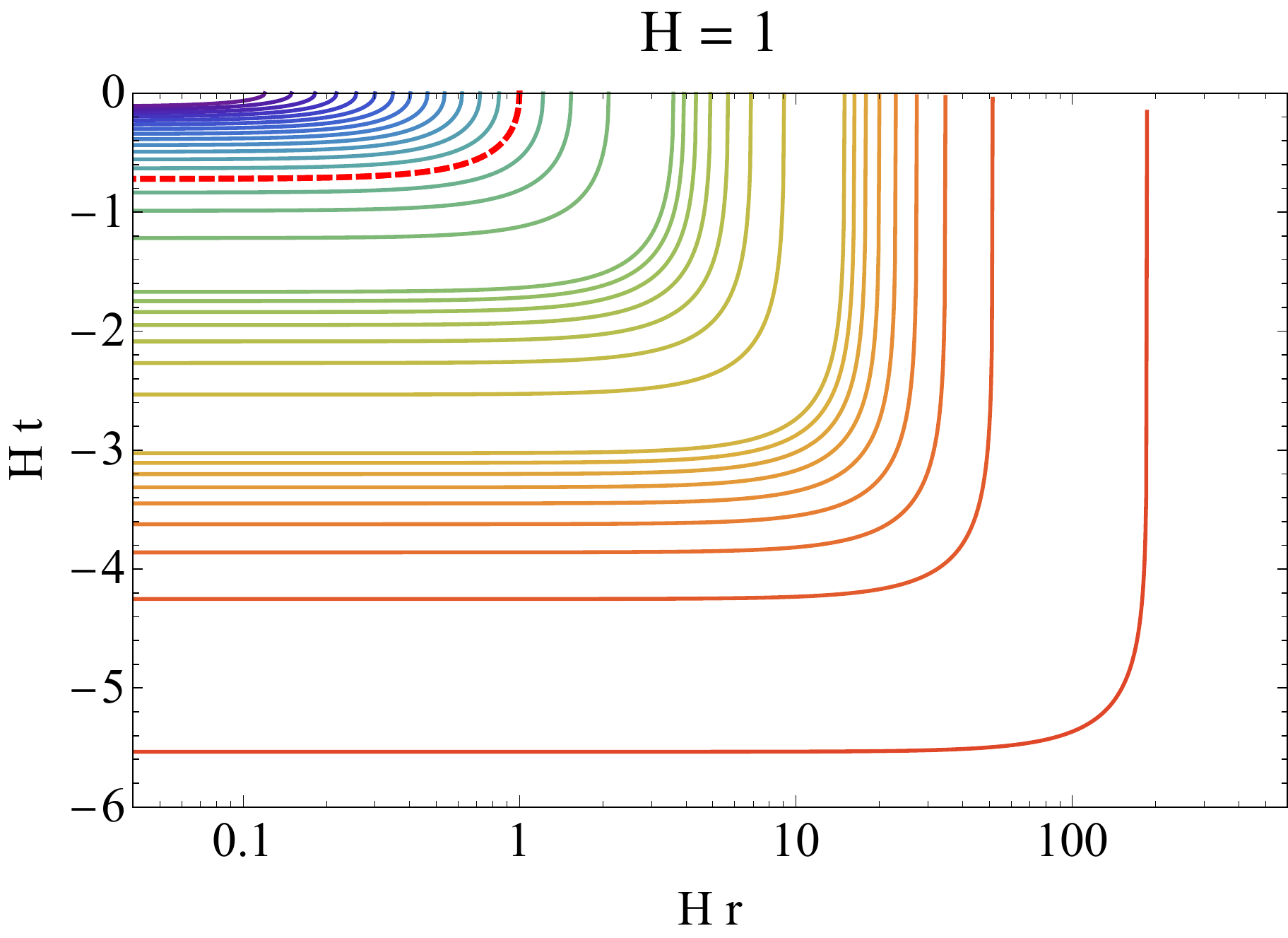}\\
 \includegraphics[width=0.49\linewidth]{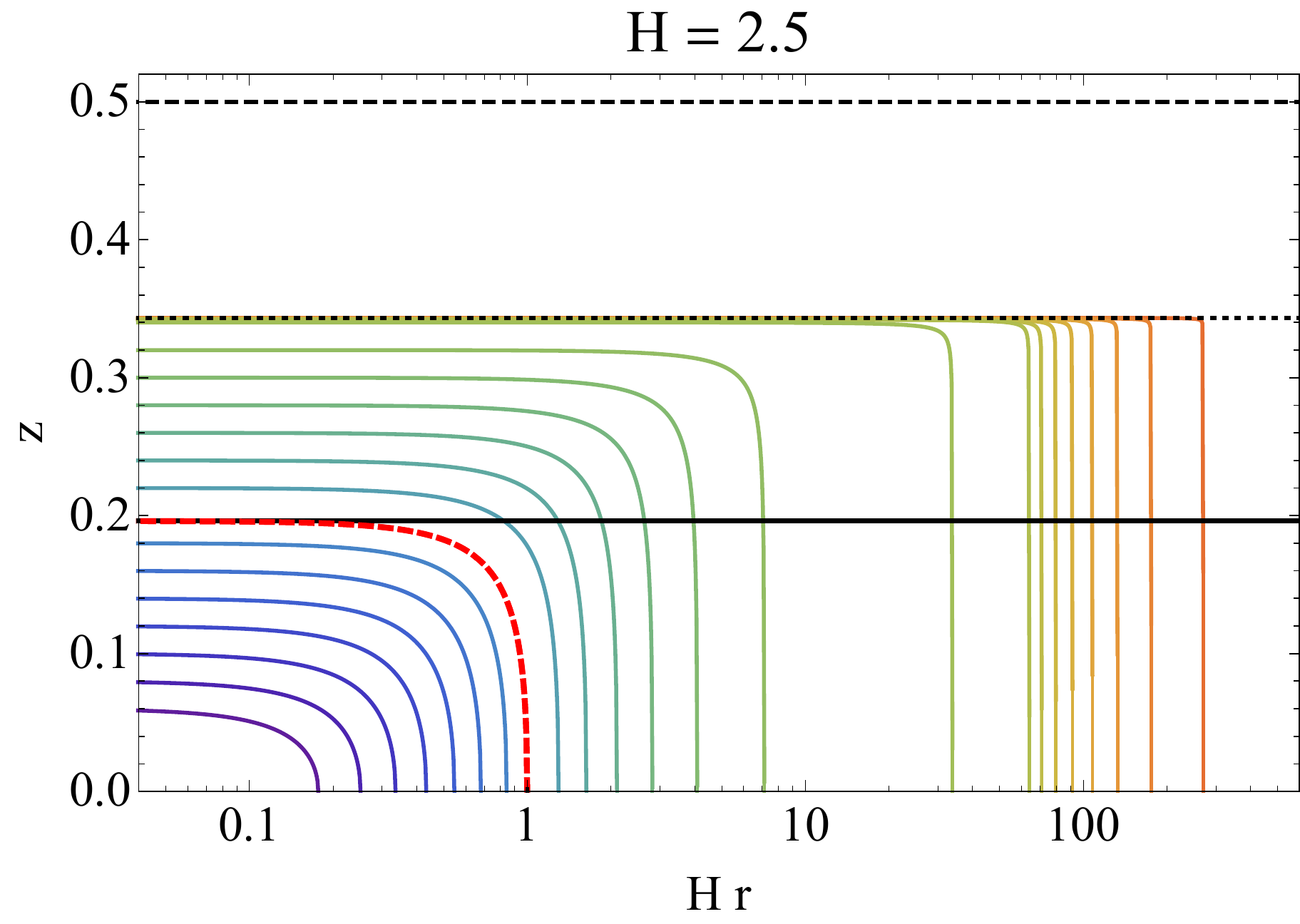}
 \includegraphics[width=0.49\linewidth]{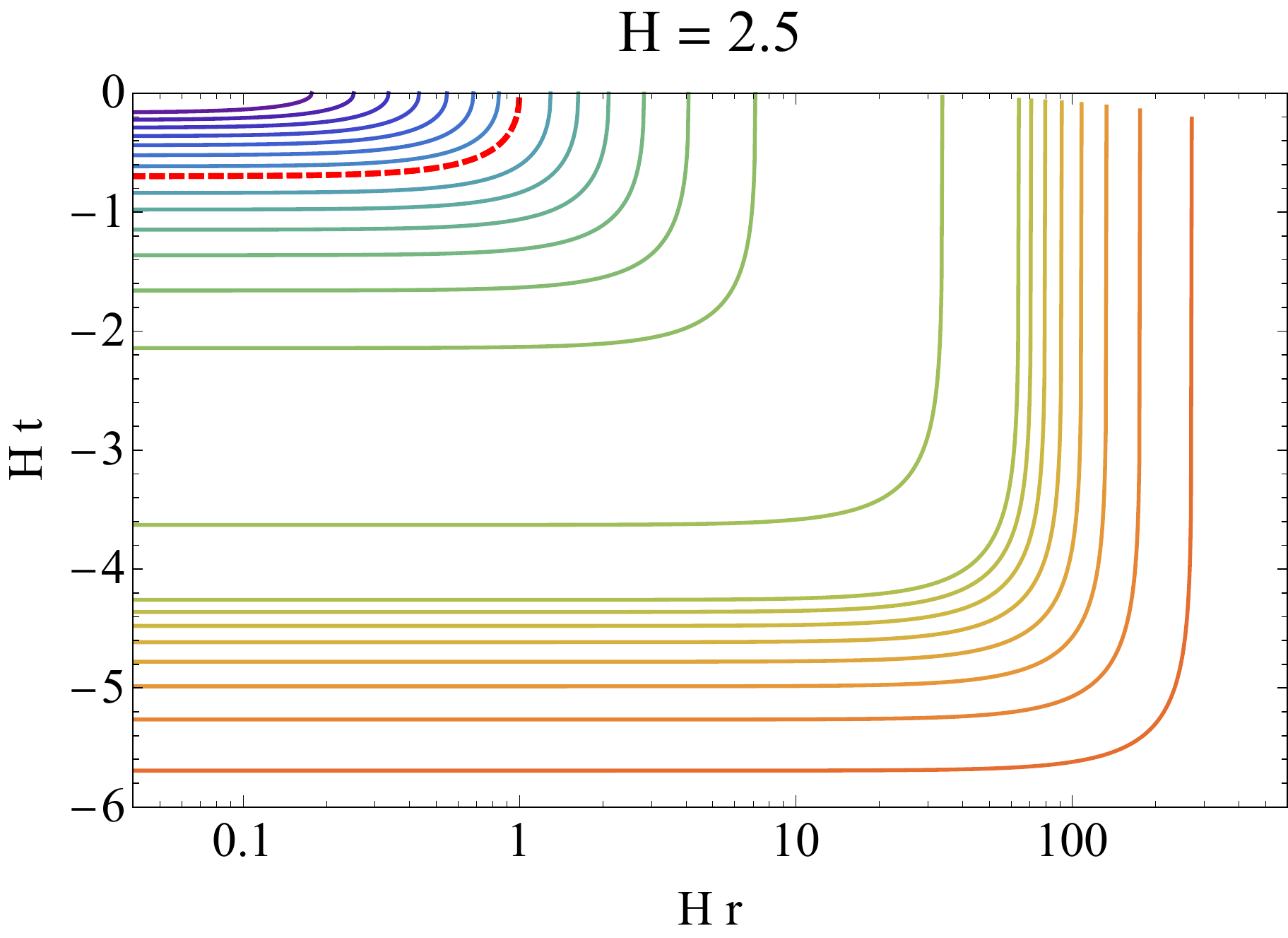}
 \caption{Extremal surfaces with various values of $z_\ast$ in the $z-r$ plane (left) and $t-r$ plane (right) for $H=1$ (top) and $H=2.5$ (bottom).
 Solid, dotted and dashed black lines indicate the radial location of event, entangling and apparent horizon, respectively. 
 Red dashed curves are extremal surfaces with $z_\ast=z_{\mathrm{EH}}$ which end on the boundary precisely at the location ($r=1/H$) of the cosmological horizon.}
\label{Fig:extremalsurfaces}
\end{figure}

In Fig.~\ref{Fig:extremalsurfaces} we show extremal surfaces in the gauge where the apparent horizon is fixed at $z_{\rm AH}=1/2$, obtained by shooting from different values of $z_\ast$, for several spherical entangling regions of different radius $\ell$.
Interestingly, surfaces with $z_\ast=z_{\mathrm{EH}}$ end on the boundary precisely at $r=1/H$ where the cosmological horizon is located.
The red dashed curves in Fig.~\ref{Fig:extremalsurfaces} are two examples for such surfaces with $z_\ast=0.3597\,(0.1961)$ for $H = 1\, (2.5)$.
Extremal surfaces of entangling regions larger than the cosmological horizon ($\ell>1/H$) probe regions behind the event horizon (solid black), but never beyond the entanglement horizon (black dotted).
In our gauge, where the apparent horizon (black dashed) is located at $z_{\mathrm{AH}}=1/2$, the entanglement horizon is located at $z = 0.4554\,(0.3434)$ and $z_{\rm EH}=0.3597\,(0.1961)$ for $H = 1\,(2.5)$.
For comparison, in the gravity dual of $\mathcal{N}=4$ SYM theory on de Sitter the event horizon is located at $z_{\mathrm{EH}}=1/(2H-\xi)$ and extremal surfaces can probe until $z=1/(H-\xi)$ in the gauge where $z_{\rm AH}=-1/\xi$.

In time independent geometries the apparent, event and entanglement horizon coincide and bound the region that is causally connected to the boundary, as well as the region that can be probed by extremal surfaces anchored at the boundary.
For de Sitter we find that extremal surfaces can penetrate the event horizon, but only if their entangling region is super-horizon, i.e. larger than the cosmological horizon.
This implies that only a `super-observer' with access to information in a region larger than the observable universe could in principle be able to reconstruct the dual spacetime behind the event horizon from field theory data.

\begin{figure}[t]
\center
\includegraphics[height=0.4\linewidth]{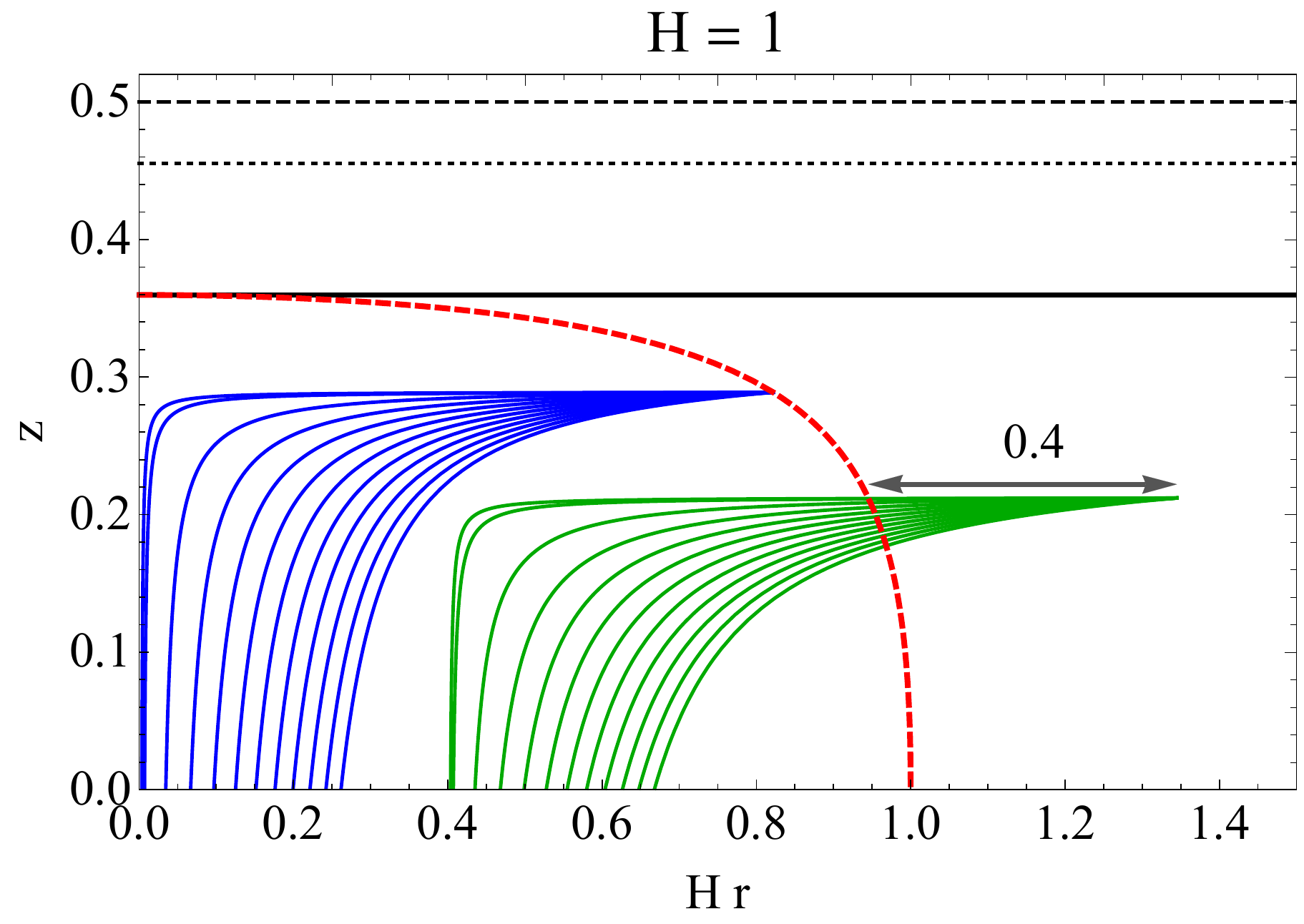}
\caption{
Blue and green curves are two families of null geodesics, originating on a generic point of the extremal surface with $z_\ast=z_{\mathrm{EH}}$ and a generic point in the bulk that is not enclosed by the extremal surface.
The fact that this family just reaches the observer at the origin implies that the extremal surface is the (bulk) cosmological horizon of an observer at the origin ($r=0$).
}
\label{Fig:CosmBulkHor}
\end{figure}

A physical explanation for this phenomenon is that the extremal surface corresponding to the observable universe in the boundary is itself a cosmological horizon in the bulk, this time of an observer at the origin on the boundary.
This is illustrated in Fig.~\ref{Fig:CosmBulkHor}, where we shoot a family of null geodesics (blue) from a generic point on the bulk extremal surface (dashed red).
This family is indeed just able to reach the origin at the boundary, which asserts our statement that this point is an element of the (bulk) cosmological horizon of an observer at the origin.
The green curves on the other hand represent a family of null geodesics that originate from a generic point beyond the cosmological horizon and are therefore unable to reach the origin. 
The starting point of this second family of geodesics is then element of a cosmological horizon in the bulk for an observer located at $H r = 0.4$.
We stress that the origin is not a special place, but only defined as the origin of the entangling region used in this example.

\begin{figure}[t]
\center
\includegraphics[height=0.32\linewidth]{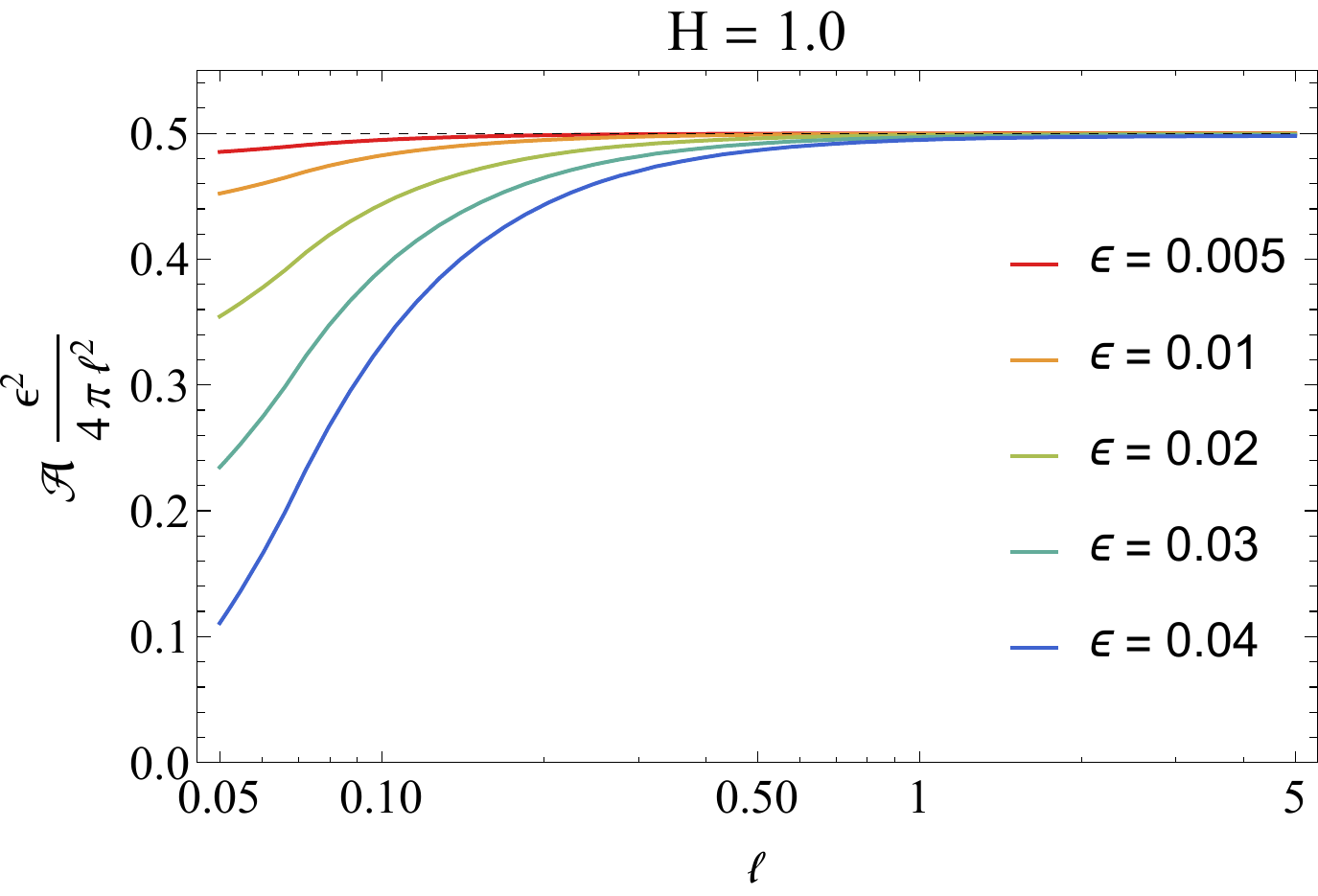}\quad
\includegraphics[height=0.32\linewidth]{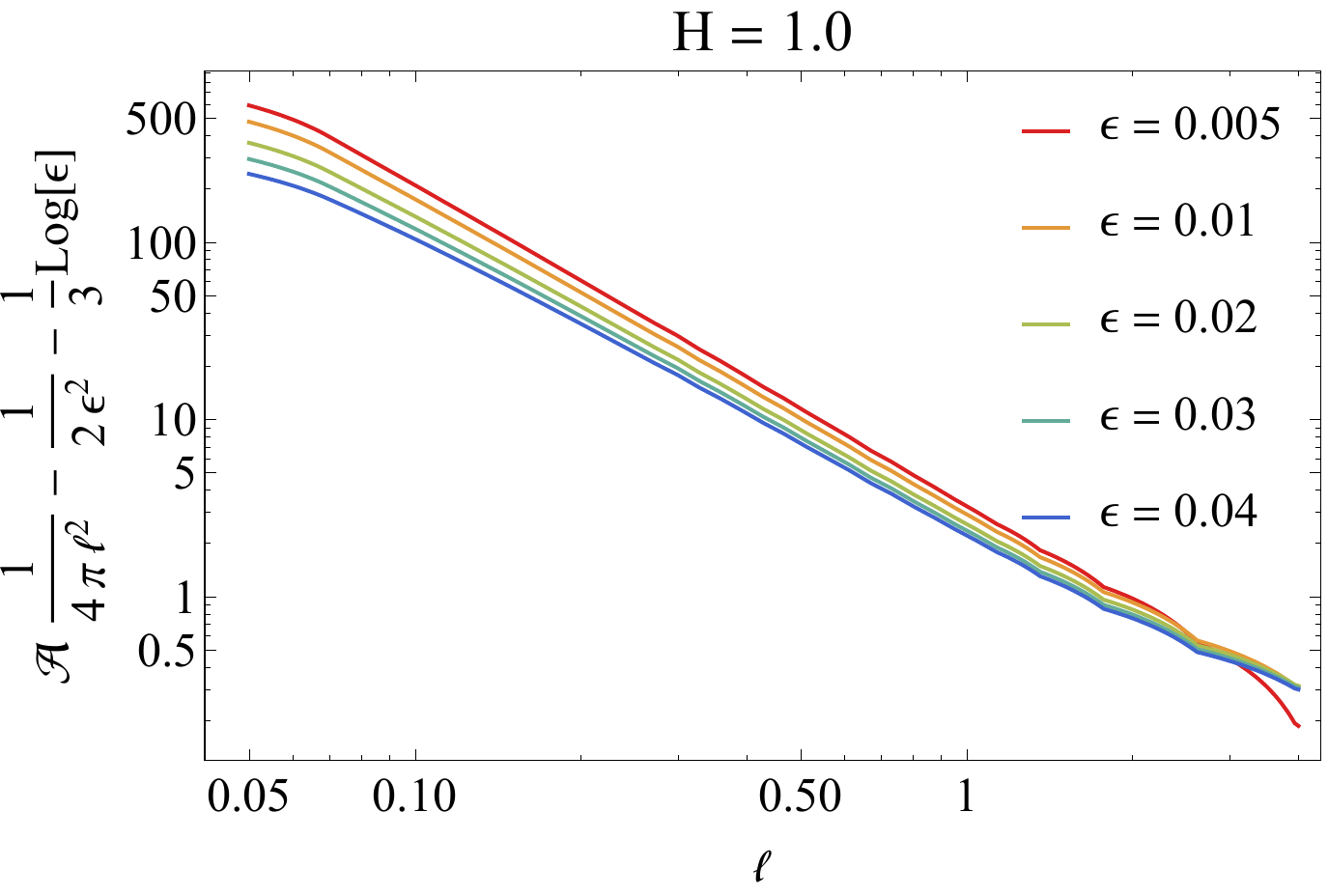}
\caption{Entanglement entropy for our $H=1.0$ (left), rescaled by the area and cut-off $\epsilon^2$, together with the leading order divergence subtracted (right), for different values of the cut-off.}
\label{Fig:EEareas}
\end{figure}

\begin{figure}[t]
\center
\includegraphics[height=0.4\linewidth]{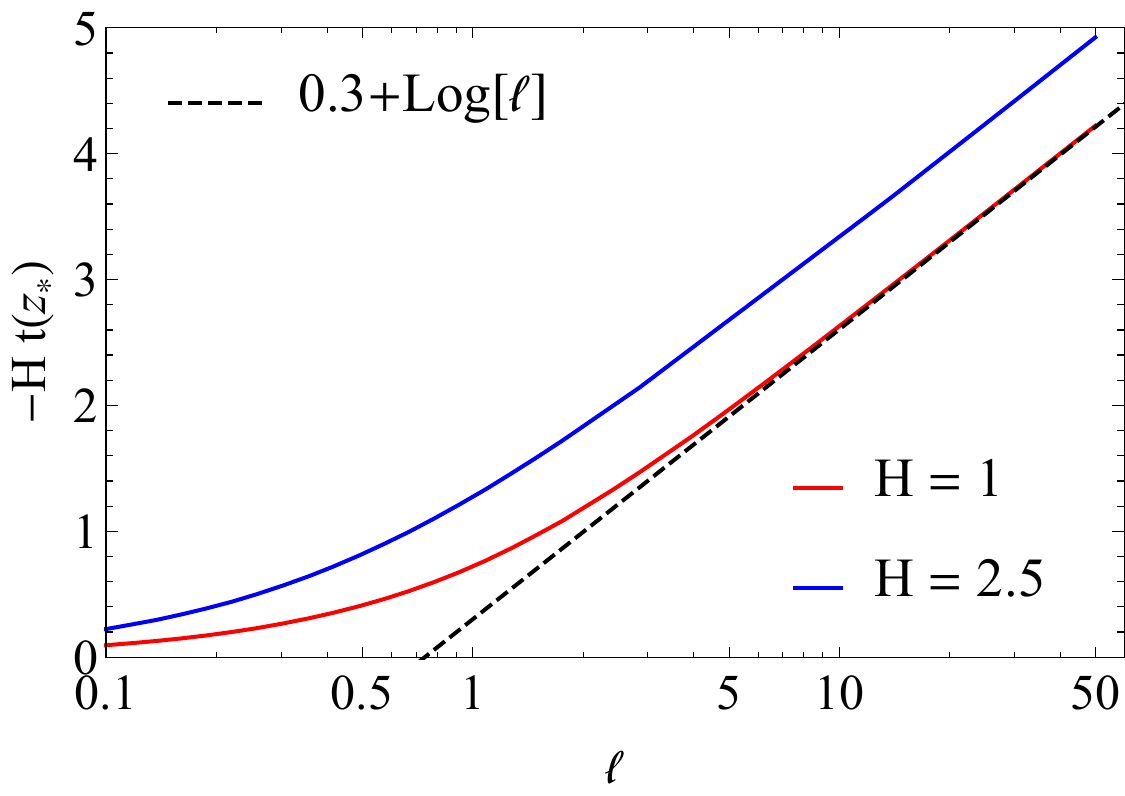}
\caption{
Minimum time reached on the extremal surface (see also Fig.~\ref{Fig:extremalsurfaces}) as a function of length. 
At large $\ell$ this time grows logarithmically with $\ell$.
}
\label{Fig:tversusL}
\end{figure}

Fig.~\ref{Fig:EEareas} shows the areas associated to the extremal surfaces for various values of the cut-off $\epsilon$.
Asymptotically the area equals \cite{Maldacena:2012xp}
\begin{equation}
    \mathcal{A} =4\pi \ell^2\left(\frac{1}{2\epsilon^2} + \frac{1}{3}\log(\epsilon) + O(1)\right) + O(\log(\ell)).
\end{equation}
The leading divergence is clear from the left figure, whereas the subtracted version (right) shows the difference with the leading order divergence.
Unfortunately it is numerically difficult to extract higher order coefficients that depend on the non-conformality of the theory. 
Naively from Fig.~\ref{Fig:extremalsurfaces} it may seem that for large $\ell$ we should obtain a volume law scaling, by the usual argument that the extremal surface hovers at the entanglement horizon and hence gets a contribution proportional to $S(z_{\rm Ent},\,t)^3$ times the volume of the region. 
For time independent geometries this is indeed the case, but interestingly our extremal surfaces have a non-trivial time dependence through $S(z,\,t)=S_0(z)a(t(r))$ and $t(r)$.
In fact, as illustrated in Fig.~\ref{Fig:extremalsurfaces} and more explicitly in Fig.~\ref{Fig:tversusL} the time as the surface hovers the horizon is for $H=1$ approximately given by $t_{\rm Ent}\approx -0.3 - \log(\ell)$. 
In combination with $a(t)=e^{H t}$ and $S_0(z_{Ent})\approx 0.69$ this implies that the part of the surface\footnote{In the following we assume that this is the entire surface, which is valid for $\ell\gg 1/H$.} hovering the entanglement horizon has an area $\frac{4}{3}\pi\ell^3 0.134 / \ell^3 \approx 0.559$ which is only a constant term as opposed to the usual volume scaling of the time independent setting.

\section{Discussion}
\label{sec:discussion}
We have used a holographic model to study the dynamics of a strongly coupled non-conformal gauge theory in four-dimensional de Sitter space.
The four-dimensional dS metric is prescribed a priori and is probed by the strongly coupled matter.
In other words, the five-dimensional gravitational model provides a dual description of the dynamics of the gauge theory matter but not of the four-dimensional gravitational field on which this matter propagates. 

We have carefully explained the holographic renormalisation of the model and the anomalies that arise in the dual field theory due to the curved boundary geometry and the scalar field.
After reviewing thermodynamic and transport properties of the model in flat space, we have presented numerical results for the fully non-linear time evolution of finite-temperature states towards the Bunch-Davis vacuum at late times.
We have shown that the approach to the de Sitter vacuum is characterised by an effective relation of the form $\calP=w \calE$.
This is different from the equilibrium equation of state in flat space, so much so that actually $w$ is negative if the ratio $H/M$ is close to unity.

We have studied in detail the properties of the de Sitter vacua of the holographic model and we have analyzed the different horizons that arise in the bulk geometry.
The connection between event horizons and thermodynamics found for black holes \cite{Hawking:1974sw} also applies to cosmological horizons \cite{PhysRevD.15.2738}.
Therefore, in analogy with the Bekenstein-Hawking temperature of black holes \cite{PhysRevD.7.2333,Hawking:1974sw}, an observer living at the boundary would associate a temperature to the cosmological horizon of de Sitter space
\begin{equation}\label{eq:TdS}
T_\mathrm{dS}=\frac{\kappa_{\mathrm{dS}}}{2\pi}=\frac{H}{2\pi}\,,
\end{equation}
where the surface gravity $\kappa_{\mathrm{dS}}$ is evaluated at the horizon and equals the Hubble rate $H$.
Interestingly, we found two temperatures in our five-dimensional dual description: one at the bulk event horizon, equal to the result by Hawking and Gibbons of $H/2\pi$, and another temperature at the deeper apparent horizon, equal to $-H/2\pi$ (also found in \cite{Buchel:2019qcq}).
In the literature there are several works \cite{Klemm:2004mb,Visser:2019muv,Jacobson:2019gco} suggesting such a negative temperature based on the first law.
In our case the apparent horizon is however causally disconnected from the boundary as well as time-slicing dependent, which suggests that indeed the positive temperature of the event horizon is the physical temperature.

In analogy with the Bekenstein-Hawking law, which relates the entropy of a black hole to the area of its event horizon, one can also associate a gravitational  entropy to de Sitter space
\begin{equation}\label{eq:SdS}
S_\mathrm{dS}=\frac{A_\mathrm{dS}}{4 G_4}=\frac{\pi}{H^2G_4}\,,
\end{equation}
where $A_\mathrm{dS}=4\pi/H^2$ is the area of the cosmological horizon and $G_4$ is the Newton's constant of the four-dimensional boundary theory.
Note that this entropy is formally infinite in our case since the boundary metric is non-dynamical and hence implicitly we are setting $G_4=0$.  
Therefore the gravitational entropy \eqref{eq:SdS} should not be confused with the entropy that one may want to assign to the area densities of the event, entanglement and apparent horizons that we studied in sections \eqref{sec:latetimesol} and \eqref{sec:entropy0}. 
In principle these would be related to the entropy of the matter in de Sitter space, but this relation is not straight-forward.
For a stationary geometry the three areas agree and can be identified with the entropy density of the boundary gauge theory.
Firstly, in our expanding geometry the horizons do not coincide with one another, and furthermore the mapping between points at the horizon and at the boundary is ambiguous.
Secondly, such an entropy density interpretation suggests a volume law, whereas we showed in Fig.~\ref{Fig:EEareas} in combination with Fig.~\ref{Fig:tversusL} that for large regions the entanglement horizon contribution to the entanglement entropy is just a constant term.
The divergent piece of the entanglement entropy satisfies an area law, which prohibits a direct extraction of the IR part of the entropy (see also \cite{Maldacena:2012xp}).
It is hence difficult to have a direct interpretation of the entropy of de Sitter itself, but we note that in a theory with dynamical gravity it is conjectured that this entropy is limited by a Bekenstein-Hawking term of $A/4 G_{4}$.
This entropy, or part thereof, can then potentially be identified with the entanglement entropy whereby $1/G_{4}$ plays the role of the UV cut-off \cite{Susskind:1994sm}.

A further result of our analysis of the entanglement entropy of boundary regions was that the extremal surfaces corresponding to entangling regions that coincide with the boundary observable universe exactly touch the bulk event horizon and are in fact a bulk cosmological horizon. 
It would be interesting to understand analytically why the extremal surface associated with the boundary cosmological horizon is itself a bulk cosmological horizon.

Our analysis of perturbations around the late-time state showed that, after a quench, the state relaxes within a time $1/T \sim 1/H$ (see  Fig.~\ref{Fig:qnm} and \ref{Fig:universalityquench}), in agreement with \cite{Buchel:2017lhu}.
In our case this time can be a parametrically different from $\Delta\mathcal{E}^{1/4}$.
This hence gives further credibility that the de Sitter temperature provides a physical temperature. 
On the holographic side this can be understood by the fact that the relaxation time is determined by the distance between the boundary and the event horizon, which is indeed proportional to $1/H$.

For our late-time solution the energy density excess over the asymptotic  late-time solution decreases exponentially. 
As a consequence, at the time when the negative excess pressure becomes relevant the energy density excess $\Delta \mathcal{E}$ will quickly become smaller than $T^4$, with $T\sim H$ the background de Sitter temperature (note that this temperature is the minimal temperature, accelerating observers will see an even higher temperature \cite{Narnhofer:1996zk}).
This raises the question of whether  the energy and pressure excesses in this regime can be measured by an actual observer, since the relevant modes will have wavelengths larger than the observable Universe.

\acknowledgments
We thank Alex Buchel and Michal Heller for their comments on the manuscript. JCS and DM are supported by grants FPA2016-76005-C2-1-P, FPA2016-76005-C2-2-P, 2014-SGR-104, 2014-SGR-1474, SGR-2017-754, MDM-2014-0369, PID2019-105614GB-C21, PID2019-105614GB-C22. They also acknowledge financial support from the State Agency for Research of the Spanish Ministry of Science and Innovation through the “Unit of Excellence María de Maeztu 2020-2023” award to the Institute of Cosmos Sciences (CEX2019-000918-M).

\bibliography{deSitter.bib}{}

\providecommand{\href}[2]{#2}\begingroup\raggedright\begin{thebibliography}{10}

\bibitem{CasalderreySolana:2011us}
J.~Casalderrey-Solana, H.~Liu, D.~Mateos, K.~Rajagopal and U.~A. Wiedemann,
  \emph{{Gauge/String Duality, Hot QCD and Heavy Ion Collisions}}. Cambridge
  University Press, 2014,
  \href{https://doi.org/10.1017/CBO9781139136747}{10.1017/CBO9781139136747},
  [\href{https://arxiv.org/abs/1101.0618}{{\ttfamily 1101.0618}}].

\bibitem{Busza:2018rrf}
W.~Busza, K.~Rajagopal and W.~van~der Schee, \emph{{Heavy Ion Collisions: The
  Big Picture, and the Big Questions}},
  \href{https://doi.org/10.1146/annurev-nucl-101917-020852}{\emph{Ann. Rev.
  Nucl. Part. Sci.} {\bfseries 68} (2018) 339}
  [\href{https://arxiv.org/abs/1802.04801}{{\ttfamily 1802.04801}}].

\bibitem{Heller:2011ju}
M.~P. Heller, R.~A. Janik and P.~Witaszczyk, \emph{{The characteristics of
  thermalization of boost-invariant plasma from holography}},
  \href{https://doi.org/10.1103/PhysRevLett.108.201602}{\emph{Phys. Rev. Lett.}
  {\bfseries 108} (2012) 201602}
  [\href{https://arxiv.org/abs/1103.3452}{{\ttfamily 1103.3452}}].

\bibitem{Chesler:2010bi}
P.~M. Chesler and L.~G. Yaffe, \emph{{Holography and colliding gravitational
  shock waves in asymptotically AdS$_{5}$ spacetime}},
  \href{https://doi.org/10.1103/PhysRevLett.106.021601}{\emph{Phys. Rev. Lett.}
  {\bfseries 106} (2011) 021601}
  [\href{https://arxiv.org/abs/1011.3562}{{\ttfamily 1011.3562}}].

\bibitem{Casalderrey-Solana:2013aba}
J.~Casalderrey-Solana, M.~P. Heller, D.~Mateos and W.~van~der Schee,
  \emph{{From full stopping to transparency in a holographic model of heavy ion
  collisions}},
  \href{https://doi.org/10.1103/PhysRevLett.111.181601}{\emph{Phys. Rev. Lett.}
  {\bfseries 111} (2013) 181601}
  [\href{https://arxiv.org/abs/1305.4919}{{\ttfamily 1305.4919}}].

\bibitem{Aoki:2006we}
Y.~Aoki, G.~Endrodi, Z.~Fodor, S.~Katz and K.~Szabo, \emph{{The Order of the
  quantum chromodynamics transition predicted by the standard model of particle
  physics}}, \href{https://doi.org/10.1038/nature05120}{\emph{Nature}
  {\bfseries 443} (2006) 675}
  [\href{https://arxiv.org/abs/hep-lat/0611014}{{\ttfamily hep-lat/0611014}}].

\bibitem{Schwaller:2015tja}
P.~Schwaller, \emph{{Gravitational Waves from a Dark Phase Transition}},
  \href{https://doi.org/10.1103/PhysRevLett.115.181101}{\emph{Phys. Rev. Lett.}
  {\bfseries 115} (2015) 181101}
  [\href{https://arxiv.org/abs/1504.07263}{{\ttfamily 1504.07263}}].

\bibitem{Caprini:2015zlo}
C.~Caprini et~al., \emph{{Science with the space-based interferometer eLISA.
  II: Gravitational waves from cosmological phase transitions}},
  \href{https://doi.org/10.1088/1475-7516/2016/04/001}{\emph{JCAP} {\bfseries
  04} (2016) 001} [\href{https://arxiv.org/abs/1512.06239}{{\ttfamily
  1512.06239}}].

\bibitem{Kribs:2016cew}
G.~D. Kribs and E.~T. Neil, \emph{{Review of strongly-coupled composite dark
  matter models and lattice simulations}},
  \href{https://doi.org/10.1142/S0217751X16430041}{\emph{Int. J. Mod. Phys. A}
  {\bfseries 31} (2016) 1643004}
  [\href{https://arxiv.org/abs/1604.04627}{{\ttfamily 1604.04627}}].

\bibitem{Tulin:2017ara}
S.~Tulin and H.-B. Yu, \emph{{Dark Matter Self-interactions and Small Scale
  Structure}}, \href{https://doi.org/10.1016/j.physrep.2017.11.004}{\emph{Phys.
  Rept.} {\bfseries 730} (2018) 1}
  [\href{https://arxiv.org/abs/1705.02358}{{\ttfamily 1705.02358}}].

\bibitem{Strominger:2001pn}
A.~Strominger, \emph{{The dS / CFT correspondence}},
  \href{https://doi.org/10.1088/1126-6708/2001/10/034}{\emph{JHEP} {\bfseries
  10} (2001) 034} [\href{https://arxiv.org/abs/hep-th/0106113}{{\ttfamily
  hep-th/0106113}}].

\bibitem{McFadden:2009fg}
P.~McFadden and K.~Skenderis, \emph{{Holography for Cosmology}},
  \href{https://doi.org/10.1103/PhysRevD.81.021301}{\emph{Phys. Rev. D}
  {\bfseries 81} (2010) 021301}
  [\href{https://arxiv.org/abs/0907.5542}{{\ttfamily 0907.5542}}].

\bibitem{Koyama:2001rf}
K.~Koyama and J.~Soda, \emph{{Strongly coupled CFT in FRW universe from AdS /
  CFT correspondence}},
  \href{https://doi.org/10.1088/1126-6708/2001/05/027}{\emph{JHEP} {\bfseries
  05} (2001) 027} [\href{https://arxiv.org/abs/hep-th/0101164}{{\ttfamily
  hep-th/0101164}}].

\bibitem{Marolf:2010tg}
D.~Marolf, M.~Rangamani and M.~Van~Raamsdonk, \emph{{Holographic models of de
  Sitter QFTs}},
  \href{https://doi.org/10.1088/0264-9381/28/10/105015}{\emph{Class. Quant.
  Grav.} {\bfseries 28} (2011) 105015}
  [\href{https://arxiv.org/abs/1007.3996}{{\ttfamily 1007.3996}}].

\bibitem{Ghoroku:2012vi}
K.~Ghoroku and A.~Nakamura, \emph{{Holographic Friedmann equation and N=4
  supersymmetric Yang-Mills theory}},
  \href{https://doi.org/10.1103/PhysRevD.87.063507}{\emph{Phys. Rev. D}
  {\bfseries 87} (2013) 063507}
  [\href{https://arxiv.org/abs/1212.2304}{{\ttfamily 1212.2304}}].

\bibitem{Buchel:2016cbj}
A.~Buchel, M.~P. Heller and J.~Noronha, \emph{{Entropy Production,
  Hydrodynamics, and Resurgence in the Primordial Quark-Gluon Plasma from
  Holography}}, \href{https://doi.org/10.1103/PhysRevD.94.106011}{\emph{Phys.
  Rev. D} {\bfseries 94} (2016) 106011}
  [\href{https://arxiv.org/abs/1603.05344}{{\ttfamily 1603.05344}}].

\bibitem{Buchel:2017lhu}
A.~Buchel, \emph{{Ringing in de Sitter spacetime}},
  \href{https://doi.org/10.1016/j.nuclphysb.2018.01.021}{\emph{Nucl. Phys. B}
  {\bfseries 928} (2018) 307}
  [\href{https://arxiv.org/abs/1707.01030}{{\ttfamily 1707.01030}}].

\bibitem{Buchel:2017pto}
A.~Buchel and A.~Karapetyan, \emph{{de Sitter Vacua of Strongly Interacting
  QFT}}, \href{https://doi.org/10.1007/JHEP03(2017)114}{\emph{JHEP} {\bfseries
  03} (2017) 114} [\href{https://arxiv.org/abs/1702.01320}{{\ttfamily
  1702.01320}}].

\bibitem{Buchel:2019qcq}
A.~Buchel, \emph{{Entanglement entropy of ${\cal N}=2^*$ de Sitter vacuum}},
  \href{https://doi.org/10.1016/j.nuclphysb.2019.114769}{\emph{Nucl. Phys. B}
  {\bfseries 948} (2019) 114769}
  [\href{https://arxiv.org/abs/1904.09968}{{\ttfamily 1904.09968}}].

\bibitem{Buchel:2019pjb}
A.~Buchel, \emph{{$\chi\rm{SB}$ of cascading gauge theory in de Sitter}},
  \href{https://doi.org/10.1007/JHEP05(2020)035}{\emph{JHEP} {\bfseries 05}
  (2020) 035} [\href{https://arxiv.org/abs/1912.03566}{{\ttfamily
  1912.03566}}].

\bibitem{Apostolopoulos_2009}
P.~S. Apostolopoulos, G.~Siopsis and N.~Tetradis, \emph{Cosmology from an
  anti-de sitter–schwarzschild black hole via holography},
  \href{https://doi.org/10.1103/physrevlett.102.151301}{\emph{Physical Review
  Letters} {\bfseries 102} (2009) }.

\bibitem{Attems:2016ugt}
M.~Attems, J.~Casalderrey-Solana, D.~Mateos, I.~Papadimitriou,
  D.~Santos-Oliv\'an, C.~F. Sopuerta et~al., \emph{{Thermodynamics, transport
  and relaxation in non-conformal theories}},
  \href{https://doi.org/10.1007/JHEP10(2016)155}{\emph{JHEP} {\bfseries 10}
  (2016) 155} [\href{https://arxiv.org/abs/1603.01254}{{\ttfamily
  1603.01254}}].

\bibitem{deHaro:2000vlm}
S.~de~Haro, S.~N. Solodukhin and K.~Skenderis, \emph{{Holographic
  reconstruction of space-time and renormalization in the AdS / CFT
  correspondence}}, \href{https://doi.org/10.1007/s002200100381}{\emph{Commun.
  Math. Phys.} {\bfseries 217} (2001) 595}
  [\href{https://arxiv.org/abs/hep-th/0002230}{{\ttfamily hep-th/0002230}}].

\bibitem{Bianchi:2001de}
M.~Bianchi, D.~Z. Freedman and K.~Skenderis, \emph{{How to go with an RG
  flow}}, \href{https://doi.org/10.1088/1126-6708/2001/08/041}{\emph{JHEP}
  {\bfseries 08} (2001) 041}
  [\href{https://arxiv.org/abs/hep-th/0105276}{{\ttfamily hep-th/0105276}}].

\bibitem{Bianchi:2001kw}
M.~Bianchi, D.~Z. Freedman and K.~Skenderis, \emph{{Holographic
  renormalization}},
  \href{https://doi.org/10.1016/S0550-3213(02)00179-7}{\emph{Nucl. Phys. B}
  {\bfseries 631} (2002) 159}
  [\href{https://arxiv.org/abs/hep-th/0112119}{{\ttfamily hep-th/0112119}}].

\bibitem{Gubser:2008ny}
S.~S. Gubser and A.~Nellore, \emph{{Mimicking the QCD equation of state with a
  dual black hole}},
  \href{https://doi.org/10.1103/PhysRevD.78.086007}{\emph{Phys. Rev.}
  {\bfseries D78} (2008) 086007}
  [\href{https://arxiv.org/abs/0804.0434}{{\ttfamily 0804.0434}}].

\bibitem{Kovtun:2004de}
P.~Kovtun, D.~T. Son and A.~O. Starinets, \emph{{Viscosity in strongly
  interacting quantum field theories from black hole physics}},
  \href{https://doi.org/10.1103/PhysRevLett.94.111601}{\emph{Phys. Rev. Lett.}
  {\bfseries 94} (2005) 111601}
  [\href{https://arxiv.org/abs/hep-th/0405231}{{\ttfamily hep-th/0405231}}].

\bibitem{Eling:2011ms}
C.~Eling and Y.~Oz, \emph{{A Novel Formula for Bulk Viscosity from the Null
  Horizon Focusing Equation}},
  \href{https://doi.org/10.1007/JHEP06(2011)007}{\emph{JHEP} {\bfseries 06}
  (2011) 007} [\href{https://arxiv.org/abs/1103.1657}{{\ttfamily 1103.1657}}].

\bibitem{romatschke_romatschke_2019}
P.~Romatschke and U.~Romatschke, \emph{Relativistic Fluid Dynamics In and Out
  of Equilibrium: And Applications to Relativistic Nuclear Collisions},
  Cambridge Monographs on Mathematical Physics. Cambridge University Press,
  2019, \href{https://doi.org/10.1017/9781108651998}{10.1017/9781108651998}.

\bibitem{Kovtun:2019hdm}
P.~Kovtun, \emph{{First-order relativistic hydrodynamics is stable}},
  \href{https://doi.org/10.1007/JHEP10(2019)034}{\emph{JHEP} {\bfseries 10}
  (2019) 034} [\href{https://arxiv.org/abs/1907.08191}{{\ttfamily
  1907.08191}}].

\bibitem{Chesler:2013lia}
P.~M. Chesler and L.~G. Yaffe, \emph{{Numerical solution of gravitational
  dynamics in asymptotically anti-de Sitter spacetimes}},
  \href{https://doi.org/10.1007/JHEP07(2014)086}{\emph{JHEP} {\bfseries 07}
  (2014) 086} [\href{https://arxiv.org/abs/1309.1439}{{\ttfamily 1309.1439}}].

\bibitem{vanderSchee:2014qwa}
W.~van~der Schee, \emph{{Gravitational collisions and the quark-gluon plasma}},
  Ph.D. thesis, Utrecht U., 2014.
\newblock \href{https://arxiv.org/abs/1407.1849}{{\ttfamily 1407.1849}}.

\bibitem{Ecker:2018jgh}
C.~Ecker, \emph{{Entanglement Entropy from Numerical Holography}}, Ph.D.
  thesis, Vienna, Tech. U., 9, 2018.
\newblock \href{https://arxiv.org/abs/1809.05529}{{\ttfamily 1809.05529}}.

\bibitem{boyd01}
J.~P. Boyd, \emph{{Chebyshev} and {Fourier} Spectral Methods}, Dover Books on
  Mathematics. Dover Publications, Mineola, NY, second~ed., 2001.

\bibitem{press2007numerical}
W.~Press, S.~Teukolsky, W.~Vetterling and B.~Flannery, \emph{Numerical Recipes:
  The Art of Scientific Computing}. Cambridge University Press, 3~ed., 2007.

\bibitem{Chesler:2009cy}
P.~M. Chesler and L.~G. Yaffe, \emph{{Boost invariant flow, black hole
  formation, and far-from-equilibrium dynamics in N = 4 supersymmetric
  Yang-Mills theory}},
  \href{https://doi.org/10.1103/PhysRevD.82.026006}{\emph{Phys. Rev. D}
  {\bfseries 82} (2010) 026006}
  [\href{https://arxiv.org/abs/0906.4426}{{\ttfamily 0906.4426}}].

\bibitem{Buchel:2015saa}
A.~Buchel, M.~P. Heller and R.~C. Myers, \emph{{Equilibration rates in a
  strongly coupled nonconformal quark-gluon plasma}},
  \href{https://doi.org/10.1103/PhysRevLett.114.251601}{\emph{Phys. Rev. Lett.}
  {\bfseries 114} (2015) 251601}
  [\href{https://arxiv.org/abs/1503.07114}{{\ttfamily 1503.07114}}].

\bibitem{Ghosh:2017big}
J.~K. Ghosh, E.~Kiritsis, F.~Nitti and L.~T. Witkowski, \emph{{Holographic RG
  flows on curved manifolds and quantum phase transitions}},
  \href{https://doi.org/10.1007/JHEP05(2018)034}{\emph{JHEP} {\bfseries 05}
  (2018) 034} [\href{https://arxiv.org/abs/1711.08462}{{\ttfamily
  1711.08462}}].

\bibitem{Bunch:1978yq}
T.~Bunch and P.~Davies, \emph{{Quantum Field Theory in de Sitter Space:
  Renormalization by Point Splitting}},
  \href{https://doi.org/10.1098/rspa.1978.0060}{\emph{Proc. Roy. Soc. Lond. A}
  {\bfseries 360} (1978) 117}.

\bibitem{Apostolopoulos:2008ru}
P.~S. Apostolopoulos, G.~Siopsis and N.~Tetradis, \emph{{Cosmology from an AdS
  Schwarzschild black hole via holography}},
  \href{https://doi.org/10.1103/PhysRevLett.102.151301}{\emph{Phys. Rev. Lett.}
  {\bfseries 102} (2009) 151301}
  [\href{https://arxiv.org/abs/0809.3505}{{\ttfamily 0809.3505}}].

\bibitem{hawking1972}
S.~W. Hawking, \emph{Black holes in general relativity}, {\emph{Comm. Math.
  Phys.} {\bfseries 25} (1972) 152}.

\bibitem{Holzhey:1994we}
C.~Holzhey, F.~Larsen and F.~Wilczek, \emph{{Geometric and renormalized entropy
  in conformal field theory}},
  \href{https://doi.org/10.1016/0550-3213(94)90402-2}{\emph{Nucl. Phys. B}
  {\bfseries 424} (1994) 443}
  [\href{https://arxiv.org/abs/hep-th/9403108}{{\ttfamily hep-th/9403108}}].

\bibitem{Maldacena:2012xp}
J.~Maldacena and G.~L. Pimentel, \emph{{Entanglement entropy in de Sitter
  space}}, \href{https://doi.org/10.1007/JHEP02(2013)038}{\emph{JHEP}
  {\bfseries 02} (2013) 038} [\href{https://arxiv.org/abs/1210.7244}{{\ttfamily
  1210.7244}}].

\bibitem{Calabrese:2004eu}
P.~Calabrese and J.~L. Cardy, \emph{{Entanglement entropy and quantum field
  theory}}, \href{https://doi.org/10.1088/1742-5468/2004/06/P06002}{\emph{J.
  Stat. Mech.} {\bfseries 0406} (2004) P06002}
  [\href{https://arxiv.org/abs/hep-th/0405152}{{\ttfamily hep-th/0405152}}].

\bibitem{Srednicki:1993im}
M.~Srednicki, \emph{{Entropy and area}},
  \href{https://doi.org/10.1103/PhysRevLett.71.666}{\emph{Phys. Rev. Lett.}
  {\bfseries 71} (1993) 666}
  [\href{https://arxiv.org/abs/hep-th/9303048}{{\ttfamily hep-th/9303048}}].

\bibitem{Ryu:2006bv}
S.~Ryu and T.~Takayanagi, \emph{{Holographic derivation of entanglement entropy
  from AdS/CFT}},
  \href{https://doi.org/10.1103/PhysRevLett.96.181602}{\emph{Phys. Rev. Lett.}
  {\bfseries 96} (2006) 181602}
  [\href{https://arxiv.org/abs/hep-th/0603001}{{\ttfamily hep-th/0603001}}].

\bibitem{Hubeny:2007xt}
V.~E. Hubeny, M.~Rangamani and T.~Takayanagi, \emph{{A Covariant holographic
  entanglement entropy proposal}},
  \href{https://doi.org/10.1088/1126-6708/2007/07/062}{\emph{JHEP} {\bfseries
  07} (2007) 062} [\href{https://arxiv.org/abs/0705.0016}{{\ttfamily
  0705.0016}}].

\bibitem{Ecker:2015kna}
C.~Ecker, D.~Grumiller and S.~A. Stricker, \emph{{Evolution of holographic
  entanglement entropy in an anisotropic system}},
  \href{https://doi.org/10.1007/JHEP07(2015)146}{\emph{JHEP} {\bfseries 07}
  (2015) 146} [\href{https://arxiv.org/abs/1506.02658}{{\ttfamily
  1506.02658}}].

\bibitem{Engelhardt:2013tra}
N.~Engelhardt and A.~C. Wall, \emph{{Extremal Surface Barriers}},
  \href{https://doi.org/10.1007/JHEP03(2014)068}{\emph{JHEP} {\bfseries 03}
  (2014) 068} [\href{https://arxiv.org/abs/1312.3699}{{\ttfamily 1312.3699}}].

\bibitem{Hubeny:2012ry}
V.~E. Hubeny, \emph{{Extremal surfaces as bulk probes in AdS/CFT}},
  \href{https://doi.org/10.1007/JHEP07(2012)093}{\emph{JHEP} {\bfseries 07}
  (2012) 093} [\href{https://arxiv.org/abs/1203.1044}{{\ttfamily 1203.1044}}].

\bibitem{Hawking:1974sw}
S.~Hawking, \emph{{Particle Creation by Black Holes}},
  \href{https://doi.org/10.1007/BF02345020}{\emph{Commun. Math. Phys.}
  {\bfseries 43} (1975) 199}.

\bibitem{PhysRevD.15.2738}
G.~W. Gibbons and S.~W. Hawking, \emph{Cosmological event horizons,
  thermodynamics, and particle creation},
  \href{https://doi.org/10.1103/PhysRevD.15.2738}{\emph{Phys. Rev. D}
  {\bfseries 15} (1977) 2738}.

\bibitem{PhysRevD.7.2333}
J.~D. Bekenstein, \emph{Black holes and entropy},
  \href{https://doi.org/10.1103/PhysRevD.7.2333}{\emph{Phys. Rev. D} {\bfseries
  7} (1973) 2333}.

\bibitem{Klemm:2004mb}
D.~Klemm and L.~Vanzo, \emph{{Aspects of quantum gravity in de Sitter spaces}},
  \href{https://doi.org/10.1088/1475-7516/2004/11/006}{\emph{JCAP} {\bfseries
  11} (2004) 006} [\href{https://arxiv.org/abs/hep-th/0407255}{{\ttfamily
  hep-th/0407255}}].

\bibitem{Visser:2019muv}
M.~R. Visser, \emph{{Emergent gravity in a holographic universe}}, Ph.D.
  thesis, Amsterdam U., 2019.
\newblock \href{https://arxiv.org/abs/1908.05469}{{\ttfamily 1908.05469}}.

\bibitem{Jacobson:2019gco}
T.~Jacobson and M.~Visser, \emph{{Spacetime Equilibrium at Negative Temperature
  and the Attraction of Gravity}},
  \href{https://doi.org/10.1142/S0218271819440164}{\emph{Int. J. Mod. Phys. D}
  {\bfseries 28} (2019) 1944016}
  [\href{https://arxiv.org/abs/1904.04843}{{\ttfamily 1904.04843}}].

\bibitem{Susskind:1994sm}
L.~Susskind and J.~Uglum, \emph{{Black hole entropy in canonical quantum
  gravity and superstring theory}},
  \href{https://doi.org/10.1103/PhysRevD.50.2700}{\emph{Phys. Rev. D}
  {\bfseries 50} (1994) 2700}
  [\href{https://arxiv.org/abs/hep-th/9401070}{{\ttfamily hep-th/9401070}}].

\bibitem{Narnhofer:1996zk}
H.~Narnhofer, I.~Peter and W.~E. Thirring, \emph{{How hot is the de Sitter
  space?}}, \href{https://doi.org/10.1142/S0217979296000611}{\emph{Int. J. Mod.
  Phys. B} {\bfseries 10} (1996) 1507}.

\end{thebibliography}\endgroup
\bibliographystyle{JHEP}
\end{document}